\documentclass[aps,pre,floats]{revtex4}
\usepackage{amssymb}
\usepackage{float}
\usepackage{makeidx}
\usepackage{amsmath}
\usepackage{graphicx}
\usepackage{amsthm}
\usepackage{hyperref}

\def\tt{\theta}

\input{tcilatex}

\begin{document}

\title{Density Matrix Equation for a Bathed Small System \linebreak\ and its
Application to Molecular Magnets}
\author{D. A. Garanin}
\affiliation{\mbox{Department of Physics and Astronomy, Lehman College, City
University of New York,} \\ \mbox{250 Bedford Park Boulevard
West, Bronx, New York 10468-1589, U.S.A.} }
\date{\today}

\begin{abstract}
The technique of density matrix equation (DME) for a small system interacting with a bath is explained in detail. Special attention is given to the nonsecular DME that is needed in the vicinity of overdamped tunnelling resonances in molecular magnets (MM). The relaxation terms of the DME for MM are represented in the universal form that does not employ any unknown spin-lattice coupling constants and absorbs the information about the spin Hamiltonian in the exact basis states and transition frequencies. This makes adding new types of anisotropy easy and error-free. The Mathematica code is available from the author.
\end{abstract}

\maketitle
\tableofcontents

\section{General theory}


Density matrix is used to describe properties of a system that is a part of
a larger system with which it interacts. Whereas isolated systems (such as
the above mentioned larger system) can be described by a Schr\"{o}dinger
equation, systems which interact with their environments cannot. Starting
from the Schr\"{o}dinger equation for the isolated whole system, small
system + environment (or \emph{bath}), and eliminating the environmental
variables, one can, in principle, construct an object that can be used to
calculate observables of the small system in a short way. This object is the
density matrix of the small system. Of course, integrating or taking matrix
elements in two steps, at first over the environment and then over the small
system is not a big simplification. This approach becomes really useful if
one obtains a closed equation of motion for the density matrix of the small
system, the density matrix equation (DME). This is possible if the
interaction between the small system and its environment is small and can be
considered as a perturbation, and the small system does not strongly perturb
the state of the environment. The derivation of the DME for a bathed small
system will be presented in Sec. \ref{Sec-DME}. Here the necessary components
 of the formalism will be introduced.

\subsection{From the wave function to the density matrix}

The general wave function or state $|\psi \rangle $ of an isolated quantum
system can be expanded over a set of complete basis states $|\Psi
_{m}\rangle $ as
\begin{equation}
|\Psi \rangle =\sum_{m}c_{m}|\Psi _{m}\rangle .  \label{DME-psiexpansion}
\end{equation}
The coefficients $c_{m}$ completely characterize the state $|\Psi \rangle $
and can be used to calculate physical quantities $A$ described by
corresponding operators $\hat{A}$:
\begin{equation}
A\equiv \langle \hat{A}\rangle \equiv \langle \Psi |\hat{A}|\Psi \rangle
=\sum_{mn}c_{m}c_{n}^{\ast }\langle \Psi _{n}|\hat{A}|\Psi _{m}\rangle
=\sum_{mn}c_{m}c_{n}^{\ast }A_{nm}.  \label{DME-Aavr}
\end{equation}
One can define the \emph{density matrix} $\mathbf{\rho }$ corresponding to
quantum-mechanical states (pure states) of our small system by its matrix
elements as
\begin{equation}
\left\{ \mathbf{\rho }\right\} _{mn}=\rho _{mn}=c_{m}c_{n}^{\ast },
\label{DME-rhomunuDefcmucnu}
\end{equation}
then Eq. (\ref{DME-Aavr}) becomes
\begin{equation}
A=\sum_{mn}\rho _{mn}A_{nm}.  \label{DME-Aavrrho}
\end{equation}
The density matrix satisfies the normalization condition
\begin{equation}
\mathrm{Tr}\left\{ \mathbf{\rho }\right\} \mathbf{\equiv }\sum_{m}\rho
_{mm}=1.  \label{DME-rhoNorm}
\end{equation}
This condition follows from Eq. (\ref{DME-rhomunuDefcmucnu}) for the pure
states but it holds in general, too, since the average of the unity operator
$A_{mn}=\delta _{mn}$ should be 1. Additionally, the condition
\begin{equation}
\sum_{mn}\left| \rho _{mn}\right| ^{2}=1  \label{DME-rho2Norm}
\end{equation}
is satisfied for the density partix of pure states, Eq. (\ref
{DME-rhomunuDefcmucnu}).

For systems interacting with their environment, observables $A$ still are
given by Eq. (\ref{DME-Aavrrho}), although the coefficients $\rho _{nm}$ in
general do not reduce to products as in Eq. (\ref{DME-rhomunuDefcmucnu}).
For the whole system including the environment, one can use basis states
that are direct products of those of the small system $|\psi _{m}\rangle $
and those of the environment $|\phi _{\varpi }\rangle $:
\begin{equation}
|\Psi _{m\varpi }\rangle =|\psi _{m}\rangle \otimes |\phi _{\varpi }\rangle
\equiv |\psi _{m}\phi _{\varpi }\rangle .  \label{DME-PsiBasis}
\end{equation}
The quantum mechanical states of the whole system (considered as isolated)
can be written, similarly to Eq. (\ref{DME-psiexpansion}), in the form
\begin{equation}
|\Psi \rangle =\sum_{m\varpi }C_{m\varpi }|\Psi _{m\varpi }\rangle .
\label{DME-PsiExpansion}
\end{equation}
The expression for the observable $A$ of the small system becomes
\begin{equation}
A=\langle \Psi |\hat{A}|\Psi \rangle =\sum_{m\varpi ,n\varpi ^{\prime
}}C_{m\varpi }^{\ast }C_{n\varpi ^{\prime }}\langle \Psi _{m\varpi }|\hat{A}%
|\Psi _{n\varpi ^{\prime }}\rangle =\sum_{m\varpi ,n\varpi ^{\prime
}}C_{m\varpi }^{\ast }C_{n\varpi ^{\prime }}\langle \psi _{m}|\hat{A}|\psi
_{n}\rangle \left\langle \phi _{\varpi }\right. |\phi _{\varpi ^{\prime
}}\rangle .  \label{DME-Aavrenv}
\end{equation}
For the orthonormal set of $|\phi _{\varpi }\rangle $ this can be rewritten
in the form of Eq. (\ref{DME-Aavrrho}) with $\rho _{mn}\Rightarrow \rho
_{mn}^{\mathrm{s}},$ where $\rho _{mn}^{\mathrm{s}}$ is the reduced density
matrix of the subsystem given by
\begin{equation}
\rho _{mn}^{\mathrm{s}}=\sum_{\varpi }C_{m\varpi }C_{n\varpi }^{\ast }\equiv
\sum_{\varpi }\rho _{m\varpi ,n\varpi }.  \label{DME-rhomunuenv}
\end{equation}
Since, in general, $C_{m\varpi }$ do not split into the factors depending on
$m$ and $\varpi ,$ the reduced DM $\rho _{mn}^{\mathrm{s}}$ is not a product
of its wave-function coefficients, $\rho _{mn}^{\mathrm{s}}\neq
c_{m}c_{n}^{\ast }.$ Obviously $\rho _{mn}^{\mathrm{s}}$ in Eq. (\ref
{DME-rhomunuenv}) depends on the state of the whole system. One can check
that it satisfies the normalization, Eq. (\ref{DME-rhoNorm}). On the other
hand, Eq. (\ref{DME-rho2Norm}) is not satisfied, in general.

\subsection{Entangled states and quantum statistics}

It would be wrong to think that when the interaction between the small
system and the environment becomes very weak, the density matrix of the
small system, Eq. (\ref{DME-rhomunuenv}), simplifies to Eq. (\ref
{DME-rhomunuDefcmucnu}). If the coupling is vanishingly weak, quantum states
of both the small system and the environment are well defined. Thus both the
small system and the environment could be in their pure quantum mechanical
states. However, there are pure states of the whole system that do not
consist of pure states of the two subsystems. Such states can be called
\emph{entangled}. For instance, if both subsystems are two-level systems
with the states $|\psi _{1}\rangle ,$ $|\psi _{2}\rangle $ and $|\phi
_{1}\rangle ,$ $|\phi _{2}\rangle ,$ respectively, and the whole system is
in the state
\begin{equation}
|\Psi \rangle =C_{11}|\psi _{1}\phi _{1}\rangle +C_{22}|\psi _{2}\phi
_{2}\rangle  \label{DME-Entanglement-example-Psi}
\end{equation}
with $\left| C_{11}\right| ^{2}+\left| C_{22}\right| ^{2}=1,$ Eq. (\ref
{DME-rhomunuenv}) with $C_{m\varpi }=C_{mm}\delta _{m\varpi }$ yields the
diagonal density matrix
\begin{equation}
\rho _{mn}^{\mathrm{s}}=\left| C_{mm}\right| ^{2}\delta _{mn}.
\label{DME-Entanglement-example-rhomunu}
\end{equation}
This density matrix satisfies Eq. (\ref{DME-rhoNorm}). On the other hand,
diagonal $\rho _{mn}^{\mathrm{s}}$ does not correspond to any pure state,
cf. Eq. (\ref{DME-rhomunuDefcmucnu}). Thus Eq. (\ref{DME-rho2Norm}) is not
satisfied,
\begin{equation}
\sum_{mn}\left| \rho _{mn}^{\mathrm{s}}\right| ^{2}=\sum_{m}\left|
C_{mm}\right| ^{4}\leq 1.  \label{DME-Entanglement-example-2Norm}
\end{equation}
In particular, for $\left| C_{11}\right| ^{2}=\left| C_{22}\right| ^{2}=1/2$
the result is 1/2 instead of 1. This is unrelated to the interaction between
the two subsystems and it alone mandates using the density matrix rather
than the wave function for a subsystem in the general case.

In a non-entangled state, in the absence of interaction, the wave function
of the whole system factorizes:
\begin{equation}
|\Psi \rangle =|\psi \rangle \otimes |\phi \rangle =\sum_{m}c_{m}|\psi
_{m}\rangle \sum_{\varpi }d_{\varpi }|\phi _{\varpi }\rangle .
\label{DME-PsiFactorizable}
\end{equation}
If the whole system and both subsystems are described by density matrices,
the total density matrix under the above conditions factorizes such as
\begin{equation}
\rho _{m\varpi ,n\varpi ^{\prime }}=\rho _{mn}^{\mathrm{s}}\rho _{\varpi
\varpi ^{\prime }}^{\mathrm{b}}  \label{DME-rhototFactorize}
\end{equation}
for a small system and the bath. This factorization means that both
subsystems are completely independent. If they are prepared at the initial
moment independently from each other, one cannot expect their entanglement
that requires a special care. As the time goes, interaction between the
subsystems can cause some entanglement that, however, should remain small if
the interaction is weak. The measure of entanglement is the mismatch in Eq. (%
\ref{DME-rhototFactorize}), where $\rho _{mn}^{\mathrm{s}}$ is defined by
Eq. (\ref{DME-rhomunuenv}) and $\rho _{\varpi \varpi ^{\prime }}^{\mathrm{b}%
} $ is defined by similar tracing out the variables of the small system.
Note that both rhs and lhs of Eq. (\ref{DME-rhototFactorize}) are properly
normalized according to Eq. (\ref{DME-rhoNorm}) and its variants. As a
number measuring the entanglement, one can use, e.g.,
\begin{equation}
\text{Ent}=\frac{n^{2}}{n^{2}-1}\left[ 1-\frac{\sum_{mn}\left| \rho _{mn}^{%
\mathrm{s}}\right| ^{2}\sum_{\varpi \varpi ^{\prime }}\left| \rho _{\varpi
\varpi ^{\prime }}^{\mathrm{b}}\right| ^{2}}{\sum_{m\varpi ,n\varpi ^{\prime
}}\left| \rho _{m\varpi ,n\varpi ^{\prime }}\right| ^{2}}\right]
\label{DME-EntDef}
\end{equation}
that is related to Eq. (\ref{DME-rho2Norm}). Here $n$ is the number of
states of each subsystem that is assumed to be common. If the whole system
is in a pure state, the denominator is equal to 1. In particular, for $\Psi $
given by Eq. (\ref{DME-Entanglement-example-Psi}), $n=2,$ one has $\rho
_{mn}^{\mathrm{s}}$ given by Eq. (\ref{DME-Entanglement-example-rhomunu})
and similarly $\rho _{\varpi \varpi ^{\prime }}^{\mathrm{b}}=\left|
C_{\varpi \varpi }\right| ^{2}\delta _{\varpi \varpi ^{\prime }},$ so that
\begin{equation}
\text{Ent}=\frac{4}{3}\left[ 1-\left( \left| C_{11}\right| ^{4}+\left|
C_{22}\right| ^{4}\right) ^{2}\right] .
\end{equation}
In the case $\left| C_{11}\right| ^{2}=\left| C_{22}\right| ^{2}=1/2$
entanglement reaches its maximal value 1.

The coupling between subsystems also results in impossibility to describe
them in terms of wave functions. The coupling causes slow transitions
between energy levels, so that the total energy of the whole system is
conserved. For the small system this means that it spends time in its
different quantum mechanical states (in the interaction is weak and they are
well defined), so that over a large time its state is a mixture of different
states rather than a pure quantum mechanical state. Alternatively one can
think about an ensemble of bathed small systems being in different pure
states. An example of a mixture state is the thermal-equilibrium state
\begin{equation}
\rho _{mn}=\frac{1}{Z_{\mathrm{s}}}\exp \left( -\frac{E_{m}}{k_{B}T}\right)
\delta _{mn},  \label{DME-Thermal}
\end{equation}
where $Z_{\mathrm{s}}=\sum_{m}e^{-E_{m}/(k_{B}T)}$ is the partition
function. This formula holds if the basis states $|\psi _{m}\rangle $ are
eigenfunctions of the small-system's Hamiltonian $\hat{H}_{\mathrm{s}}$,
that is a natural \ choice. We will see below that in the course of temporal
evolution described by the DME, the initial density matrix approaches the
form above.

In fact, the environment of the small system also can be weakly coupled to
some \emph{super-bath}, so that the small system + environment is not an
isolated system and it also should be described by the density matrix
instead of the wave function. The role of the super-bath is just is to
ensure that the whole system is not in a pure quantum state but in a mixture
state, whereas the coupling between the bath and super-bath is assumed to be
very weak and is never considered explicitly. In most cases\ the state of
the environment (bath) is the thermal equilibrium that is described by the
density matrix similar to Eq. (\ref{DME-Thermal}). Since the bath is much
larger than the small system, the weak coupling to the latter won't drive it
out of the equilibrium.

\subsection{Density matrix as an operator}

The formalism of quantum mechanics allows one to consider the density matrix
as an operator. More precisely, one can introduce the density operator
\begin{equation}
\hat{\rho}=\sum_{m n }\rho _{m n }|\psi _{m }\rangle \langle \psi _{n }|,
\label{DME-rhoOperatorDef}
\end{equation}
where $|\psi _{m }\rangle $ is a complete orthogonal set of states. The
density matrix consists of matrix elements of the density operator:
\begin{equation}
\rho _{m n }=\langle \psi _{m }|\hat{\rho}|\psi _{n }\rangle ,
\label{DME-DMviaDO}
\end{equation}
so that both DM and DO contain the same information and are equivalent.
Using the density operator allows one to put formulas in a more compact form
without subscripts.

In particular, the expectation value of an operator $\hat{A}$ can be
obtained as a trace of $\hat{A}\hat{\rho}$ over any complete orthogonal set
of states $|\psi _{n }\rangle $. The calculation is especially simple if one
uses the set of states in the definition of $\hat{\rho}$:
\begin{equation}
\langle \hat{A}\rangle =\mathrm{Tr}\left\{ \hat{A}\hat{\rho}\right\}
=\sum_{n }\langle \psi _{n }|\hat{A}\hat{\rho}|\psi _{n }\rangle =\sum_{m n
}\langle \psi _{n }|\hat{A}|\psi _{m }\rangle \langle \psi _{m }|\hat{\rho}%
|\psi _{n }\rangle =\sum_{m n }A_{n m }\rho _{m n },  \label{DME-ASp}
\end{equation}
and the result coincides with Eq. (\ref{DME-Aavrrho}). It can be proven that
the trace of an operator is independent of the choice of the basis in $%
\mathrm{Tr}\left\{ \ldots \right\} $ and operators can be cyclically
permuted under the trace symbol, so that $\mathrm{Tr}\left\{ \hat{A}\hat{\rho%
}\right\} =\mathrm{Tr}\left\{ \hat{\rho}\hat{A}\right\} .$

If the system interacts with its environment described by the basis $|\phi
_{\varpi }\rangle $, the total DO has the form
\begin{equation}
\hat{\rho}=\sum_{m\varpi ,n\varpi ^{\prime }}\rho _{m\varpi ,n\varpi
^{\prime }}|\psi _{m}\phi _{\varpi }\rangle \langle \psi _{n}\phi _{\varpi
^{\prime }}|\equiv \sum_{m\varpi ,n\varpi ^{\prime }}\rho _{m\varpi ,n\varpi
^{\prime }}|\psi _{m}\rangle \langle \psi _{n}|\otimes |\phi _{\varpi
}\rangle \langle \phi _{\varpi ^{\prime }}|.  \label{DME-DOtot}
\end{equation}
Calculating any observable $A$ that corresponds to the small system can be
done in two steps: First calculating the trace over the variables of the
bath and then calculating the trace over the basis states of the small
system using Eq. (\ref{DME-ASp}). The first step yields the reduced density
operator for the small system
\begin{equation}
\hat{\rho}^{\mathrm{s}}=\mathrm{Tr}_{\mathrm{b}}\hat{\rho}\equiv \mathrm{Tr}%
_{\varpi }\hat{\rho}  \label{DME-DOredTrEnv}
\end{equation}
that has the form of Eq. (\ref{DME-rhoOperatorDef}) with
\begin{equation}
\rho _{mn}^{\mathrm{s}}=\sum_{\varpi }\rho _{m\varpi ,n\varpi }.
\label{DME-DMred}
\end{equation}
If the whole system is in a pure state, this formula coincides with Eq. (\ref
{DME-rhomunuenv}).

The density operator corresponding to the thermal equilibrium of the small
subsystem in contact with the bath has the beautiful form
\begin{equation}
\hat{\rho}^{\mathrm{s,eq}}=\frac{1}{Z_{\mathrm{s}}}\exp \left( -\frac{\hat{H}%
_{\mathrm{s}}}{k_{B}T}\right) .  \label{DME-rhoOperEq}
\end{equation}
Because of applications in quantum statistics density operator is also
called \emph{statistical operator}. Defining the density matrix $\rho _{mn}^{%
\mathrm{s}}$ with respect to the basis of eigenstates of $\hat{H}$ in Eq. (%
\ref{DME-DMviaDO}), one arrives at Eq. (\ref{DME-Thermal}). The eigenstate
basis is the most convenient for calculating thermal averages of physical
quantities. However, as pointed out above, this could be done with the help
of any other basis. If the whole system consisting of the small subsystem
and the bath is in contact with a super-bath, the equilibrium statistical
operator of the whole system has the form similar to Eq. (\ref{DME-rhoOperEq}%
).

For a system in a pure state, the density operator defined by Eq. (\ref
{DME-rhoOperatorDef}) can be with the help of Eqs. (\ref
{DME-rhomunuDefcmucnu}) and (\ref{DME-psiexpansion}) rewritten in the form
\begin{equation}
\hat{\rho}=|\psi \rangle \langle \psi |.  \label{DME-DOpure}
\end{equation}

\subsection{Temporal evolution and interaction representation}

\label{Sec-DME-Intro-Temporal}

Temporal evolution of the DM\ or DO of an isolated system such as the small
system + bath obeys the equation that follows from the Schr\"{o}dinger
equation. If, for instance, the whole system is in a pure state $|\Psi
\rangle ,$ its density operator is given by
\begin{equation}
\hat{\rho}=|\Psi \rangle \langle \Psi |  \label{DME-DOPsipure}
\end{equation}
c.f. Eq. (\ref{DME-DOpure}). Then with the help of the Schr\"{o}dinger
equation and its conjugate
\begin{equation}
i\hbar \frac{\partial }{\partial t}|\Psi \rangle =\hat{H}|\Psi \rangle
,\qquad -i\hbar \frac{\partial }{\partial t}\langle \Psi |=\langle \Psi |%
\hat{H}  \label{DME-SchrEq}
\end{equation}
one obtains the quantum Liouville equation
\begin{equation}
i\hbar \frac{\partial \hat{\rho}}{\partial t}=\left[ \hat{H},\hat{\rho}%
\right] .  \label{DME-Liouville}
\end{equation}
It should be noted that this equation is \emph{not} an equation of motion
for an operator in the Heisenberg representation that has another sign. The
DO consists of states that have their \emph{own} time dependence, unlike
Heisenberg operators whose time dependence is \emph{borrowed} from the
states. In the presence of the super-bath the system is not in a pure state
but rather in a mixture of different states $|\Psi \rangle $,
\begin{equation}
\hat{\rho}=\sum_{\Psi }A_{\Psi }|\Psi \rangle \langle \Psi |.
\label{DME-DOPsiMixture}
\end{equation}
Fortunately, Eq. (\ref{DME-Liouville}) is valid also in this case. Remember
that Eq. (\ref{DME-Liouville}) describes the whole system, small system +
bath, although we dropped the index ``tot'' for brevity. Our final aim is,
however, to obtain a density matrix equation (DME) for the density matrix of
the small system alone with the bath degrees of freedom eliminated.

Let us break up the Hamiltonian of the system into a ``simple'' Hamiltonian $%
\hat{H}_{0}$ and perturbation or interaction $\hat{V}$:
\begin{equation}
\hat{H}=\hat{H}_{0}+\hat{V}.  \label{DMA-HH0V}
\end{equation}
In the absence of $\hat{V},$ evolution of quantum states $|\Psi \rangle $ is
described by the unitary evolution operator $\hat{U}_{0}(t)$:
\begin{equation}
|\Psi \rangle _{t}=\hat{U}_{0}(t)|\Psi \rangle _{0},\qquad \langle \Psi
|_{t}=\langle \Psi |_{0}\hat{U}_{0}^{\dagger }(t),  \label{DME-U0Def}
\end{equation}
where $|\Psi \rangle _{0}$ corresponds to the starting moment $t=0.$ We call
$\hat{U}_{0}(t)$ \emph{bare} evolution operator. From the Schr\"{o}dinger
equation (\ref{DME-SchrEq}) the equations of motion for $\hat{U}_{0}$
and its conjugate follow :
\begin{equation}
i\hbar \frac{\partial }{\partial t}\hat{U}_{0}=\hat{H}_{0}\hat{U}_{0},\qquad
-i\hbar \frac{\partial }{\partial t}\hat{U}_{0}^{\dagger }=\hat{U}%
_{0}^{\dagger }\hat{H}_{0},\qquad \hat{U}_{0}^{\dagger }(t)=\hat{U}%
_{0}^{-1}(t).  \label{DME-U0Eq}
\end{equation}
If $\hat{H}_{0}$ is time independent, the solution of these equations is
\begin{equation}
\hat{U}_{0}(t)=e^{-i\hat{H}_{0}t/\hbar },\qquad \hat{U}_{0}^{\dagger
}(t)=e^{i\hat{H}_{0}t/\hbar }.  \label{DME-U0Sol}
\end{equation}
Of course, one can write similar formulas with the total Hamiltonian $\hat{H}
$ as well, $\hat{U}(t)$ being the full evolution operator. The idea of using
$\hat{U}_{0}(t)$ is to split the nontrivial part of the evolution due to the
perturbation $\hat{V}$ \ from the trivial evolution due to $\hat{H}_{0}.$ To
effectuate this, one can introduce the density matrix in the \emph{%
interaction representation}
\begin{equation}
\hat{\rho}(t)_{I}=\hat{U}_{0}^{\dagger }(t)\hat{\rho}(t)\hat{U}_{0}(t).
\label{DME-RhoIDef}
\end{equation}
The equation of motion for $\hat{\rho}(t)_{I}$ follows from Eqs. (\ref
{DME-Liouville}) and (\ref{DME-U0Eq}):
\begin{eqnarray}
i\hbar \frac{\partial \hat{\rho}_{I}}{\partial t} &=&i\hbar \frac{\partial
\hat{U}_{0}^{\dagger }(t)}{\partial t}\hat{\rho}(t)\hat{U}_{0}(t)+\hat{U}%
_{0}^{\dagger }(t)i\hbar \frac{\partial \hat{\rho}}{\partial t}\hat{U}%
_{0}(t)+\hat{U}_{0}^{\dagger }(t)\hat{\rho}(t)i\hbar \frac{\partial \hat{U}%
_{0}(t)}{\partial t}  \nonumber \\
&=&-\hat{U}_{0}^{\dagger }\hat{H}_{0}\hat{\rho}(t)\hat{U}_{0}(t)+\hat{U}%
_{0}^{\dagger }(t)\left[ \hat{H},\hat{\rho}\right] \hat{U}_{0}(t)+\hat{U}%
_{0}^{\dagger }(t)\hat{\rho}(t)\hat{H}_{0}\hat{U}_{0}  \nonumber \\
&=&\hat{U}_{0}^{\dagger }(t)\left[ \hat{V},\hat{\rho}\right] \hat{U}_{0}(t).
\end{eqnarray}
Inserting $\hat{U}_{0}(t)\hat{U}_{0}^{\dagger }(t)=1$ [see Eq. (\ref
{DME-U0Eq})] between the operators and defining
\begin{equation}
\hat{V}_{I}(t)\equiv \hat{U}_{0}^{\dagger }(t)\hat{V}\hat{U}_{0}(t),
\label{DME-VIDef}
\end{equation}
this equation can be brought into the form
\begin{equation}
i\hbar \frac{\partial \hat{\rho}_{I}}{\partial t}=\left[ \hat{V}_{I},\hat{%
\rho}_{I}\right] .  \label{DME-rhoIEq}
\end{equation}
One can see that the temporal evolution of the density matrix in the
interaction representation is governed by the interaction only. This
facilitates constructing the perturbation theory in $\hat{V}.$
%

\newpage
\section{The density matrix equation}


\label{Sec-DME}

\subsection{From the full to reduced DOE}

In this section the equation of motion for the reduces density operator of a
small system $s$ weakly interacting with a bath will be obtained. We will be
following the method of Karl Blum \cite{blu81} that is most practical,
although not rigorous. Rigorous methods using the projection operator
technique \cite{gra82,romorsopp89} lead to the same result with much greater
efforts.

The Hamiltonian can be written in the form
\begin{equation}
\hat{H}=\hat{H}_{0}+\hat{V},\qquad \hat{H}_{0}\equiv \hat{H}_{\mathrm{s}}+%
\hat{H}_{\mathrm{b}},\qquad \hat{V}\equiv \hat{H}_{\mathrm{s-b}}.
\label{DME-HamTot}
\end{equation}
Eq. (\ref{DME-rhoIEq}) for the DO of the whole system can be integrated over
the time resulting in the integral equation
\begin{equation}
\hat{\rho}(t)_{I}=\hat{\rho}(0)_{I}-\frac{i}{\hbar }\int_{0}^{t}dt^{\prime }%
\left[ \hat{V}(t^{\prime })_{I},\hat{\rho}(t^{\prime })_{I}\right] .
\label{DME-rhotot1stIter}
\end{equation}
Inserting it back into Eq. (\ref{DME-rhoIEq}) one obtains the
integro-differential equation
\begin{equation}
i\hbar \frac{d}{dt}\hat{\rho}(t)_{I}=\left[ \hat{V}(t)_{I},\hat{\rho}(0)_{I}%
\right] -\frac{i}{\hbar }\int_{0}^{t}dt^{\prime }\left[ \hat{V}(t)_{I},\left[
\hat{V}(t^{\prime })_{I},\hat{\rho}(t^{\prime })_{I}\right] \right]
\label{DME-rhotot2Iter}
\end{equation}
which is still an exact relation. Although this equation seems to be more
complicated than the initial equation (\ref{DME-rhoIEq}), it is convenient
for perturbative treatment of the interaction $\hat{V}.$ If the small system
is not entangled with the bath in the initial state, that is a natural
assumption, a small interaction cannot cause a significant entanglement as
well. Thus the density operator of the whole system nearly factorizes, see
Eq. (\ref{DME-rhototFactorize}). Here we additionally use that the bath is
at thermal equilibrium that cannot be noticeably distorted by a weak
interaction with the small subsystem:

\begin{equation}
\hat{\rho}(t)_{I}\cong \hat{\rho}_{\mathrm{s}}(t)_{I}\hat{\rho}_{\mathrm{b}%
}^{\mathrm{eq}}=\hat{\rho}_{\mathrm{s}}(t)\frac{1}{Z_{\mathrm{b}}}\exp
\left( -\frac{\hat{H}_{\mathrm{b}}}{k_{B}T}\right)  \label{DME-DOFactorize}
\end{equation}
Note that this approximation cannot be done in Eq. (\ref{DME-rhotot1stIter})
since it leads to disappearance of the effect of interaction. To properly
describe this effect, one has then to account for the corrections to Eq. (%
\ref{DME-DOFactorize}) in the first order in $\hat{V}.$ This way is
inconvenient. To the contrary, Eq. (\ref{DME-rhotot2Iter}) already contains
quadratic $\hat{V}$ terms that capture the main effect of interaction. Here
taking into account small corrections to Eq. (\ref{DME-DOFactorize}) can
bring only irrelevant small terms. Now using Eq. (\ref{DME-DOFactorize}) one
can transform Eq. (\ref{DME-rhotot2Iter}) into the density operator equation
for the small system by making a trace over the bath variables according to
Eq. (\ref{DME-DOredTrEnv}):
\begin{equation}
\frac{d}{dt}\hat{\rho}_{\mathrm{s}}(t)_{I}=\frac{i}{\hbar }\mathrm{Tr}_{%
\mathrm{b}}\left[ \hat{V}(t)_{I},\hat{\rho}_{\mathrm{s}}(0)_{I}\hat{\rho}_{%
\mathrm{b}}^{\mathrm{eq}}\right] -\frac{1}{\hbar ^{2}}\int_{0}^{t}dt^{\prime
}\mathrm{Tr}_{\mathrm{b}}\left[ \hat{V}(t)_{I},\left[ \hat{V}(t^{\prime
})_{I},\hat{\rho}_{\mathrm{s}}(t^{\prime })_{I}\hat{\rho}_{\mathrm{b}}^{%
\mathrm{eq}}\right] \right] .  \label{DME-DOEqReduced}
\end{equation}
For the couplings $\hat{V}$ we consider here, the first linear-$\hat{V}$ term
disappears.

The equation above is an integro-differential equation with integration over
preceding times, $t^{\prime }\leq t,$ in the rhs. Such equations are called
equations with memory and they are difficult to solve directly. In our case,
however, the problem simplifies. The time evolution of $\hat{\rho}_{\mathrm{s%
}}(t)_{I}$ In Eq. (\ref{DME-DOEqReduced}) is slow since it is governed by
the weak interaction between the system and the bath. On the other hand, $%
t^{\prime }$ dependences of the other terms in the integrand (the kernel)
are governed by $\hat{H}_{0}$ and thus they are fast at the scale of $\hat{%
\rho}_{\mathrm{s}}(t)_{I}.$ The analysis shows
that the kernel in Eq. (\ref{DME-DOEqReduced}) is localized in the region $%
\left| t-t^{\prime }\right| \lesssim 1/\omega _{\max },$ where $\omega
_{\max }$ \ is the maximal frequency of the bath excitations. Thus in the
integral over $t^{\prime }$ one can make the short-memory approximation $%
\hat{\rho}_{\mathrm{s}}(t^{\prime })_{I}\Rightarrow \hat{\rho}_{\mathrm{s}%
}(t)_{I}$ after which the time integral can be calculated explicitly.

Returning to the original reduced density operator
\begin{equation}
\hat{\rho}_{\mathrm{s}}(t)=\hat{U}_{0}(t)\hat{\rho}_{\mathrm{s}}(t)_{I}\hat{U%
}_{0}^{\dagger }(t),  \label{DME-rhoOriginal}
\end{equation}
c.f. Eq. (\ref{DME-RhoIDef}), and computing the derivative with the help of
Eq. (\ref{DME-U0Eq}),
\begin{equation}
\frac{d}{dt}\hat{\rho}_{\mathrm{s}}(t)=-\frac{i}{\hbar }\left[ \hat{H}_{%
\mathrm{s}},\hat{\rho}_{\mathrm{s}}(t)\right] +\hat{U}_{0}(t)\left( \frac{d}{%
dt}\hat{\rho}_{\mathrm{s}}(t)_{I}\right) \hat{U}_{0}^{\dagger }(t),
\label{DME-rhoDerback}
\end{equation}
one obtains
\begin{equation}
\frac{d}{dt}\hat{\rho}_{\mathrm{s}}(t)=-\frac{i}{\hbar }\left[ \hat{H}_{%
\mathrm{s}},\hat{\rho}_{\mathrm{s}}(t)\right] -\frac{1}{\hbar ^{2}}%
\int_{0}^{t}dt^{\prime }\mathrm{Tr}_{\mathrm{b}}\left[ \hat{V},\left[ \hat{V}%
(t^{\prime }-t)_{I},\hat{\rho}_{\mathrm{s}}(t)\hat{\rho}_{\mathrm{b}}^{%
\mathrm{eq}}\right] \right]  \label{DME-RhoEq}
\end{equation}
or, with $\tau \equiv t-t^{\prime }$ and dropping the subscript \textrm{s}, $%
\hat{\rho}^{\mathrm{s}}(t)\Rightarrow \hat{\rho}(t),$
\begin{equation}
\frac{d}{dt}\hat{\rho}(t)=-\frac{i}{\hbar }\left[ \hat{H}_{\mathrm{s}},\hat{%
\rho}(t)\right] +\hat{R}\hat{\rho}(t),  \label{DME-rhoEqtau}
\end{equation}
where the first term with commutator is the conservative term and
\begin{equation}
\hat{R}\hat{\rho}(t)\equiv -\frac{1}{\hbar ^{2}}\int_{0}^{t}d\tau \frac{1}{%
Z_{\mathrm{b}}}\mathrm{Tr}_{\mathrm{b}}\left[ \hat{V},\left[ \hat{V}(-\tau
)_{I},\hat{\rho}(t)\exp \left( -\frac{\hat{H}_{\mathrm{b}}}{k_{B}T}\right) %
\right] \right]  \label{DME-ROperDef}
\end{equation}
describes the relaxation of the small system. Here, for time-independent
problems,
\begin{equation}
\hat{V}(\tau )_{I}=\hat{U}_{0}^{\dagger }(\tau )\hat{V}\hat{U}_{0}(\tau
)=e^{i\hat{H}_{0}\tau }\hat{V}e^{-i\hat{H}_{0}\tau }.  \label{DME-Vttpr}
\end{equation}
If the Hamiltonian of the small system $\hat{H}_{\mathrm{s}}$ depends on
time, this time dependence is typically slow in comparizon to the frequency
of the bath excitations $\omega _{\max },$ so that during the short times $%
\tau \sim 1/\omega _{\max }$ the change of $\hat{H}_{\mathrm{s}}$ is
negligibly small. Thus one can simply use Eq. (\ref{DME-Vttpr}) with $\hat{H}%
_{\mathrm{s}}=\hat{H}_{\mathrm{s}}(t).$

\subsection{From the DOE\ to DME}

Now one can go over from the density operator $\hat{\rho}$ to the density
matrix $\rho _{mn}$ using Eq. (\ref{DME-DMviaDO}) and the notations
\begin{equation}
H_{\mathrm{s},mn}\equiv \langle \psi _{m}|\hat{H}_{\mathrm{s}}|\psi
_{n}\rangle ,\qquad V_{m\varpi ,n\varpi ^{\prime }}\equiv \langle \psi
_{m}\phi _{\varpi }|\hat{V}|\psi _{n}\phi _{\varpi ^{\prime }}\rangle ,
\label{DME-Vmatreleme}
\end{equation}
where $\left| \phi _{\varpi }\right\rangle $ are eigenfunctions of $\hat{H}_{%
\mathrm{b}}$. Then the partition function of the bath becomes
\begin{equation}
Z_{\mathrm{b}}=\sum_{\varpi }\langle \phi _{\varpi }|e^{-\hat{H}%
_{b}/(k_{B}T)}|\phi _{\varpi }\rangle =\sum_{\varpi }e^{-E_{\varpi
}/(k_{B}T)}.  \label{DEM-Zb}
\end{equation}
The conservative term of the resulting DME has three different forms in
three different cases. If the small-system states $\left| \psi
_{m}\right\rangle $ are time independent and form a so-called \emph{natural
basis} unrelated to $\hat{H}_{\mathrm{s}},$ the DME has the form
\begin{equation}
\frac{d}{dt}\rho _{mn}=-\frac{i}{\hbar }\sum_{l}\left( H_{\mathrm{s},ml}\rho
_{ln}-\rho _{ml}H_{\mathrm{s},ln}\right) +\langle \psi _{m}|\hat{R}\hat{\rho}%
|\psi _{n}\rangle .  \label{DME-RhoEqNatural}
\end{equation}
The natural basis is inconvenient for the evaluation of the relaxation term
since the relaxation of the small system takes place not between the states $%
\left| \psi _{m}\right\rangle $ but between the eigenstates of $\hat{H}_{%
\mathrm{s}}$:
\begin{equation}
\hat{H}_{\mathrm{s}}|\chi _{\alpha }\rangle =\varepsilon _{\alpha }|\chi
_{\alpha }\rangle ,\qquad e^{-i\hat{H}_{\mathrm{s}}\tau /\hbar }|\chi
_{\alpha }\rangle =e^{-i\varepsilon _{\alpha }\tau /\hbar }|\chi _{\alpha
}\rangle .  \label{DME-chieigen}
\end{equation}
The basis of $|\chi _{\alpha }\rangle $ will be called \emph{diagonal basis}
since the Hamiltonian matrix $\langle \chi _{\alpha }|\hat{H}_{\mathrm{s}%
}|\chi _{\beta }\rangle =\varepsilon _{\alpha }\delta _{\alpha \beta }$ is
diagonal. In the diagonal basis the DME has the form
\begin{equation}
\frac{d}{dt}\rho _{\alpha \beta }=-i\omega _{\alpha \beta }\rho _{\alpha
\beta }+\langle \chi _{\alpha }|\hat{R}\hat{\rho}|\chi _{\beta }\rangle ,
\label{DME-rhoEq-diagbasis}
\end{equation}
where $\omega _{\alpha \beta }$ are transition frequencies between the
energy levels of the small system,
\begin{equation}
\hbar \omega _{\alpha \beta }\equiv \varepsilon _{\alpha }-\varepsilon
_{\beta },  \label{DME-omegaalphabetaDef}
\end{equation}
and the relaxation term can be conveniently evaluated (see next
section). If $\hat{H}_{\mathrm{s}}$ depends on time, one can use the \emph{%
adiabatic basis} of the states $|\chi _{\alpha }(t)\rangle $ defined as
\begin{equation}
\hat{H}_{\mathrm{s}}(t)|\chi _{\alpha }(t)\rangle =\varepsilon _{\alpha
}(t)|\chi _{\alpha }(t)\rangle .  \label{DME-chiAdiabatic}
\end{equation}
In this basis $\dot{\rho}_{\alpha \beta }$ acquires additional \emph{%
non-adiabatic} terms:
\begin{equation}
\frac{d}{dt}\rho _{\alpha \beta }=\frac{d}{dt}\langle \chi _{\alpha }|\hat{%
\rho}|\chi _{\beta }\rangle =\langle \dot{\chi}_{\alpha }|\hat{\rho}|\chi
_{\beta }\rangle +\langle \chi _{\alpha }|\hat{\rho}|\dot{\chi}_{\beta
}\rangle -\frac{i}{\hbar }\langle \chi _{\alpha }|\left[ \hat{H}_{\mathrm{s}%
},\hat{\rho}\right] |\chi _{\beta }\rangle +\langle \chi _{\alpha }|\hat{R}%
\hat{\rho}|\chi _{\beta }\rangle
\end{equation}
i.e.,
\begin{equation}
\frac{d}{dt}\rho _{\alpha \beta }=\sum_{\gamma }\left( \langle \dot{\chi}%
_{\alpha }\left| \chi _{\gamma }\right\rangle \rho _{\gamma \beta }+\rho
_{\alpha \gamma }\langle \chi _{\gamma }\left| \dot{\chi}_{\beta
}\right\rangle \right) -i\omega _{\alpha \beta }\rho _{\alpha \beta
}+\langle \chi _{\alpha }|\hat{R}\hat{\rho}|\chi _{\beta }\rangle .
\label{DME-rhoEq-diagbasist}
\end{equation}
As argued at the end of the preceding section, calculation of the relaxation
term $\langle \chi _{\alpha }|\hat{R}\hat{\rho}|\chi _{\beta }\rangle $ is
not complicated by the time dependence of $\hat{H}_{\mathrm{s}}.$ This
calculation will be done in the next section where Eq. (\ref
{DME-rhoEq-diagbasis}) will be used for brevity. The terms due to the time
dependence of $\hat{H}_{\mathrm{s}}$ in Eq. (\ref{DME-rhoEq-diagbasist}) can
be added if needed.

\subsection{DME in the diagonal basis}

Using Eqs. (\ref{DME-rhoEq-diagbasis}) and (\ref{DME-ROperDef}) and
inserting summation over intermediate states yields
\begin{eqnarray}
\frac{d}{dt}\rho _{\alpha \beta } &=&-i\omega _{\alpha \beta }\rho _{\alpha
\beta }-\frac{1}{\hbar ^{2}}\int_{0}^{t}d\tau \frac{1}{Z_{\mathrm{b}}}%
\sum_{\alpha ^{\prime }\beta ^{\prime }\varpi \varpi ^{\prime }\varpi
^{\prime \prime }}  \nonumber \\
&&\left\{ \langle \chi _{\alpha }\phi _{\varpi }|\hat{V}|\chi _{\alpha
^{\prime }}\phi _{\varpi ^{\prime }}\rangle \langle \chi _{\alpha ^{\prime
}}\phi _{\varpi ^{\prime }}|e^{-i\hat{H}_{0}\tau /\hbar }\hat{V}e^{i\hat{H}%
_{0}\tau /\hbar }|\chi _{\beta ^{\prime }}\phi _{\varpi ^{\prime \prime
}}\rangle \langle \chi _{\beta ^{\prime }}\phi _{\varpi ^{\prime \prime }}|%
\hat{\rho}e^{-\hat{H}_{\mathrm{b}}/(k_{B}T)}|\chi _{\beta }\phi _{\varpi
}\rangle \right.  \nonumber \\
&&-\left. \langle \chi _{\alpha }\phi _{\varpi }|\hat{V}|\chi _{\alpha
^{\prime }}\phi _{\varpi ^{\prime }}\rangle \langle \chi _{\alpha ^{\prime
}}\phi _{\varpi ^{\prime }}|\hat{\rho}e^{-\hat{H}_{\mathrm{b}%
}/(k_{B}T)}|\chi _{\beta ^{\prime }}\phi _{\varpi ^{\prime \prime }}\rangle
\langle \chi _{\beta ^{\prime }}\phi _{\varpi ^{\prime \prime }}|e^{-i\hat{H}%
_{0}\tau /\hbar }\hat{V}e^{i\hat{H}_{0}\tau /\hbar }|\chi _{\beta }\phi
_{\varpi }\rangle \right.  \nonumber \\
&&-\left. \langle \chi _{\alpha }\phi _{\varpi }|e^{-i\hat{H}_{0}\tau /\hbar
}\hat{V}e^{i\hat{H}_{0}\tau /\hbar }|\chi _{\alpha ^{\prime }}\phi _{\varpi
^{\prime }}\rangle \langle \chi _{\alpha ^{\prime }}\phi _{\varpi ^{\prime
}}|\hat{\rho}e^{-\hat{H}_{\mathrm{b}}/(k_{B}T)}|\chi _{\beta ^{\prime }}\phi
_{\varpi ^{\prime \prime }}\rangle \langle \chi _{\beta ^{\prime }}\phi
_{\varpi ^{\prime \prime }}|\hat{V}|\chi _{\beta }\phi _{\varpi }\rangle
\right.  \nonumber \\
&&+\left. \langle \chi _{\alpha }\phi _{\varpi }|\hat{\rho}e^{-\hat{H}_{%
\mathrm{b}}/(k_{B}T)}|\chi _{\alpha ^{\prime }}\phi _{\varpi ^{\prime
}}\rangle \langle \chi _{\alpha ^{\prime }}\phi _{\varpi ^{\prime }}|e^{-i%
\hat{H}_{0}\tau /\hbar }\hat{V}e^{i\hat{H}_{0}\tau /\hbar }|\chi _{\beta
^{\prime }}\phi _{\varpi ^{\prime \prime }}\rangle \langle \chi _{\beta
^{\prime }}\phi _{\varpi ^{\prime \prime }}|\hat{V}|\chi _{\beta }\phi
_{\varpi }\rangle \right\}.
\end{eqnarray}
Since the states are eigenfunctions of $\hat{H}_{\mathrm{s}}$ and $\hat{H}_{%
\mathrm{b}},$ this simplifies to
\begin{eqnarray}
\frac{d}{dt}\rho _{\alpha \beta } &=&-i\omega _{\alpha \beta }\rho _{\alpha
\beta }-\frac{1}{\hbar ^{2}}\int_{0}^{t}d\tau \frac{1}{Z_{\mathrm{b}}}%
\sum_{\alpha ^{\prime }\beta ^{\prime }\varpi \varpi ^{\prime }\varpi
^{\prime \prime }}  \nonumber \\
&&\left\{ e^{i\left( -\varepsilon _{\alpha ^{\prime }}-E_{\varpi ^{\prime
}}+\varepsilon _{\beta ^{\prime }}+E_{\varpi ^{\prime \prime }}\right) \tau
/\hbar }e^{-E_{\varpi }/(k_{B}T)}V_{\alpha \varpi ,\alpha ^{\prime }\varpi
^{\prime }}V_{\alpha ^{\prime }\varpi ^{\prime },\beta ^{\prime }\varpi
^{\prime \prime }}\rho _{\beta ^{\prime }\beta }\delta _{\varpi ^{\prime
\prime }\varpi }\right.  \nonumber \\
&&-\left. e^{i\left( -\varepsilon _{\beta ^{\prime }}-E_{\varpi ^{\prime
\prime }}+\varepsilon _{\beta }+E_{\varpi }\right) \tau /\hbar
}e^{-E_{\varpi ^{\prime \prime }}/(k_{B}T)}V_{\alpha \varpi ,\alpha ^{\prime
}\varpi ^{\prime }}\rho _{\alpha ^{\prime }\beta ^{\prime }}\delta _{\varpi
^{\prime }\varpi ^{\prime \prime }}V_{\beta ^{\prime }\varpi ^{\prime \prime
},\beta \varpi }\right.  \nonumber \\
&&-\left. e^{i\left( -\varepsilon _{\alpha }-E_{\varpi }+\varepsilon
_{\alpha ^{\prime }}+E_{\varpi ^{\prime }}\right) \tau /\hbar }e^{-E_{\varpi
^{\prime \prime }}/(k_{B}T)}V_{\alpha \varpi ,\alpha ^{\prime }\varpi
^{\prime }}\rho _{\alpha ^{\prime }\beta ^{\prime }}\delta _{\varpi ^{\prime
}\varpi ^{\prime \prime }}V_{\beta ^{\prime }\varpi ^{\prime \prime },\beta
\varpi }\right.  \nonumber \\
&&+\left. e^{i\left( -\varepsilon _{\alpha ^{\prime }}-E_{\varpi ^{\prime
}}+\varepsilon _{\beta ^{\prime }}+E_{\varpi ^{\prime \prime }}\right) \tau
/\hbar }e^{-E_{\varpi ^{\prime }}/(k_{B}T)}\rho _{\alpha \alpha ^{\prime
}}\delta _{\varpi \varpi ^{\prime }}V_{\alpha ^{\prime }\varpi ^{\prime
},\beta ^{\prime }\varpi ^{\prime \prime }}V_{\beta ^{\prime }\varpi
^{\prime \prime },\beta \varpi }\right\}
\end{eqnarray}
and further to
\begin{eqnarray}
\frac{d}{dt}\rho _{\alpha \beta } &=&-i\omega _{\alpha \beta }\rho _{\alpha
\beta }-\frac{1}{\hbar ^{2}}\int_{0}^{t}d\tau \frac{1}{Z_{\mathrm{b}}}%
\sum_{\alpha ^{\prime }\beta ^{\prime }\varpi \varpi ^{\prime }}  \nonumber
\\
&&\left\{ e^{i\left( -\varepsilon _{\alpha ^{\prime }}-E_{\varpi ^{\prime
}}+\varepsilon _{\beta ^{\prime }}+E_{\varpi }\right) \tau /\hbar
}e^{-E_{\varpi }/(k_{B}T)}V_{\alpha \varpi ,\alpha ^{\prime }\varpi ^{\prime
}}V_{\alpha ^{\prime }\varpi ^{\prime },\beta ^{\prime }\varpi }\rho _{\beta
^{\prime }\beta }\right.  \nonumber \\
&&-\left. e^{i\left( -\varepsilon _{\beta ^{\prime }}-E_{\varpi ^{\prime
}}+\varepsilon _{\beta }+E_{\varpi }\right) \tau /\hbar }e^{-E_{\varpi
^{\prime }}/(k_{B}T)}V_{\alpha \varpi ,\alpha ^{\prime }\varpi ^{\prime
}}\rho _{\alpha ^{\prime }\beta ^{\prime }}V_{\beta ^{\prime }\varpi
^{\prime },\beta \varpi }\right.  \nonumber \\
&&-\left. e^{i\left( -\varepsilon _{\alpha }-E_{\varpi }+\varepsilon
_{\alpha ^{\prime }}+E_{\varpi ^{\prime }}\right) \tau /\hbar }e^{-E_{\varpi
^{\prime }}/(k_{B}T)}V_{\alpha \varpi ,\alpha ^{\prime }\varpi ^{\prime
}}\rho _{\alpha ^{\prime }\beta ^{\prime }}V_{\beta ^{\prime }\varpi
^{\prime },\beta \varpi }\right.  \nonumber \\
&&+\left. e^{i\left( -\varepsilon _{\alpha ^{\prime }}-E_{\varpi
}+\varepsilon _{\beta ^{\prime }}+E_{\varpi ^{\prime }}\right) \tau /\hbar
}e^{-E_{\varpi }/(k_{B}T)}\rho _{\alpha \alpha ^{\prime }}V_{\alpha ^{\prime
}\varpi ,\beta ^{\prime }\varpi ^{\prime }}V_{\beta ^{\prime }\varpi
^{\prime },\beta \varpi }\right\} .  \label{DME-rhoEqnoKron}
\end{eqnarray}
Since the time kernel is sharply localized, the integration over $\tau $ can
be extended to the interval $\left( 0,\infty \right) .$ Further, the
relaxation of the small system that we are mainly interested in is due to
the real part of the bath coupling term in Eq. (\ref{DME-rhoEqnoKron}). Its
imaginary part is in most cases only a small correction to the first
(conservative) term in this equation. Thus taking into account the time
symmetry of $F_{ij}(\tau )$ one can replace
\begin{equation}
\int_{0}^{t}d\tau e^{i\left( -\varepsilon _{\alpha ^{\prime }}-E_{\varpi
^{\prime }}+\varepsilon _{\beta ^{\prime }}+E_{\varpi }\right) \tau /\hbar
}\Rightarrow \hbar \pi \delta \left( -\varepsilon _{\alpha ^{\prime
}}-E_{\varpi ^{\prime }}+\varepsilon _{\beta ^{\prime }}+E_{\varpi }\right)
\end{equation}
etc. After renaming indices in Eq. (\ref{DME-rhoEqnoKron}) ($\alpha ^{\prime
}\Rightarrow \lambda ,$ $\beta ^{\prime }\Rightarrow \alpha ^{\prime }$ in
the first term and $\beta ^{\prime }\Rightarrow \lambda ,$ $\alpha ^{\prime
}\Rightarrow \beta ^{\prime }$ in the forth term) one obtains the DME in the
form
\begin{equation}
\frac{d}{dt}\rho _{\alpha \beta }=-i\omega _{\alpha \beta }\rho _{\alpha
\beta }+\sum_{\alpha ^{\prime }\beta ^{\prime }}R_{\alpha \beta ,\alpha
^{\prime }\beta ^{\prime }}\rho _{\alpha ^{\prime }\beta ^{\prime }},
\label{DME-rhoEqTensor}
\end{equation}
where
\begin{eqnarray}
R_{\alpha \beta ,\alpha ^{\prime }\beta ^{\prime }} &=&\frac{\pi }{\hbar Z_{%
\mathrm{b}}}\sum_{\varpi \varpi ^{\prime }}\left\{ -\sum_{\gamma
}e^{-E_{\varpi }/(k_{B}T)}\delta \left( \varepsilon _{\alpha ^{\prime
}}-\varepsilon _{\gamma }+E_{\varpi }-E_{\varpi ^{\prime }}\right) V_{\alpha
\varpi ,\gamma \varpi ^{\prime }}V_{\gamma \varpi ^{\prime },\alpha ^{\prime
}\varpi }\delta _{\beta ^{\prime }\beta }\right.  \nonumber \\
&&-\sum_{\gamma }e^{-E_{\varpi }/(k_{B}T)}\delta \left( \varepsilon _{\beta
^{\prime }}-\varepsilon _{\gamma }+E_{\varpi }-E_{\varpi ^{\prime }}\right)
\delta _{\alpha \alpha ^{\prime }}V_{\beta ^{\prime }\varpi ,\gamma \varpi
^{\prime }}V_{\gamma \varpi ^{\prime },\beta \varpi }  \nonumber \\
&&+\left. e^{-E_{\varpi ^{\prime }}/(k_{B}T)}\left[ \delta \left(
\varepsilon _{\beta }-\varepsilon _{\beta ^{\prime }}+E_{\varpi }-E_{\varpi
^{\prime }}\right) +\delta \left( \varepsilon _{\alpha }-\varepsilon
_{\alpha ^{\prime }}+E_{\varpi }-E_{\varpi ^{\prime }}\right) \right]
V_{\alpha \varpi ,\alpha ^{\prime }\varpi ^{\prime }}V_{\beta ^{\prime
}\varpi ^{\prime },\beta \varpi }\right\} .  \label{DME-Rdiag}
\end{eqnarray}
Eq. (\ref{DME-rhoEqTensor}) can be written in a more compact form
\begin{equation}
\frac{d}{dt}\rho _{\alpha \beta }=\sum_{\alpha ^{\prime }\beta ^{\prime
}}\Phi _{\alpha \beta ,\alpha ^{\prime }\beta ^{\prime }}\rho _{\alpha
^{\prime }\beta ^{\prime }},  \label{DME-rhoEq-Phi}
\end{equation}
where
\begin{equation}
\Phi _{\alpha \beta ,\alpha ^{\prime }\beta ^{\prime }}\equiv -i\omega
_{\alpha \beta }\delta _{\alpha \alpha ^{\prime }}\delta _{\beta \beta
^{\prime }}+R_{\alpha \beta ,\alpha ^{\prime }\beta ^{\prime }}.
\label{DME-PhiDef}
\end{equation}
In the case of time-dependent Hamiltonian one has to add the non-adiabatic
terms from Eq. (\ref{DME-rhoEq-diagbasist}) to the above DME.

The asymptotic solution of Eq. (\ref{DME-rhoEqTensor}) is
\begin{equation}
\rho _{\alpha \beta }^{\mathrm{eq}}=\frac{1}{Z_{\mathrm{s}}}\exp \left( -%
\frac{\varepsilon _{\alpha }}{k_{B}T}\right) \delta _{\alpha \beta }
\label{DME-rhoalphabetaEquil}
\end{equation}
that describes thermal equilibrium. This will be shown in Sec. \ref
{Sec-NonSec-Analysis}.

\subsection{Secular approximation and Fermi golden rule}

\label{Sec-DME-SecAppr}

One can transform Eq. (\ref{DME-rhoEqTensor}) back into the interaction
representation using Eqs. (\ref{DME-RhoIDef}) and (\ref{DME-DMviaDO}). In
the diagonal basis, the relation between $\rho _{\alpha \beta }(t)$ and $%
\rho _{\alpha \beta }(t)_{I}$ has the simple form
\begin{equation}
\rho _{\alpha \beta }(t)_{I}=\langle \chi _{\alpha }|e^{i\hat{H}_{s}t/\hbar }%
\hat{\rho}(t)e^{-i\hat{H}_{s}t/\hbar }|\chi _{\beta }\rangle =e^{i\omega
_{\alpha \beta }t}\rho _{\alpha \beta }(t).  \label{DME-rhotIrhotRelation}
\end{equation}
Computing the time derivative of $\rho _{\alpha \beta }(t)_{I}$ and using
Eq. (\ref{DME-rhoEqTensor}), one arrives at the equation
\begin{equation}
\frac{d}{dt}\rho _{\alpha \beta }(t)_{I}=\sum_{\alpha ^{\prime }\beta
^{\prime }}e^{i\left( \omega _{\alpha \beta }-\omega _{\alpha ^{\prime
}\beta ^{\prime }}\right) t}R_{\alpha \beta ,\alpha ^{\prime }\beta ^{\prime
}}\rho _{\alpha ^{\prime }\beta ^{\prime }}(t)_{I},  \label{DME-rhodiagIEq}
\end{equation}
where the conservative term disappeared and the relaxation term has an
explicit time dependence. While the change of the density matrix due to the
relaxation is slow, the oscillation in the terms with $\omega _{\alpha \beta
}\neq \omega _{\alpha ^{\prime }\beta ^{\prime }}$ are generally fast. These
fast oscillating terms average out and make a negligible contribution into
the dynamics of the small system. In general, all transition frequencies are
nondegenerate, so that one can drop all terms with $\alpha \neq \alpha
^{\prime }$ and $\beta \neq \beta ^{\prime },$ if $\alpha \neq \beta .$ In
the equations for the diagonal terms $\rho _{\alpha \alpha }(t)_{I}$ one can
keep only diagonal terms with $\alpha ^{\prime }=\beta ^{\prime }.$ This is
the \emph{secular approximation} that tremendously simplifies the DME. In
the secular approximation one has
\begin{eqnarray}
\frac{d}{dt}\rho _{\alpha \beta }(t)_{I} &=&\delta _{\alpha \beta
}\sum_{\alpha ^{\prime }}R_{\alpha \alpha ,\alpha ^{\prime }\alpha ^{\prime
}}\rho _{\alpha ^{\prime }\alpha ^{\prime }}(t)_{I}+\left( 1-\delta _{\alpha
\beta }\right) R_{\alpha \beta ,\alpha \beta }\rho _{\alpha \beta }(t)_{I}
\nonumber \\
&=&\delta _{\alpha \beta }\sum_{\alpha ^{\prime }\neq \alpha }R_{\alpha
\alpha ,\alpha ^{\prime }\alpha ^{\prime }}\rho _{\alpha ^{\prime }\alpha
^{\prime }}(t)_{I}+R_{\alpha \beta ,\alpha \beta }\rho _{\alpha \beta
}(t)_{I}.  \label{DME-Secular-I}
\end{eqnarray}
Simplifying Eq. (\ref{DME-Rdiag}) and using the Hermiticity
\begin{equation}
V_{\alpha ^{\prime }\varpi ^{\prime },\alpha \varpi }=V_{\alpha \varpi
,\alpha ^{\prime }\varpi ^{\prime }}^{\ast }
\end{equation}
one obtains
\begin{equation}
\left. R_{\alpha \alpha ,\alpha ^{\prime }\alpha ^{\prime }}\right| _{\alpha
^{\prime }\neq \alpha }=\frac{2\pi }{\hbar Z_{b}}\sum_{\varpi \varpi
^{\prime }}e^{-E_{\varpi ^{\prime }}/(k_{B}T)}\delta \left( \varepsilon
_{\alpha }-\varepsilon _{\alpha ^{\prime }}+E_{\varpi }-E_{\varpi ^{\prime
}}\right) \left| V_{\alpha \varpi ,\alpha ^{\prime }\varpi ^{\prime
}}\right| ^{2}\equiv \Gamma _{\alpha \alpha ^{\prime }}
\label{DME-WaplaalphaprDef}
\end{equation}
and
\begin{eqnarray}
R_{\alpha \beta ,\alpha \beta } &=&\frac{\pi }{\hbar Z_{b}}\sum_{\varpi
\varpi ^{\prime }}\left\{ -\sum_{\gamma }e^{-E_{\varpi }/(k_{B}T)}\delta
\left( \varepsilon _{\alpha }-\varepsilon _{\gamma }+E_{\varpi }-E_{\varpi
^{\prime }}\right) \left| V_{\alpha \varpi ,\gamma \varpi ^{\prime }}\right|
^{2}\right.  \nonumber \\
&&-\sum_{\gamma }e^{-E_{\varpi }/(k_{B}T)}\delta \left( \varepsilon _{\beta
}-\varepsilon _{\gamma }+E_{\varpi }-E_{\varpi ^{\prime }}\right) \left|
V_{\beta \varpi ,\gamma \varpi ^{\prime }}\right| ^{2}  \nonumber \\
&&+2\left. e^{-E_{\varpi ^{\prime }}/(k_{B}T)}\delta \left( E_{\varpi
}-E_{\varpi ^{\prime }}\right) V_{\alpha \varpi ,\alpha \varpi ^{\prime
}}V_{\beta \varpi ^{\prime },\beta \varpi }\right\} .
\end{eqnarray}
Rearranging the terms one obtains
\begin{equation}
R_{\alpha \beta ,\alpha \beta }=-\tilde{\Gamma}_{\alpha \beta },
\label{DME-outgoing}
\end{equation}
where
\begin{equation}
\tilde{\Gamma}_{\alpha \beta }=\bar{\Gamma}_{\alpha \beta }+\frac{1}{2}%
\left( \sum_{\alpha ^{\prime }\neq \alpha }\Gamma _{\alpha ^{\prime }\alpha
}+\sum_{\beta ^{\prime }\neq \beta }\Gamma _{\beta ^{\prime }\beta }\right) .
\label{DME-GammaalphabetaDef}
\end{equation}
Here $\Gamma _{\alpha ^{\prime }\alpha }$ is defined by Eq. (\ref
{DME-WaplaalphaprDef}) and
\begin{equation}
\bar{\Gamma}_{\alpha \beta }=\frac{\pi }{\hbar Z_{b}}\sum_{\varpi \varpi
^{\prime }}e^{-E_{\varpi }/(k_{B}T)}\delta \left( E_{\varpi }-E_{\varpi
^{\prime }}\right) \left| V_{\alpha \varpi ,\alpha \varpi ^{\prime
}}-V_{\beta \varpi ^{\prime },\beta \varpi }\right| ^{2}.
\label{DME-DephasingRate}
\end{equation}
Using these results in Eq. (\ref{DME-Secular-I}) and changing to $\rho
_{\alpha \beta },$ one obtains the secular DME in the diagonal basis in the
form
\begin{equation}
\frac{d}{dt}\rho _{\alpha \beta }=-\left( i\omega _{\alpha \beta }+\tilde{%
\Gamma}_{\alpha \beta }\right) \rho _{\alpha \beta }+\delta _{\alpha \beta
}\sum_{\alpha ^{\prime }\neq \alpha }\Gamma _{\alpha \alpha ^{\prime }}\rho
_{\alpha ^{\prime }\alpha ^{\prime }}.  \label{DME-DME-final}
\end{equation}

In Eq. (\ref{DME-WaplaalphaprDef}), $\Gamma _{\alpha \alpha ^{\prime }}$ is
the rate of quantum transitions $\alpha ^{\prime }\rightarrow \alpha $ in
the small system, accompanied by an appropriate transition in the bath so
that the total energy is conserved. To the contrary, $\Gamma _{\alpha
^{\prime }\alpha }$ in Eq. (\ref{DME-outgoing}) is the rate of quantum
transitions $\alpha \rightarrow \alpha ^{\prime }$ in the small system. One
can relate both rates as
\begin{eqnarray}
\Gamma _{\alpha ^{\prime }\alpha } &=&\frac{\pi }{\hbar Z_{b}}\sum_{\varpi
\varpi ^{\prime }}e^{-E_{\varpi }/(k_{B}T)}\delta \left( \varepsilon
_{\alpha }-\varepsilon _{\alpha ^{\prime }}+E_{\varpi }-E_{\varpi ^{\prime
}}\right) \left| V_{\alpha \varpi ,\alpha ^{\prime }\varpi ^{\prime
}}\right| ^{2}  \nonumber \\
&=&\frac{\pi }{\hbar Z_{b}}\sum_{\varpi \varpi ^{\prime }}e^{\left(
-E_{\varpi ^{\prime }}+\varepsilon _{\alpha }-\varepsilon _{\alpha ^{\prime
}}\right) /(k_{B}T)}\delta \left( \varepsilon _{\alpha }-\varepsilon
_{\alpha ^{\prime }}+E_{\varpi }-E_{\varpi ^{\prime }}\right) \left|
V_{\alpha \varpi ,\alpha ^{\prime }\varpi ^{\prime }}\right| ^{2}
\end{eqnarray}
or
\begin{equation}
\Gamma _{\alpha ^{\prime }\alpha }=e^{\left( \varepsilon _{\alpha
}-\varepsilon _{\alpha ^{\prime }}\right) /(k_{B}T)}\Gamma _{\alpha \alpha
^{\prime }}.  \label{DME-detailedbalance}
\end{equation}
This is the so-called \emph{detailed-balance relation} that ensures that asymptotically the small system reaches the thermal equilibrium
described by Eq. (\ref{DME-rhoalphabetaEquil}). If $\varepsilon _{\alpha
}<\varepsilon _{\alpha ^{\prime }},$ then at low temperatures the rate of
transitions $\alpha \rightarrow \alpha ^{\prime }$ (with increasing energy)
is exponentially small. Note that the transition rates $\Gamma _{\alpha
^{\prime }\alpha }$ and $\Gamma _{\alpha \alpha ^{\prime }}$ correspond to
the Fermi golden rule.

The quantity $\bar{\Gamma}_{\alpha \beta }$ of Eq. (\ref{DME-DephasingRate})
is the dephasing rate that turns to zero for the diagonal elements of the
density matrix
\begin{equation}
n_{\alpha }\equiv \rho _{\alpha \alpha },  \label{DME-nalphaDef}
\end{equation}
the populations of states $\alpha$. The dephasing rate is not related to any transitions of the small system.
Its origin is modulating its transition frequencies $\omega _{\alpha \beta }$
by fluctuations of the bath.

One can see that in the decoupled equations for nondiagonal terms of the
density matrix $\alpha \neq \beta $ in Eq. (\ref{DME-DME-final}) there are
only outgoing terms, so that the nondiagonal terms tend to zero
asymptotically. The diagonal terms of the DME satisfy the system of rate
equations
\begin{equation}
\frac{d}{dt}n_{\alpha }=\sum_{\alpha ^{\prime }\neq \alpha }\left( \Gamma
_{\alpha \alpha ^{\prime }}n_{\alpha ^{\prime }}-\Gamma _{\alpha ^{\prime
}\alpha }n_{\alpha }\right) .  \label{DME-rateEqs}
\end{equation}
The asymptotic solution satisfies
\begin{equation}
\frac{n_{\alpha ^{\prime }}}{n_{\alpha }}=\frac{\Gamma _{\alpha ^{\prime
}\alpha }}{\Gamma _{\alpha \alpha ^{\prime }}}=\frac{e^{-\varepsilon
_{\alpha ^{\prime }}/(k_{B}T)}}{e^{-\varepsilon _{\alpha }/(k_{B}T)}}
\label{DME-nalphaEquilibr}
\end{equation}
that corresponds to the thermal equilibrium. The equation for nondiagonal
elements can be written in the form
\begin{equation}
\frac{d}{dt}\rho _{\alpha \beta }=-\left( i\omega _{\alpha \beta }+\tilde{%
\Gamma}_{\alpha \beta }\right) \rho _{\alpha \beta }
\label{DME-nondiagelementsEq}
\end{equation}
with $\Gamma _{\alpha \beta }$ given by Eq. (\ref{DME-GammaalphabetaDef}).

\subsection{Analysis of the non-secular DME}

\label{Sec-NonSec-Analysis}

We have seen above that within the secular approximation nondiagonal DM
elements are decoupled from diagonal ones and oscillate with decay
independently of each other. In the full DME of Eq. (\ref{DME-rhoEqTensor}),
nondiagonal DM elements are coupled to the diagonal elements. Nevertheless,
they approach zero at equilibrium, in spite of diagonal elements being
nonzero. This points at an interesting feature of the full DME that has to
be worked out in more detail. Separating diagonal and nondiagonal elements
in the relaxation term of Eq. (\ref{DME-rhoEqTensor}), one obtains
\begin{eqnarray}
\frac{d}{dt}\rho _{\alpha \beta } &=&-i\omega _{\alpha \beta }\rho _{\alpha
\beta }+\sum_{\alpha ^{\prime }\beta ^{\prime }}R_{\alpha \beta ,\alpha
^{\prime }\beta ^{\prime }}\left( 1-\delta _{\alpha ^{\prime }\beta ^{\prime
}}\right) \rho _{\alpha ^{\prime }\beta ^{\prime }}+\sum_{\alpha ^{\prime
}}R_{\alpha \beta ,\alpha ^{\prime }\alpha ^{\prime }}\rho _{\alpha ^{\prime
}\alpha ^{\prime }}  \nonumber \\
&=&-i\omega _{\alpha \beta }\rho _{\alpha \beta }+\sum_{\alpha ^{\prime
}\beta ^{\prime }}R_{\alpha \beta ,\alpha ^{\prime }\beta ^{\prime }}\left(
1-\delta _{\alpha ^{\prime }\beta ^{\prime }}\right) \rho _{\alpha ^{\prime
}\beta ^{\prime }}  \nonumber \\
&&+\frac{\pi }{\hbar Z_{\mathrm{b}}}\sum_{\varpi \varpi ^{\prime }}\left\{
-\sum_{\gamma }e^{-E_{\varpi }/(k_{B}T)}\delta \left( \varepsilon _{\beta
}-\varepsilon _{\gamma }+E_{\varpi }-E_{\varpi ^{\prime }}\right) V_{\alpha
\varpi ,\gamma \varpi ^{\prime }}V_{\gamma \varpi ^{\prime },\beta \varpi
}\rho _{\beta \beta }\right.  \nonumber \\
&&-\sum_{\gamma }e^{-E_{\varpi }/(k_{B}T)}\delta \left( \varepsilon _{\alpha
}-\varepsilon _{\gamma }+E_{\varpi }-E_{\varpi ^{\prime }}\right) V_{\alpha
\varpi ,\gamma \varpi ^{\prime }}V_{\gamma \varpi ^{\prime },\beta \varpi
}\rho _{\alpha \alpha }  \nonumber \\
&&+\sum_{\gamma }e^{-E_{\varpi ^{\prime }}/(k_{B}T)}\delta \left(
\varepsilon _{\beta }-\varepsilon _{\gamma }+E_{\varpi }-E_{\varpi ^{\prime
}}\right) V_{\alpha \varpi ,\gamma \varpi ^{\prime }}V_{\gamma \varpi
^{\prime },\beta \varpi }\rho _{\gamma \gamma }  \nonumber \\
&&+\left. \sum_{\gamma }e^{-E_{\varpi ^{\prime }}/(k_{B}T)}\delta \left(
\varepsilon _{\alpha }-\varepsilon _{\gamma }+E_{\varpi }-E_{\varpi ^{\prime
}}\right) V_{\alpha \varpi ,\gamma \varpi ^{\prime }}V_{\gamma \varpi
^{\prime },\beta \varpi }\rho _{\gamma \gamma }\right\} .
\end{eqnarray}
With the use of the energy conservation this can be rewritten as
\begin{eqnarray}
\frac{d}{dt}\rho _{\alpha \beta } &=&-i\omega _{\alpha \beta }\rho _{\alpha
\beta }+\sum_{\alpha ^{\prime }\beta ^{\prime }}R_{\alpha \beta ,\alpha
^{\prime }\beta ^{\prime }}\left( 1-\delta _{\alpha ^{\prime }\beta ^{\prime
}}\right) \rho _{\alpha ^{\prime }\beta ^{\prime }}  \nonumber \\
&&+\frac{\pi }{\hbar Z_{\mathrm{b}}}\sum_{\gamma }\sum_{\varpi \varpi
^{\prime }}e^{-E_{\varpi ^{\prime }}/(k_{B}T)}V_{\alpha \varpi ,\gamma
\varpi ^{\prime }}V_{\gamma \varpi ^{\prime },\beta \varpi }  \nonumber \\
&&\times \left[ \delta \left( \varepsilon _{\beta }-\varepsilon _{\gamma
}+E_{\varpi }-E_{\varpi ^{\prime }}\right) \left( \rho _{\gamma \gamma
}-e^{(\varepsilon _{\beta }-\varepsilon _{\gamma })/(k_{B}T)}\rho _{\beta
\beta }\right) \right.  \nonumber \\
&&+\left. \delta \left( \varepsilon _{\alpha }-\varepsilon _{\gamma
}+E_{\varpi }-E_{\varpi ^{\prime }}\right) \left( \rho _{\gamma \gamma
}-e^{(\varepsilon _{\alpha }-\varepsilon _{\gamma })/(k_{B}T)}\rho _{\alpha
\alpha }\right) \right] .  \label{DME-RhoEqNonSecDetailed}
\end{eqnarray}
One can see that as the diagonal DM elements approach their equilibrium
values, they cease to drive nondiagonal elements because of the detailed
balance relation. Thus the equilibrium solution of the full DME is Eq. (\ref
{DME-rhoalphabetaEquil}).

\subsection{Semi-secular approximation}

\label{Sec-semisecular0}

While the secular approximation neglects the interaction between diagonal and slow
nondiagonal DM elements, the full non-secular formalism involves a big $%
N^{2}\times N^{2}$ matrix that has to be diagonalized. In important
particular cases such as thermal activation over a barrier or tunneling, the
eigenvalues of the DM span a broad range from very fast to very slow, the
latter being of a primary importance in relaxation. Because of this, one has
to do numerical calculations with increased precision that makes them very
slow. This difficulty can be overcome with the help of the semi-secular
approximation that considers coupled equations for diagonal and slow
nondiagonal DM elements plus decoupled equations for the fast DM elements. The easiest way to
implement it is to include the diagonal and subdiagonal terms $\rho _{\alpha
\alpha }$ and $\rho _{\alpha ,\alpha \pm 1}$ into the slow group, because in
most situations there are only two levels that come close to each other,
making $\rho _{\alpha ,\alpha +1}$ or $\rho _{\alpha ,\alpha -1}$ slow.
Implementation of the semi-secular DME in the case of time-independent $%
\hat{H}_{\mathrm{s}}$ will be done in Sec. \ref{Sec-semisecular}.

\subsection{Transformation to the natural basis}

\label{Sec-Trans-to-nat-bas}

One can transform the DME, Eq. (\ref{DME-DME-final}) to the natural basis
using Eqs. (\ref{DME-DMviaDO}) and (\ref{DME-rhoOperatorDef}) in the form
\begin{equation}
\rho _{m n }=\langle \psi _{m }|\hat{\rho}|\psi _{n }\rangle =\sum_{\alpha
\beta }\langle \psi _{m }|\chi _{\alpha }\rangle \rho _{\alpha \beta
}\langle \chi _{\beta }|\psi _{n }\rangle .  \label{DME-rgoviarhodiag}
\end{equation}
The inverse transformation has the form
\begin{equation}
\rho _{\alpha \beta }=\langle \chi _{\alpha }|\hat{\rho}|\chi _{\beta
}\rangle =\sum_{m n }\langle \chi _{\alpha }|\psi _{m }\rangle \rho _{m n
}\langle \psi _{n }|\chi _{\beta }\rangle .  \label{DME-rhodiagviarho}
\end{equation}

The general DME in the diagonal basis, Eq. (\ref{DME-rhoEqTensor}), can be
transformed to the natural basis with the help of Eq. (\ref
{DME-rgoviarhodiag}) as follows
\begin{eqnarray}
\frac{d}{dt}\rho _{mn} &=&\sum_{\alpha \beta }\langle \psi _{m}|\chi
_{\alpha }\rangle \frac{d}{dt}\rho _{\alpha \beta }\langle \chi _{\beta
}|\psi _{n}\rangle  \nonumber \\
&=&-\frac{i}{\hbar }\sum_{\alpha \beta }\langle \psi _{m}|\chi _{\alpha
}\rangle \left( \varepsilon _{\alpha }-\varepsilon _{\beta }\right) \rho
_{\alpha \beta }\langle \chi _{\beta }|\psi _{n}\rangle +\sum_{\alpha \beta
}\langle \psi _{m}|\chi _{\alpha }\rangle \sum_{\alpha ^{\prime }\beta
^{\prime }}R_{\alpha \beta ,\alpha \beta ^{\prime }}\rho _{\alpha ^{\prime
}\beta ^{\prime }}\langle \chi _{\beta }|\psi _{n}\rangle  \nonumber \\
&=&-\frac{i}{\hbar }\sum_{m^{\prime }n^{\prime }}\sum_{\alpha \beta }\langle
\psi _{m}|\chi _{\alpha }\rangle \varepsilon _{\alpha }\langle \chi _{\alpha
}|\psi _{m^{\prime }}\rangle \rho _{m^{\prime }n^{\prime }}\langle \psi
_{n^{\prime }}|\chi _{\beta }\rangle \langle \chi _{\beta }|\psi _{n}\rangle
\nonumber \\
&&+\frac{i}{\hbar }\sum_{m^{\prime }n^{\prime }}\sum_{\alpha \beta }\langle
\psi _{m}|\chi _{\alpha }\rangle \langle \chi _{\alpha }|\psi _{m^{\prime
}}\rangle \rho _{m^{\prime }n^{\prime }}\langle \psi _{n^{\prime }}|\chi
_{\beta }\rangle \varepsilon _{\beta }\langle \chi _{\beta }|\psi _{n}\rangle
\nonumber \\
&&+\sum_{m^{\prime }n^{\prime }}\sum_{\alpha \beta }\langle \psi _{m}|\chi
_{\alpha }\rangle \sum_{\alpha ^{\prime }\beta ^{\prime }}R_{\alpha \beta
,\alpha ^{\prime }\beta ^{\prime }}\langle \chi _{\alpha ^{\prime }}|\psi
_{m^{\prime }}\rangle \rho _{m^{\prime }n^{\prime }}\langle \psi _{n^{\prime
}}|\chi _{\beta ^{\prime }}\rangle \langle \chi _{\beta }|\psi _{n}\rangle .
\end{eqnarray}
This yields
\begin{equation}
\frac{d}{dt}\rho _{mn}=-\frac{i}{\hbar }\sum_{m^{\prime }}H_{s,mm^{\prime
}}\rho _{m^{\prime }n}+\frac{i}{\hbar }\sum_{n^{\prime }}\rho _{mn^{\prime
}}H_{s,n^{\prime }n}+\sum_{m^{\prime }n^{\prime }}R_{mn,m^{\prime }n^{\prime
}}\rho _{m^{\prime }n^{\prime }},  \label{DME-rhoEqTensorNatural}
\end{equation}
where $H_{s,mm^{\prime }}\equiv \left\langle \psi _{m}\left| \hat{H}%
_{s}\right| \psi _{m^{\prime }}\right\rangle $ and
\begin{equation}
R_{mn,m^{\prime }n^{\prime }}\equiv \sum_{\alpha \beta ,\alpha ^{\prime
}\beta ^{\prime }}T_{mn,m^{\prime }n^{\prime };\alpha \beta ,\alpha ^{\prime
}\beta ^{\prime }}R_{\alpha \beta ,\alpha ^{\prime }\beta ^{\prime }},
\label{DME-RNatural}
\end{equation}
where
\begin{equation}
T_{mn,m^{\prime }n^{\prime };\alpha \beta ,\alpha ^{\prime }\beta ^{\prime
}}\equiv \langle \psi _{m}|\chi _{\alpha }\rangle \langle \chi _{\beta
}|\psi _{n}\rangle \langle \chi _{\alpha ^{\prime }}|\psi _{m^{\prime
}}\rangle \langle \psi _{n^{\prime }}|\chi _{\beta ^{\prime }}\rangle .
\label{DME-TTransfDef}
\end{equation}

Additionally, the matrix elements in the Fermi-golden-rule transition rate,
Eq. (\ref{DME-WaplaalphaprDef}), that are defined with respect to the
diagonal basis, can be expressed through those with respect to the natural
basis. Similarly to Eq. (\ref{DME-rhodiagviarho}) one obtains
\begin{equation}
V_{\alpha \varpi ,\alpha ^{\prime }\varpi ^{\prime }}=\sum_{mm^{\prime
}}\langle \chi _{\alpha }|\psi _{m}\rangle V_{m\varpi ,m^{\prime }\varpi
^{\prime }}\langle \psi _{m^{\prime }}|\chi _{\alpha ^{\prime }}\rangle .
\label{DME-VTransformation}
\end{equation}
%

\newpage
\section{Time-dependent problems}


In this section we consider the DME and its solution in three important
cases: i) free evolution for time-independent $\hat{H}_{\mathrm{s}}$; ii)
fast resonance perturbation and iii) periodic or nonperiodic slow
perturbation. In the second case it is sufficient to keep the perturbation
in the conservative part of the DME only, while in the third case
modification of the relaxation terms is required, as well. In all these
cases the solution can be obtained by matrix algebra. To the contrast,
problems with a large temporal change of $\hat{H}_{\mathrm{s}}$ cannot be
solved by matrix algebra. The secular, non-secular, and semi-secular
versions of the DME yield the same results except for the case of
anomalously close energy levels (e.g., tunnel split levels). In the latter
case the semi-secular DME is preferred, while in general the fastest and
easiest secular DME is the best choice.

\subsection{Free evolution}

\label{Sec-Free-evolution}

\subsubsection{Non-secular DME}

The solution of time-independent DME, Eq. (\ref{DME-rhoEqTensor}), that can
be rewritten as
\begin{equation}
\frac{d}{dt}\rho _{\alpha \beta }=\sum_{\alpha ^{\prime }\beta ^{\prime
}}\Phi _{\alpha \beta ,\alpha ^{\prime }\beta ^{\prime }}\rho _{\alpha
^{\prime }\beta ^{\prime }},\qquad \Phi _{\alpha \beta ,\alpha ^{\prime
}\beta ^{\prime }}=-i\omega _{\alpha \beta }\delta _{\alpha \alpha ^{\prime
}}\delta _{\beta \beta ^{\prime }}+R_{\alpha \beta ,\alpha ^{\prime }\beta
^{\prime }},  \label{DME-LambdaDef}
\end{equation}
is a linear combination of time exponentials with exponents being
eigenvalues of the matrix $\mathbf{\Phi }$ building the DME. To bring the
DME into a standard form, it is convenient to introduce the compound index $%
a $ defined by
\begin{equation}
a=\alpha +N(\beta -1),  \label{DME-aIndexDef}
\end{equation}
where $N$ is the size of the density matrix, i.e., $\alpha ,\beta
=1,2,\ldots ,N.$ Then $a=1,2,\ldots ,N^{2}.$ Inversion of Eq. (\ref
{DME-aIndexDef}) yields
\begin{equation}
\alpha =1+N\mathrm{Frac}\left( \frac{a-1}{N}\right) ,\qquad \beta =1+\mathrm{%
Int}\left( \frac{a-1}{N}\right) .  \label{DME-alphabetaviaa}
\end{equation}
With the index $a,$ the density matrix $\rho _{\alpha \beta }$ becomes a
vector with the components $\rho _{a}$ while $\Phi _{\alpha \beta ,\alpha
^{\prime }\beta ^{\prime }}$ becomes a matrix with the elements $\Phi
_{aa^{\prime }}:$%
\begin{equation}
\frac{d}{dt}\mathbf{\rho =\Phi \cdot \rho ,\qquad }\frac{d}{dt}\rho
_{a}=\sum_{a^{\prime }}\Phi _{aa^{\prime }}\rho _{a^{\prime }}.
\label{DME-DME-vectorized}
\end{equation}

The eigenvalue problem for the DME can be written as
\begin{equation}
\mathbf{\Phi \cdot R}_{\mu }\mathbf{=-}\Lambda _{\mu }\mathbf{R}_{\mu
},\qquad \Lambda _{\mu }=\Gamma _{\mu }+i\Omega _{\mu },
\label{DME-DME-eigenproblem}
\end{equation}
where $\mathbf{R}_{\mu }$ is the right eigenvector corresponding to the
eigenvalue $\Lambda _{\mu }$ and $\mu =1,2,\ldots ,(2S+1)^{2}.$ Since $%
\mathbf{\Phi }$ is a non-Hermitean matrix, right eigenvectors differ from
left eigenvectors that satisfy $\mathbf{L}_{\mu }\mathbf{\cdot \Phi =-}%
\Lambda _{\mu }\mathbf{L}_{\mu }.$ Left and right eigenvectors satisfy
orthonormality and completeness relations
\begin{equation}
\sum_{a}L_{\mu a}R_{\nu a}=\delta _{\mu \nu },\qquad \sum_{\mu }L_{\mu
a}R_{\mu a^{\prime }}=\delta _{aa^{\prime }}.  \label{DME-LRorthocompl}
\end{equation}
In general, $\mathbf{L}_{\mu }$ and $\mathbf{R}_{\mu }$ are not Hermitean
conjugate, see, e.g., Eq. (\ref{DME-L0R0}). All real parts of the
eigenvalues are positive, $\Gamma _{\mu }>0.$ There are $N$ purely real
eigenvalues, one of which is zero and corresponds to thermal equilibrium. We
assign the zero eigenvalue the index $\mu =1.$ Complex eigenvalues occur in
complex conjugate pairs.

The solution of the DME with the initial conditions can be written in the
form
\begin{equation}
\mathbf{\rho }(t)=\sum_{\mu }\mathbf{R}_{\mu }e^{-\Lambda _{\mu }t}\mathbf{L}%
_{\mu }\cdot \mathbf{\rho }(0).  \label{DME-rhoSol}
\end{equation}
The fully vectorized form of this equation is
\begin{equation}
\mathbf{\rho }(t)=\mathbf{E\cdot W}(t)\mathbf{\cdot E}^{-1}\cdot \mathbf{%
\rho }(0),  \label{DME-rhoSolVect}
\end{equation}
where $\mathbf{E}$ is right-eigenvector matrix composed of all eigenvectors $%
\mathbf{R}_{\mu }$ standing vertically, $\mathbf{E}^{-1}$ is the
left-eigenvector matrix, composed of all left eigenvectors lying
horizontally, and $\mathbf{W}(t)$ is the diagonal matrix with the elements $%
e^{-\Lambda _{\mu }t}.$ In fact,
\begin{equation}
\mathbf{E\cdot W}(t)\mathbf{\cdot E}^{-1}=\exp \left( \mathbf{\Phi }t\right)
.  \label{DME-EWEExpPsi}
\end{equation}
The asymptotic value of $\mathbf{\rho }(t)$ is described by the zero
eigenvalue $\Lambda _{1}=0$,
\begin{equation}
\mathbf{\rho }(\infty )=\mathbf{R}_{1}\left( \mathbf{L}_{1}\cdot \mathbf{%
\rho }(0)\right) .  \label{DME-rhoinf}
\end{equation}
Here $\mathbf{\rho }(\infty )=\mathbf{\rho }^{\mathrm{eq}}$ should be
satisfied, where $\mathbf{\rho }^{\mathrm{eq}}$ follows from Eq. (\ref
{DME-rhoalphabetaEquil}), and this result should be independent of $\mathbf{%
\rho }(0).$ Thus one concludes that
\begin{equation}
L_{1a}=\delta _{\alpha (a)\beta (a)},\qquad R_{1a}=\frac{1}{Z_{\mathrm{s}}}%
\exp \left( -\frac{\varepsilon _{\alpha (a)}}{k_{B}T}\right) \delta _{\alpha
(a)\beta (a)},  \label{DME-L0R0}
\end{equation}
where $\alpha (a)$ and $\beta (a)$ are given by Eq. (\ref{DME-alphabetaviaa}%
). This means that $\mathbf{L}_{1}$ is related to the normalization of the
DM while $\mathbf{R}_{1}$ contains the information about the equilibrium
state. One obtains
\begin{equation}
\mathbf{L}_{1}\cdot \mathbf{\rho }(0)=\sum_{a}L_{1a}\rho
_{a}(0)=\sum_{\alpha \beta }\delta _{\alpha \beta }\rho _{\alpha \beta
}(0)=\sum_{\alpha }\rho _{\alpha \alpha }(0)=1
\end{equation}
and $\mathbf{\rho }(\infty )=\mathbf{R}_{1}=\mathbf{\rho }^{\mathrm{eq}},$
as it should be. Note that $\mathbf{R}_{1}$ and $\mathbf{L}_{1}$ satisfy the
orthonormality condition in Eq. (\ref{DME-LRorthocompl}),
\begin{equation}
\sum_{a}L_{1a}R_{1a}=\frac{1}{Z_{\mathrm{s}}}\sum_{a}\exp \left( -\frac{%
\varepsilon _{\alpha (a)}}{k_{B}T}\right) \delta _{\alpha (a)\beta (a)}=%
\frac{1}{Z_{\mathrm{s}}}\sum_{\alpha \beta }\exp \left( -\frac{\varepsilon
_{\alpha }}{k_{B}T}\right) \delta _{\alpha \beta }=1.
\end{equation}

The time dependence of any physical quantity $A$ is given by Eq. (\ref
{DME-Aavrrho}) that can be rewritten in the form
\begin{equation}
A(t)=\sum_{\alpha \beta }A_{\beta \alpha }\rho _{\alpha \beta
}(t)=\sum_{a}A_{a}\rho _{a}(t),  \label{DME-AtDependenceDef}
\end{equation}
where
\begin{equation}
A_{a}\equiv A_{\beta (a)\alpha (a)},\qquad A_{\beta \alpha }\equiv
\left\langle \beta \left| \hat{A}\right| \alpha \right\rangle .
\label{DME-AaDef}
\end{equation}
Writing Eq. (\ref{DME-AtDependenceDef}) in the vector form as $A(t)=\mathbf{%
A\cdot \rho }(t)$ one obtains
\begin{equation}
A(t)=\sum_{\mu }\mathbf{A\cdot R}_{\mu }e^{-\Lambda _{\mu }t}\mathbf{L}_{\mu
}\cdot \mathbf{\rho }(0)  \label{DME-AtDependence}
\end{equation}
or, in the fully vectorized form,
\begin{equation}
A(t)=\mathbf{A\cdot E\cdot W}(t)\mathbf{\cdot E}^{-1}\cdot \mathbf{\rho }(0).
\label{DME-AtDependenceVec}
\end{equation}

Since the time dependence of observables in the course of evolution of the
density matrix is described by more than one exponential, one needs an
appropriate definition of the relaxation rate or relaxation time. A
convenient way is to use the integral relaxation time defined as the area
under the relaxation curve
\begin{equation}
\tau _{\mathrm{int}}\equiv \frac{\int_{0}^{\infty }dt\,\left[ A(t)-A(\infty )%
\right] }{A(0)-A(\infty )}.  \label{DME-tauintDef}
\end{equation}
One can check that in the case of a single exponential, $A(t)=A(\infty )+%
\left[ A(0)-A(\infty )\right] e^{-\Gamma t},$ the result is $\tau _{\mathrm{%
int}}=1/\Gamma .$ From Eq. (\ref{DME-AtDependence}) one obtains
\begin{equation}
A(t)-A(\infty )=\sum_{\mu \neq 1}\mathbf{A\cdot R}_{\mu }e^{-\Lambda _{\mu
}t}\mathbf{L}_{\mu }\cdot \mathbf{\rho }(0)
\end{equation}
and thus
\begin{equation}
\tau _{\mathrm{int}}=\frac{\sum_{\mu =2}^{N^{2}}\left( \mathbf{A\cdot R}%
_{\mu }\right) \Lambda _{\mu }^{-1}\left( \mathbf{L}_{\mu }\cdot \mathbf{%
\rho }(0)\right) }{\sum_{\mu =2}^{N^{2}}\left( \mathbf{A\cdot R}_{\mu
}\right) \left( \mathbf{L}_{\mu }\cdot \mathbf{\rho }(0)\right) }.
\label{DME-tauintGenFin}
\end{equation}
This formula cannot be fully vectorized since summation skips the static
eigenvalue $\mu =1.$

\subsubsection{Secular DME}

\label{Sec-FE-secular}

Within the secular approximation, one has to consider the dynamics of
diagonal and nondiagonal components of the density matrix separately. The
former is described by Eqs. (\ref{DME-DME-vectorized})--(\ref{DME-rhoinf})
where the vector $\mathbf{\rho }$ is replaced by the vector of the diagonal
components $\mathbf{n}=\{n_{\alpha }\}=\{\rho _{\alpha \alpha }\}$ and the $%
\left( N\right) ^{2}\times \left( N\right) ^{2}$ matrix $\mathbf{\Phi }
$ is replaced by the $\left( N\right) \times \left( N\right) $ matrix $%
\mathbf{\Phi }^{\sec }$ having matrix elements
\begin{equation}
\Phi _{\alpha \alpha ^{\prime }}^{\sec }=\left( 1-\delta _{\alpha \alpha
^{\prime }}\right) \Gamma _{\alpha \alpha ^{\prime }}-\delta _{\alpha \alpha
^{\prime }}\sum_{\gamma }\Gamma _{\gamma \alpha },  \label{DME-PsisecDeg}
\end{equation}
as follows from Eqs. (\ref{DME-DME-final}) or (\ref{DME-rateEqs}). All
eigenvalues of $\mathbf{\Phi }^{\sec }$ are positive reals, except for one
zero eigenvalue, $\Lambda _{1}=0.$ Eq. (\ref{DME-L0R0}) becomes simply
\begin{equation}
L_{1\alpha }=1,\qquad R_{1\alpha }=\frac{1}{Z_{\mathrm{s}}}\exp \left( -%
\frac{\varepsilon _{\alpha }}{k_{B}T}\right) .  \label{DME-L0R0sec}
\end{equation}
If the initial condition is a diagonal matrix, the non-diagonal elements do
not arise dynamically and hence they can be dropped. Then the time
dependence of any quantity $A$ is described by
\begin{equation}
A(t)=\sum_{\alpha }A_{\alpha \alpha }n_{\alpha }(t)=\mathbf{A\cdot E\cdot W}%
(t)\mathbf{\cdot E}^{-1}\cdot \mathbf{n}(0),
\end{equation}
c.f. Eqs. (\ref{DME-AtDependenceDef})--(\ref{DME-AtDependenceVec}). For the
integral relaxation time in the case of a purely diagonal evolution one
obtains
\begin{equation}
\tau _{\mathrm{int}}=\frac{\sum_{\mu =2}^{N}\left( \mathbf{A\cdot R}_{\mu
}\right) \Lambda _{\mu }^{-1}\left( \mathbf{L}_{\mu }\cdot \mathbf{n}%
(0)\right) }{\sum_{\mu =2}^{N}\left( \mathbf{A\cdot R}_{\mu }\right) \left(
\mathbf{L}_{\mu }\cdot \mathbf{n}(0)\right) },  \label{DME-tauintSecDiag}
\end{equation}
c.f. Eq. (\ref{DME-tauintGenFin}).

If the initial state is a non-diagonal density matrix, one has to add the
corresponding trivial terms following from Eq. (\ref{DME-nondiagelementsEq}),
\begin{equation}
A(t)=\mathbf{A\cdot E\cdot W}(t)\mathbf{\cdot E}^{-1}\cdot \mathbf{n}%
(0)+\sum_{\alpha \neq \beta }A_{\beta \alpha }\rho _{\alpha \beta
}(0)e^{-(i\omega _{\alpha \beta }+\tilde{\Gamma}_{\alpha \beta })t}.
\label{DME-At-nonsecular}
\end{equation}
Then Eq. (\ref{DME-tauintSecDiag}) is generalized to
\begin{equation}
\tau _{\mathrm{int}}=\frac{\sum_{\mu =2}^{N}\left( \mathbf{A\cdot R}_{\mu
}\right) \Lambda _{\mu }^{-1}\left( \mathbf{L}_{\mu }\cdot \mathbf{n}%
(0)\right) +\sum_{\alpha \neq \beta }A_{\beta \alpha }\rho _{\alpha \beta
}(0)(i\omega _{\alpha \beta }+\tilde{\Gamma}_{\alpha \beta })^{-1}}{%
\sum_{\mu =2}^{N}\left( \mathbf{A\cdot R}_{\mu }\right) \left( \mathbf{L}%
_{\mu }\cdot \mathbf{n}(0)\right) +\sum_{\alpha \neq \beta }A_{\beta \alpha
}\rho _{\alpha \beta }(0)}.  \label{DME-tauintSecGeneral}
\end{equation}
Obviously this expression is real.

\subsubsection{Semi-secular DME}

\label{Sec-semisecular}

Within the semisecular approximation introduced in Sec. \ref
{Sec-semisecular0}, the slow group being formed by diagonal and subdiagonal
DM elements, $\left| \alpha -\beta \right| \leq 1$, the equations of motion
for the latter have the form
\begin{equation}
\frac{d}{dt}\rho _{\alpha \beta }=\sum_{\alpha ^{\prime }}\left( \Phi
_{\alpha \beta ;\alpha ^{\prime },\alpha ^{\prime }-1}\rho _{\alpha ^{\prime
},\alpha ^{\prime }-1}+\Phi _{\alpha \beta ;\alpha ^{\prime }\alpha ^{\prime
}}\rho _{\alpha ^{\prime }\alpha ^{\prime }}+\Phi _{\alpha \beta ;\alpha
^{\prime },\alpha ^{\prime }+1}\rho _{\alpha ^{\prime },\alpha ^{\prime
}+1}\right)   \label{DME-DME-semisec}
\end{equation}
that is a subset of Eq. (\ref{DME-LambdaDef}). In labeling matrix elements,
one can introduce the compound index
\begin{equation}
a=2(\alpha -1)+\beta .
\end{equation}
Here $\alpha =1,2,\ldots ,N$ and $\beta =\alpha -1,\alpha ,\alpha +1,$ so
that $a$ takes the values $a=1,\ldots ,3N-2.$ In terms of $a$ one has
\begin{equation}
\alpha =1+\mathrm{Int}\left( \frac{a}{3}\right) ,\qquad \beta =\alpha -1+3%
\mathrm{Frac}\left( \frac{a}{3}\right) ,
\end{equation}
and Eq. (\ref{DME-DME-semisec}) can be rewritten as
\begin{equation}
\frac{d}{dt}\rho _{a}^{\mathrm{slow}}=\sum_{a^{\prime }}\tilde{\Phi}%
_{aa^{\prime }}\rho _{a^{\prime }}^{\mathrm{slow}},
\label{DME-PhitildeMatrEq}
\end{equation}
where $\tilde{\Phi}_{aa^{\prime }}=\Phi _{\alpha (a),\beta (a);\alpha
(a^{\prime }),\beta (a^{\prime })}.$

The solution of Eq. (\ref{DME-PhitildeMatrEq}) is similar to that of Eq. (%
\ref{DME-DME-vectorized}). On the other hand, there are uncoupled DM elemens
with $\left| \alpha -\beta \right| \geq 2,$ like in the secular
approximation. Instead of Eq. (\ref{DME-At-nonsecular}) one has
\begin{equation}
A(t)=\mathbf{A\cdot E\cdot W}(t)\mathbf{\cdot E}^{-1}\cdot \mathbf{\rho }^{%
\mathrm{slow}}(0)+\sum_{\left| \alpha -\beta \right| >1}A_{\beta \alpha
}\rho _{\alpha \beta }(0)e^{-(i\omega _{\alpha \beta }+\tilde{\Gamma}%
_{\alpha \beta })t}
\end{equation}
and instead of Eq. (\ref{DME-tauintSecGeneral}) one has
\begin{equation}
\tau _{\mathrm{int}}=\frac{\sum_{\mu =2}^{3N-2}\left( \mathbf{A\cdot R}_{\mu
}\right) \Lambda _{\mu }^{-1}\left( \mathbf{L}_{\mu }\cdot \mathbf{\rho }^{%
\mathrm{slow}}(0)\right) +\sum_{\left| \alpha -\beta \right| >1}A_{\beta
\alpha }\rho _{\alpha \beta }(0)(i\omega _{\alpha \beta }+\tilde{\Gamma}%
_{\alpha \beta })^{-1}}{\sum_{\mu =2}^{3N-2}\left( \mathbf{A\cdot R}_{\mu
}\right) \left( \mathbf{L}_{\mu }\cdot \mathbf{\rho }^{\mathrm{slow}%
}(0)\right) +\sum_{\left| \alpha -\beta \right| >1}A_{\beta \alpha }\rho
_{\alpha \beta }(0)}.  \label{DME-tauint-semisec}
\end{equation}

\subsection{Resonant perturbation}

Let $\hat{H}_{s}$ contain a periodic perturbation
\begin{equation}
\hat{V}(t)=\hat{V}_{0}e^{-i\omega t}+\hat{V}_{0}^{\dagger }e^{i\omega t}
\label{DME-VperiodicDef}
\end{equation}
with the frequency $\omega >0$ close to one of transition frequences $\omega
_{\eta \eta ^{\prime }}>0$. For the resonance to occur, the latter should
exceed the relaxation rates, so that one can use the secular DME, Eq. (\ref
{DME-DME-final}). Since $\hat{V}_{0}$ is small, one can disregard it in the
relaxation terms and use the diagonal basis corresponding to the
time-independent part of $\hat{H}_{s}.$ Combining Eqs. (\ref
{DME-RhoEqNatural}) and (\ref{DME-DME-final}), one obtains
\begin{eqnarray}
\frac{d}{dt}\rho _{\alpha \beta } &=&-\frac{i}{\hbar }\left( \sum_{\alpha
^{\prime }}V(t)_{\alpha \alpha ^{\prime }}\rho _{\alpha ^{\prime }\beta
}-\sum_{\beta ^{\prime }}\rho _{\alpha \beta ^{\prime }}V(t)_{\beta ^{\prime
}\beta }\right)  \nonumber \\
&&-\left( i\omega _{\alpha \beta }+\tilde{\Gamma}_{\alpha \beta }\right)
\rho _{\alpha \beta }+\delta _{\alpha \beta }\sum_{\gamma \neq \alpha
}\Gamma _{\alpha \gamma }\rho _{\gamma \gamma }.  \label{DME-RP-DME}
\end{eqnarray}
Here one should drop all nondiagonal DM elements but $\rho _{\eta \eta
^{\prime }}$ and $\rho _{\eta ^{\prime }\eta }$ since the former are not
excited by the resonance perturbation and vanish with time. As a result, one
obtains a coupled system of equations for the diagonal elements $\rho
_{\alpha \alpha }$ and the elements $\rho _{\eta \eta ^{\prime }}$ and $\rho
_{\eta ^{\prime }\eta }=\rho _{\eta \eta ^{\prime }}^{\ast }$ of the form
\begin{eqnarray}
\frac{d}{dt}\rho _{\eta \eta ^{\prime }} &=&-\frac{i}{\hbar }\left(
e^{-i\omega t}V_{0,\eta \eta ^{\prime }}\rho _{\eta ^{\prime }\eta ^{\prime
}}-e^{-i\omega t}\rho _{\eta \eta }V_{0,\eta \eta ^{\prime }}\right) -\left(
i\omega _{\eta \eta ^{\prime }}+\tilde{\Gamma}_{\eta \eta ^{\prime }}\right)
\rho _{\eta \eta ^{\prime }}  \nonumber \\
\frac{d}{dt}\rho _{\eta \eta } &=&-\frac{i}{\hbar }\left( e^{-i\omega
t}V_{0,\eta \eta ^{\prime }}\rho _{\eta ^{\prime }\eta }-e^{i\omega t}\rho
_{\eta \eta ^{\prime }}\left( \hat{V}_{0}^{\dagger }\right) _{\eta ^{\prime
}\eta }\right) +\sum_{\gamma \neq \eta }\left( \Gamma _{\eta \gamma }\rho
_{\gamma \gamma }-\Gamma _{\gamma \eta }\rho _{\eta \eta }\right)  \nonumber
\\
\frac{d}{dt}\rho _{\eta ^{\prime }\eta ^{\prime }} &=&-\frac{i}{\hbar }%
\left( e^{i\omega t}\left( \hat{V}_{0}^{\dagger }\right) _{\eta ^{\prime
}\eta }\rho _{\eta \eta ^{\prime }}-e^{-i\omega t}\rho _{\eta ^{\prime }\eta
}V_{0,\eta \eta ^{\prime }}\right) +\sum_{\gamma \neq \eta ^{\prime }}\left(
\Gamma _{\eta ^{\prime }\gamma }\rho _{\gamma \gamma }-\Gamma _{\gamma \eta
^{\prime }}\rho _{\eta ^{\prime }\eta ^{\prime }}\right) ,
\end{eqnarray}
plus Eq. (\ref{DME-rateEqs}) for all diagonal elements with $\alpha \neq
\eta ,\eta ^{\prime }.$ Also terms such as $e^{i\omega t}\left( \hat{V}%
_{0}^{\dagger }\right) _{\eta \eta ^{\prime }}\rho _{\eta ^{\prime }\eta
}\sim e^{i\left( \omega +\omega _{\eta \eta ^{\prime }}\right) t}$ in the
second equation have been dropped. Neglecting such terms that oscillate out
is the so-called rotating-wave approximation that is similar to the secular
approximation. It is convenient to introduce the slow nondiagonal elements $%
\tilde{\rho}_{\eta \eta ^{\prime }}$ and $\tilde{\rho}_{\eta ^{\prime }\eta
} $ via
\begin{equation}
\rho _{\eta \eta ^{\prime }}=\tilde{\rho}_{\eta \eta ^{\prime }}e^{-i\omega
t},\qquad \rho _{\eta ^{\prime }\eta }=\tilde{\rho}_{\eta ^{\prime }\eta
}e^{i\omega t}.  \label{DME-rhoetaetaprslowDef}
\end{equation}
With the use of $\left( \hat{V}_{0}^{\dagger }\right) _{\eta ^{\prime }\eta
}=V_{0,\eta \eta ^{\prime }}^{\ast }$ the DME acquires the form
\begin{eqnarray}
\frac{d}{dt}\tilde{\rho}_{\eta \eta ^{\prime }} &=&-\frac{i}{\hbar }%
V_{0,\eta \eta ^{\prime }}\left( \rho _{\eta ^{\prime }\eta ^{\prime }}-\rho
_{\eta \eta }\right) +\left[ i\left( \omega -\omega _{\eta \eta ^{\prime
}}\right) -\tilde{\Gamma}_{\eta \eta ^{\prime }}\right] \tilde{\rho}_{\eta
\eta ^{\prime }}  \nonumber \\
\frac{d}{dt}\rho _{\eta \eta } &=&-\frac{2}{\hbar }\func{Im}\left( \tilde{%
\rho}_{\eta \eta ^{\prime }}V_{0,\eta \eta ^{\prime }}^{\ast }\right)
+\sum_{\gamma \neq \eta }\left( \Gamma _{\eta \gamma }\rho _{\gamma \gamma
}-\Gamma _{\gamma \eta }\rho _{\eta \eta }\right)  \nonumber \\
\frac{d}{dt}\rho _{\eta ^{\prime }\eta ^{\prime }} &=&\frac{2}{\hbar }\func{%
Im}\left( \tilde{\rho}_{\eta \eta ^{\prime }}V_{0,\eta \eta ^{\prime
}}^{\ast }\right) +\sum_{\gamma \neq \eta ^{\prime }}\left( \Gamma _{\eta
^{\prime }\gamma }\rho _{\gamma \gamma }-\Gamma _{\gamma \eta ^{\prime
}}\rho _{\eta ^{\prime }\eta ^{\prime }}\right)  \nonumber \\
\frac{d}{dt}\rho _{\alpha \alpha } &=&\sum_{\gamma \neq \alpha }\left(
\Gamma _{\alpha \gamma }\rho _{\gamma \gamma }-\Gamma _{\gamma \alpha }\rho
_{\alpha \alpha }\right) ,\qquad \alpha \neq \eta ,\eta ^{\prime }.
\label{DME-Resonance}
\end{eqnarray}
Here $\tilde{\Gamma}_{\eta \eta ^{\prime }}$ is given by Eq. (\ref
{DME-GammaalphabetaDef}).

In the strong-dephasing case $\bar{\Gamma}_{\eta \eta ^{\prime }}\gg \Gamma
_{\gamma \eta ^{\prime }},\hat{V}_{0,\eta \eta ^{\prime }},$ the nondiagonal
element $\tilde{\rho}_{\eta \eta ^{\prime }}$ quickly reaches its
quasistationary value that can be obtained from the first equation of Eq. (%
\ref{DME-Resonance}) by setting $d\tilde{\rho}_{\eta \eta ^{\prime }}/dt=0.$
This results in
\begin{equation}
\tilde{\rho}_{\eta \eta ^{\prime }}=\frac{1}{\hbar }\frac{V_{0,\eta \eta
^{\prime }}\left( \rho _{\eta ^{\prime }\eta ^{\prime }}-\rho _{\eta \eta
}\right) }{\omega -\omega _{\eta \eta ^{\prime }}+i\tilde{\Gamma}_{\eta \eta
^{\prime }}}.
\end{equation}
Substituting this into the equations for the diagonal DM elements, one
obtains
\begin{eqnarray}
\frac{d}{dt}\rho _{\eta \eta } &=&\frac{\left| \hat{V}_{0,\eta \eta ^{\prime
}}\right| ^{2}}{\hbar ^{2}}\left( \rho _{\eta ^{\prime }\eta ^{\prime
}}-\rho _{\eta \eta }\right) \frac{2\tilde{\Gamma}_{\eta \eta ^{\prime }}}{%
\left( \omega -\omega _{\eta \eta ^{\prime }}\right) ^{2}+\tilde{\Gamma}%
_{\eta \eta ^{\prime }}^{2}}+\sum_{\gamma \neq \eta }\left( \Gamma _{\eta
\gamma }\rho _{\gamma \gamma }-\Gamma _{\gamma \eta }\rho _{\eta \eta
}\right)  \nonumber \\
\frac{d}{dt}\rho _{\eta ^{\prime }\eta ^{\prime }} &=&-\frac{\left| \hat{V}%
_{0,\eta \eta ^{\prime }}\right| ^{2}}{\hbar ^{2}}\left( \rho _{\eta
^{\prime }\eta ^{\prime }}-\rho _{\eta \eta }\right) \frac{2\tilde{\Gamma}%
_{\eta \eta ^{\prime }}}{\left( \omega -\omega _{\eta \eta ^{\prime
}}\right) ^{2}+\tilde{\Gamma}_{\eta \eta ^{\prime }}^{2}}+\sum_{\gamma \neq
\eta ^{\prime }}\left( \Gamma _{\eta ^{\prime }\gamma }\rho _{\gamma \gamma
}-\Gamma _{\gamma \eta ^{\prime }}\rho _{\eta ^{\prime }\eta ^{\prime
}}\right)  \nonumber \\
\frac{d}{dt}\rho _{\alpha \alpha } &=&\sum_{\gamma \neq \alpha }\left(
\Gamma _{\alpha \gamma }\rho _{\gamma \gamma }-\Gamma _{\gamma \alpha }\rho
_{\alpha \alpha }\right) ,\qquad \alpha \neq \eta ,\eta ^{\prime }.
\label{DME-Resonance-Quasist}
\end{eqnarray}
Note that the absorbed power of the field acting on the system is given by
\begin{equation}
P=\hbar \omega _{\eta \eta ^{\prime }}\frac{\left| \hat{V}_{0,\eta \eta
^{\prime }}\right| ^{2}}{\hbar ^{2}}\left( \rho _{\eta ^{\prime }\eta
^{\prime }}-\rho _{\eta \eta }\right) \frac{2\tilde{\Gamma}_{\eta \eta
^{\prime }}}{\left( \omega -\omega _{\eta \eta ^{\prime }}\right) ^{2}+%
\tilde{\Gamma}_{\eta \eta ^{\prime }}^{2}},  \label{DME-Resonance-Power}
\end{equation}
since every transition $\left| \eta ^{\prime }\right\rangle \rightarrow
\left| \eta \right\rangle $ absorbs the quant of energy equal to $\hbar
\omega _{\eta \eta ^{\prime }}.$ Eq. (\ref{DME-Resonance-Quasist}) can be
simplified by introducing the experimentally measured power $P$ and thus
eliminating the matrix element $\hat{V}_{0,\eta \eta ^{\prime }}$ and the
dephasing rate $\tilde{\Gamma}_{\eta \eta ^{\prime }}.$ Still the equations
are rather complicated so that one cannot find a simple general solution
even for the stationary state.

Let us consider at first the model with only two levels, $\eta $ and $\eta
^{\prime }.$ Here with $\rho _{\eta \eta }=1-\rho _{\eta ^{\prime }\eta
^{\prime }}$ and the induced-transition rate
\begin{equation}
\Lambda \equiv \frac{\left| \hat{V}_{0,\eta \eta ^{\prime }}\right| ^{2}}{%
\hbar ^{2}}\frac{2\tilde{\Gamma}_{\eta \eta ^{\prime }}}{\left( \omega
-\omega _{\eta \eta ^{\prime }}\right) ^{2}+\tilde{\Gamma}_{\eta \eta
^{\prime }}^{2}}
\end{equation}
Eq. (\ref{DME-Resonance-Quasist}) reduces to a single equation
\begin{eqnarray}
\frac{d}{dt}\rho _{\eta ^{\prime }\eta ^{\prime }} &=&-\Lambda \left( \rho
_{\eta ^{\prime }\eta ^{\prime }}-\rho _{\eta \eta }\right) +\Gamma _{\eta
^{\prime }\eta }\rho _{\eta \eta }-\Gamma _{\eta \eta ^{\prime }}\rho _{\eta
^{\prime }\eta ^{\prime }}  \nonumber \\
&=&-\Lambda \left( 2\rho _{\eta ^{\prime }\eta ^{\prime }}-1\right) +\Gamma
_{\eta ^{\prime }\eta }-\left( \Gamma _{\eta ^{\prime }\eta }+\Gamma _{\eta
\eta ^{\prime }}\right) \rho _{\eta ^{\prime }\eta ^{\prime }}  \nonumber \\
&=&\Lambda +\Gamma _{\eta ^{\prime }\eta }-\left( 2\Lambda +\Gamma _{\eta
^{\prime }\eta }+\Gamma _{\eta \eta ^{\prime }}\right) \rho _{\eta ^{\prime
}\eta ^{\prime }}
\end{eqnarray}
that describes relaxation with the rate $2\Lambda +\Gamma _{\eta ^{\prime
}\eta }+\Gamma _{\eta \eta ^{\prime }}$ towards the stationary state
described by
\begin{equation}
\rho _{\eta ^{\prime }\eta ^{\prime }}=\frac{\Lambda +\Gamma _{\eta ^{\prime
}\eta }}{2\Lambda +\Gamma _{\eta ^{\prime }\eta }+\Gamma _{\eta \eta
^{\prime }}},\qquad \Gamma _{\eta \eta ^{\prime }}=\Gamma _{\eta ^{\prime
}\eta }\exp \left( -\frac{\hbar \omega _{\eta \eta ^{\prime }}}{k_{B}T}%
\right)  \label{DME-Resonance-rhoGround}
\end{equation}
for the ground-state population. The limiting cases of this formula are
\begin{equation}
\rho _{\eta ^{\prime }\eta ^{\prime }}=\left\{
\begin{array}{cc}
\rho _{\eta ^{\prime }\eta ^{\prime }}^{(\mathrm{eq})}=\left[ 1+\exp \left( -%
\frac{\hbar \omega _{\eta \eta ^{\prime }}}{k_{B}T}\right) \right] ^{-1}, &
\Lambda =0 \\
1/2, & \Lambda \rightarrow \infty ,
\end{array}
\right.
\end{equation}
as it should be. For the population difference from Eq. (\ref
{DME-Resonance-rhoGround}) one obtains
\begin{equation}
\rho _{\eta ^{\prime }\eta ^{\prime }}-\rho _{\eta \eta }=2\rho _{\eta
^{\prime }\eta ^{\prime }}-1=\frac{\Gamma _{\eta ^{\prime }\eta }-\Gamma
_{\eta \eta ^{\prime }}}{2\Lambda +\Gamma _{\eta ^{\prime }\eta }+\Gamma
_{\eta \eta ^{\prime }}}=\frac{\rho _{\eta ^{\prime }\eta ^{\prime }}^{(%
\mathrm{eq})}-\rho _{\eta \eta }^{(\mathrm{eq})}}{1+2\Lambda /(\Gamma _{\eta
^{\prime }\eta }+\Gamma _{\eta \eta ^{\prime }})}.
\end{equation}
One can see that the population difference is reduced by the resonance
perturbation. The absorbed power is then given by
\begin{equation}
P=\hbar \omega _{\eta \eta ^{\prime }}\left( \rho _{\eta ^{\prime }\eta
^{\prime }}-\rho _{\eta \eta }\right) \Lambda =\frac{\hbar \omega _{\eta
\eta ^{\prime }}\left( \rho _{\eta ^{\prime }\eta ^{\prime }}^{(\mathrm{eq}%
)}-\rho _{\eta \eta }^{(\mathrm{eq})}\right) \Lambda }{1+2\Lambda /(\Gamma
_{\eta ^{\prime }\eta }+\Gamma _{\eta \eta ^{\prime }})}.
\label{DME-Resonance-PviaLambda}
\end{equation}
It increases with $\Lambda $ and reaches an asymptotic maximal value.

To relate the populations directly to the measured absorbed power $P,$ it is
most convenient to rewrite Eq. (\ref{DME-Resonance-Quasist}) for the two-level system
in the form
\begin{eqnarray}
\frac{d}{dt}\rho _{\eta ^{\prime }\eta ^{\prime }} &=&-\frac{P}{\hbar \omega
_{\eta \eta ^{\prime }}}+\Gamma _{\eta ^{\prime }\eta }\rho _{\eta \eta
}-\Gamma _{\eta \eta ^{\prime }}\rho _{\eta ^{\prime }\eta ^{\prime }}
\nonumber \\
&=&-\frac{P}{\hbar \omega _{\eta \eta ^{\prime }}}+\Gamma _{\eta ^{\prime
}\eta }-\left( \Gamma _{\eta ^{\prime }\eta }+\Gamma _{\eta \eta ^{\prime
}}\right) \rho _{\eta ^{\prime }\eta ^{\prime }}.
\end{eqnarray}
This equation has a stationary solution
\begin{equation}
\rho _{\eta ^{\prime }\eta ^{\prime }}=\frac{\Gamma _{\eta ^{\prime }\eta
}-P/(\hbar \omega _{\eta \eta ^{\prime }})}{\Gamma _{\eta ^{\prime }\eta
}+\Gamma _{\eta \eta ^{\prime }}}=\rho _{\eta ^{\prime }\eta ^{\prime }}^{(%
\mathrm{eq})}-\frac{P/(\hbar \omega _{\eta \eta ^{\prime }})}{\Gamma _{\eta
^{\prime }\eta }+\Gamma _{\eta \eta ^{\prime }}}.
\label{DME-Resonance-rhoStationary}
\end{equation}
Note that the maximal absorbed power corresponds to the saturation, $\rho
_{\eta ^{\prime }\eta ^{\prime }}=\rho _{\eta \eta },$ i.e., $\rho _{\eta
^{\prime }\eta ^{\prime }}=1/2,$ wherefrom follows
\begin{equation}
P_{\max }=\frac{\hbar \omega _{\eta \eta ^{\prime }}}{2}\left( \Gamma _{\eta
^{\prime }\eta }-\Gamma _{\eta \eta ^{\prime }}\right) =\frac{\hbar \omega
_{\eta \eta ^{\prime }}}{2}\Gamma _{\eta ^{\prime }\eta }\left[ 1-\exp
\left( -\frac{\hbar \omega _{\eta \eta ^{\prime }}}{k_{B}T}\right) \right]
\label{DME-Pmax}
\end{equation}
that also can be obtained from Eq. (\ref{DME-Resonance-PviaLambda}).

Consider now a physical quantity $\hat{A}$ that has the expectation value
given by Eq. (\ref{DME-Aavrrho}). The contributions from the nondiagonal DM
elements oscillate in time and thus average out. The result is due to the
diagonal elements only,
\begin{equation}
A=A_{\eta \eta }\rho _{\eta \eta }+A_{\eta ^{\prime }\eta ^{\prime }}\rho
_{\eta ^{\prime }\eta ^{\prime }}=A_{\eta \eta }+\left( A_{\eta ^{\prime
}\eta ^{\prime }}-A_{\eta \eta }\right) \rho _{\eta ^{\prime }\eta ^{\prime
}}
\end{equation}
With the help of Eq. (\ref{DME-Resonance-rhoStationary}) it can be rewritten
as
\begin{equation}
A=A^{(\mathrm{eq})}+\Delta A,
\end{equation}
where
\begin{equation}
A^{(\mathrm{eq})}=A_{\eta \eta }+\left( A_{\eta ^{\prime }\eta ^{\prime
}}-A_{\eta \eta }\right) \rho _{\eta ^{\prime }\eta ^{\prime }}^{(\mathrm{eq}%
)}=\frac{A_{\eta ^{\prime }\eta ^{\prime }}+A_{\eta \eta }\exp \left( -\frac{%
\hbar \omega _{\eta \eta ^{\prime }}}{k_{B}T}\right) }{1+\exp \left( -\frac{%
\hbar \omega _{\eta \eta ^{\prime }}}{k_{B}T}\right) }  \label{DME-Aequil}
\end{equation}
is the equilibrium value and the deviation from the equilibrium is given by
\begin{equation}
\Delta A=-\frac{A_{\eta ^{\prime }\eta ^{\prime }}-A_{\eta \eta }}{\Gamma
_{\eta ^{\prime }\eta }+\Gamma _{\eta \eta ^{\prime }}}\frac{P}{\hbar \omega
_{\eta \eta ^{\prime }}}.
\end{equation}
Thus the total relaxation rate between the states $\left| \eta \right\rangle
$ and $\left| \eta ^{\prime }\right\rangle $ can be expressed by the formula
\begin{equation}
\Gamma =\Gamma _{\eta ^{\prime }\eta }+\Gamma _{\eta \eta ^{\prime }}=\frac{%
A_{\eta \eta }-A_{\eta ^{\prime }\eta ^{\prime }}}{\Delta A}\frac{P}{\hbar
\omega _{\eta \eta ^{\prime }}}  \label{DME-Resonance-Rate}
\end{equation}
through the measured $P$ and $\Delta A.$

If the system has more than two levels but the levels $\eta $ and $\eta
^{\prime }$ are the lowest two levels and the temperature is low, so that
the populations of all other levels are small, one can still consider an
effective two-level model in which relaxation between $\eta $ and $\eta
^{\prime }$can be assisted by the upper levels. Instead of the direct
relaxation $\left| \eta \right\rangle \rightarrow \left| \eta ^{\prime
}\right\rangle $ (that can have a small rate) the system can be thermally
excited from $\left| \eta \right\rangle $ to some high level $\left| \alpha
\right\rangle $ and then fall down to $\left| \eta ^{\prime }\right\rangle .$
This is the Orbach mechanism that has the characteristic Arrhenius
temperature dependence of the rate. Using the effective two-level model is
justified by the fact that the upper levels do not contribute to physical
quantities $A$ at low temperatures.

\subsection{Linear response}

\label{Sec-linear-response}

In this section we consider a small harmonic perturbation of Eq. (\ref
{DME-VperiodicDef}) acting on the small system, similarly to the preceding
section. However, the frequency $\omega $ of the perturbation does not need
to be close to the resonance with any transition $\omega _{\alpha \beta }.$
If $\omega \ll \omega _{\alpha \beta }$ for all $\alpha ,\beta ,$ the
response of the small system is due to the relaxation and it depends on the
relation between $\omega $ and the relaxation rate $\Gamma .$ To make an
account of this effect, one has to include $\hat{V}(t)$ into the relaxation
terms of the density-matrix equation. Temporal change of $\hat{V}(t)$
changes the instantaneous equilibrium to which the system relaxes that gives
rise to a dissipative dynamics. On the other hand, one has to include $\hat{V%
}(t)$ into the conservative term of the DME as well, where it can work as a
resonance perturbation. The secular
approximation will be used below for simplicity.

Using the adiabatic basis of Eq. (\ref{DME-chiAdiabatic}) and combining Eq. (%
\ref{DME-rhoEq-diagbasist}) with Eqs. (\ref{DME-rateEqs}) and (\ref
{DME-nondiagelementsEq}), one can write the secular DME in the form
\begin{equation}
\frac{d}{dt}\rho _{\alpha \alpha }=\sum_{\gamma }\left( \langle \dot{\chi}%
_{\alpha }\left| \chi _{\gamma }\right\rangle \rho _{\gamma \alpha }+\rho
_{\alpha \gamma }\langle \chi _{\gamma }\left| \dot{\chi}_{\alpha
}\right\rangle \right) +\sum_{\alpha ^{\prime }\neq \alpha }\Gamma _{\alpha
^{\prime }\alpha }\left( e^{(\varepsilon _{\alpha ^{\prime }}-\varepsilon
_{\alpha })/(k_{\mathrm{B}}T)}\rho _{\alpha ^{\prime }\alpha ^{\prime
}}-\rho _{\alpha \alpha }\right)
\end{equation}
for diagonal terms and
\begin{equation}
\frac{d}{dt}\rho _{\alpha \beta }=\sum_{\gamma }\left( \langle \dot{\chi}%
_{\alpha }\left| \chi _{\gamma }\right\rangle \rho _{\gamma \beta }+\rho
_{\alpha \gamma }\langle \chi _{\gamma }\left| \dot{\chi}_{\beta
}\right\rangle \right) -\left( i\omega _{\alpha \beta }+\tilde{\Gamma}%
_{\alpha \beta }\right) \rho _{\alpha \beta }
\end{equation}
for nondiagonal terms. If time derivatives of the adiabatic states are
small, the density matrix in the adiabatic basis $\rho _{\alpha \beta }$ is
close to its instantaneous quasiequilibrium form given by Eq. (\ref
{DME-rhoalphabetaEquil}). The time dependence of the adiabatic states drives
the DM out of the instantaneous equilibrim, $\rho _{\alpha \beta }$ lagging
behind time-dependent $\rho _{\alpha \beta }^{\mathrm{eq}}.$ On the other
hand, since $\hat{V}(t)$ is a small perturbation, $\rho _{\alpha \beta }^{%
\mathrm{eq}}$ deviates from the full equilibrium $\rho _{\alpha \beta }^{(0)%
\mathrm{eq}}$ in the absense of $\hat{V}(t)$ at linear order in $\hat{V}(t).$
Thus the deviation from the full equilibrium,
\begin{equation}
\delta \rho _{\alpha \beta }\equiv \rho _{\alpha \beta }-\rho _{\alpha \beta
}^{(0)\mathrm{eq}}
\end{equation}
is also linear in $\hat{V}(t).$ Since the difference $e^{(\varepsilon
_{\alpha ^{\prime }}-\varepsilon _{\alpha })/(k_{\mathrm{B}}T)}\rho _{\alpha
^{\prime }\alpha ^{\prime }}-\rho _{\alpha \alpha }$ and the nondiagonal
elements $\rho _{\alpha \beta }$ are small, one does not have to expand $%
\omega _{\alpha \beta }$ and the relaxation rates. The only term to expand
is $e^{(\varepsilon _{\alpha ^{\prime }}-\varepsilon _{\alpha })/(k_{\mathrm{%
B}}T)},$ where one can use, at linear order in $\hat{V}(t),$
\begin{equation}
\varepsilon _{\alpha }(t)=\varepsilon _{\alpha }^{(0)}+\delta \varepsilon
_{\alpha }^{(0)}(t),\qquad \delta \varepsilon _{\alpha }(t)=\left\langle
\chi _{\alpha }^{(0)}\left| \hat{V}(t)\right| \chi _{\alpha
}^{(0)}\right\rangle =V(t)_{\alpha \alpha }^{(0)}.  \label{DME-epsilonalphat}
\end{equation}
Thus
\begin{equation}
e^{(\varepsilon _{\alpha ^{\prime }}-\varepsilon _{\alpha })/(k_{\mathrm{B}%
}T)}\cong e^{(\varepsilon _{\alpha ^{\prime }}^{(0)}-\varepsilon _{\alpha
}^{(0)})/(k_{\mathrm{B}}T)}\left( 1+\frac{V(t)_{\alpha ^{\prime }\alpha
^{\prime }}^{(0)}-V(t)_{\alpha \alpha }^{(0)}}{k_{\mathrm{B}}T}\right) .
\end{equation}
The driving term expands as follows:
\begin{eqnarray}
&&\sum_{\gamma }\left( \langle \dot{\chi}_{\alpha }\left| \chi _{\gamma
}\right\rangle \rho _{\gamma \beta }+\rho _{\alpha \gamma }\langle \chi
_{\gamma }\left| \dot{\chi}_{\beta }\right\rangle \right)  \nonumber \\
&\cong &\sum_{\gamma }\left( \langle \dot{\chi}_{\alpha }\left| \chi
_{\gamma }\right\rangle \rho _{\gamma \beta }^{(0)\mathrm{eq}}+\rho _{\alpha
\gamma }^{(0)\mathrm{eq}}\langle \chi _{\gamma }\left| \dot{\chi}_{\beta
}\right\rangle \right)  \nonumber \\
&=&\langle \dot{\chi}_{\alpha }\left| \chi _{\beta }\right\rangle \rho
_{\beta \beta }^{(0)\mathrm{eq}}+\rho _{\alpha \alpha }^{(0)\mathrm{eq}%
}\langle \chi _{\alpha }\left| \dot{\chi}_{\beta }\right\rangle =\langle
\chi _{\alpha }\left| \dot{\chi}_{\beta }\right\rangle \left( \rho _{\alpha
\alpha }^{(0)\mathrm{eq}}-\rho _{\beta \beta }^{(0)\mathrm{eq}}\right) ,
\end{eqnarray}
where $\langle \chi _{\alpha }\left| \chi _{\beta }\right\rangle =\delta
_{\alpha \beta }$ was used. One can see that in the diagonal equations this
term disappears. Thus for the diagonal equations the linearized DME has the
form
\begin{eqnarray}
\frac{d}{dt}\delta \rho _{\alpha \alpha } &=&\sum_{\alpha ^{\prime }\neq
\alpha }\Gamma _{\alpha ^{\prime }\alpha }\left( e^{(\varepsilon _{\alpha
^{\prime }}^{(0)}-\varepsilon _{\alpha }^{(0)})/(k_{\mathrm{B}}T)}\delta
\rho _{\alpha ^{\prime }\alpha ^{\prime }}-\delta \rho _{\alpha \alpha
}\right)  \nonumber \\
&&+\sum_{\alpha ^{\prime }\neq \alpha }\Gamma _{\alpha ^{\prime }\alpha
}e^{(\varepsilon _{\alpha ^{\prime }}^{(0)}-\varepsilon _{\alpha
}^{(0)})/(k_{\mathrm{B}}T)}\rho _{\alpha ^{\prime }\alpha ^{\prime }}^{(0)%
\mathrm{eq}}\frac{V(t)_{\alpha ^{\prime }\alpha ^{\prime
}}^{(0)}-V(t)_{\alpha \alpha }^{(0)}}{k_{\mathrm{B}}T}
\end{eqnarray}
or
\begin{equation}
\frac{d}{dt}\rho _{\alpha \alpha }=\sum_{\alpha ^{\prime }}\left( \Gamma
_{\alpha \alpha ^{\prime }}\rho _{\alpha ^{\prime }\alpha ^{\prime }}-\Gamma
_{\alpha ^{\prime }\alpha }\rho _{\alpha \alpha }\right) +f_{\alpha \alpha },
\label{DME-xbfhf-diag}
\end{equation}
where
\begin{equation}
f_{\alpha \alpha }\equiv \frac{e^{-\varepsilon _{\alpha }^{(0)}/(k_{\mathrm{B%
}}T)}}{Z_{\mathrm{s}}}\sum_{\alpha ^{\prime }}\Gamma _{\alpha ^{\prime
}\alpha }\frac{V(t)_{\alpha ^{\prime }\alpha ^{\prime }}^{(0)}-V(t)_{\alpha
\alpha }^{(0)}}{k_{\mathrm{B}}T}.
\end{equation}
Linearization of nondiagonal equations yields
\begin{equation}
\frac{d}{dt}\rho _{\alpha \beta }=-\left( i\omega _{\alpha \beta }+\tilde{%
\Gamma}_{\alpha \beta }\right) \rho _{\alpha \beta }+f_{\alpha \beta },
\label{DME-xnsltr}
\end{equation}
where
\begin{equation}
f_{\alpha \beta }\equiv \langle \chi _{\alpha }\left| \dot{\chi}_{\beta
}\right\rangle \left( \rho _{\alpha \alpha }^{(0)\mathrm{eq}}-\rho _{\beta
\beta }^{(0)\mathrm{eq}}\right) .
\end{equation}
One can see that within the secular approximation the driving term is
conservative in the nondiagonal equations and dissipative in the diagonal
equations.

The density matrix $\rho _{\alpha \beta }$ in the equations above is still
defined with respect to the adiabatic basis $\left| \chi _{\alpha
}\right\rangle ,$ although the relaxation term is already expanded around
the unperturbed states $\left| \chi _{\alpha }^{(0)}\right\rangle .$ Let us
now make transformation to $\rho _{\alpha \beta }^{(0)}$ defined with
respect to the unperturbed diagonal basis $\left| \chi _{\alpha
}^{(0)}\right\rangle $. The relations between the two density matrices have
the form
\begin{equation}
\rho _{\alpha \beta }^{(0)}=\sum_{\alpha ^{\prime }\beta ^{\prime }}\langle
\chi _{\alpha }^{(0)}\left| \chi _{\alpha ^{\prime }}\right\rangle \rho
_{\alpha ^{\prime }\beta ^{\prime }}\langle \chi _{\beta ^{\prime }}\left|
\chi _{\beta }^{(0)}\right\rangle ,
\end{equation}
c.f. Eq. (\ref{DME-rgoviarhodiag}), and
\begin{equation}
\rho _{\alpha \beta }=\sum_{\alpha ^{\prime }\beta ^{\prime }}\langle \chi
_{\alpha }\left| \chi _{\alpha ^{\prime }}^{(0)}\right\rangle \rho _{\alpha
^{\prime }\beta ^{\prime }}^{(0)}\langle \chi _{\beta ^{\prime
}}^{(0)}\left| \chi _{\beta }\right\rangle .
\end{equation}
Differentiating the first equation over time one obtains, at linear order in
$\hat{V}(t),$
\begin{eqnarray}
\frac{d}{dt}\rho _{\alpha \beta }^{(0)} &=&\sum_{\alpha ^{\prime }\beta
^{\prime }}\langle \chi _{\alpha }^{(0)}\left| \dot{\chi}_{\alpha ^{\prime
}}\right\rangle \rho _{\alpha ^{\prime }\beta ^{\prime }}\langle \chi
_{\beta ^{\prime }}\left| \chi _{\beta }^{(0)}\right\rangle +\sum_{\alpha
^{\prime }\beta ^{\prime }}\langle \chi _{\alpha }^{(0)}\left| \chi _{\alpha
^{\prime }}\right\rangle \rho _{\alpha ^{\prime }\beta ^{\prime }}\langle
\dot{\chi}_{\beta ^{\prime }}\left| \chi _{\beta }^{(0)}\right\rangle
+\sum_{\alpha ^{\prime }\beta ^{\prime }}\langle \chi _{\alpha }^{(0)}\left|
\chi _{\alpha ^{\prime }}\right\rangle \dot{\rho}_{\alpha ^{\prime }\beta
^{\prime }}\langle \chi _{\beta ^{\prime }}\left| \chi _{\beta
}^{(0)}\right\rangle  \nonumber \\
&\cong &\sum_{\alpha ^{\prime }}\langle \chi _{\alpha }^{(0)}\left| \dot{\chi%
}_{\alpha ^{\prime }}\right\rangle \rho _{\alpha ^{\prime }\beta }^{(0)%
\mathrm{eq}}+\sum_{\beta ^{\prime }}\rho _{\alpha \beta ^{\prime }}^{(0)%
\mathrm{eq}}\langle \dot{\chi}_{\beta ^{\prime }}\left| \chi _{\beta
}^{(0)}\right\rangle +\sum_{\alpha ^{\prime }\beta ^{\prime }}\langle \chi
_{\alpha }^{(0)}\left| \chi _{\alpha ^{\prime }}\right\rangle \dot{\rho}%
_{\alpha ^{\prime }\beta ^{\prime }}\langle \chi _{\beta ^{\prime }}\left|
\chi _{\beta }^{(0)}\right\rangle  \nonumber \\
&=&-\langle \chi _{\alpha }^{(0)}\left| \dot{\chi}_{\beta }\right\rangle
\left( \rho _{\alpha \alpha }^{(0)\mathrm{eq}}-\rho _{\beta \beta }^{(0)%
\mathrm{eq}}\right) +\sum_{\alpha ^{\prime }\beta ^{\prime }}\langle \chi
_{\alpha }^{(0)}\left| \chi _{\alpha ^{\prime }}\right\rangle \dot{\rho}%
_{\alpha ^{\prime }\beta ^{\prime }}\langle \chi _{\beta ^{\prime }}\left|
\chi _{\beta }^{(0)}\right\rangle .
\end{eqnarray}
For $\alpha =\beta $ the first term in this equation disappears and for $%
\alpha \neq \beta $ it cancels a similar term in Eq. (\ref{DME-xnsltr}).
Thus Eq. (\ref{DME-xnsltr}) with $\alpha \neq \beta $ transforms as
\begin{eqnarray*}
\frac{d}{dt}\rho _{\alpha \beta }^{(0)} &=&-\sum_{\alpha ^{\prime }\beta
^{\prime }}\langle \chi _{\alpha }^{(0)}\left| \chi _{\alpha ^{\prime
}}\right\rangle \left( i\omega _{\alpha ^{\prime }\beta ^{\prime }}+\tilde{%
\Gamma}_{\alpha ^{\prime }\beta ^{\prime }}\right) \rho _{\alpha ^{\prime
}\beta ^{\prime }}\langle \chi _{\beta ^{\prime }}\left| \chi _{\beta
}^{(0)}\right\rangle \\
&=&-\sum_{\alpha ^{\prime }\beta ^{\prime }}\sum_{\alpha ^{\prime \prime
}\beta ^{\prime \prime }}\langle \chi _{\alpha }^{(0)}\left| \chi _{\alpha
^{\prime }}\right\rangle \left( i\omega _{\alpha ^{\prime }\beta ^{\prime }}+%
\tilde{\Gamma}_{\alpha ^{\prime }\beta ^{\prime }}\right) \langle \chi
_{\alpha ^{\prime }}\left| \chi _{\alpha ^{\prime \prime
}}^{(0)}\right\rangle \rho _{\alpha ^{\prime \prime }\beta ^{\prime \prime
}}^{(0)}\langle \chi _{\beta ^{\prime \prime }}^{(0)}\left| \chi _{\beta
^{\prime }}\right\rangle \langle \chi _{\beta ^{\prime }}\left| \chi _{\beta
}^{(0)}\right\rangle .
\end{eqnarray*}
In this expression the projectors such as $\langle \chi _{\alpha
}^{(0)}\left| \chi _{\alpha ^{\prime }}\right\rangle $ are linear in $\hat{V}%
(t),$ if the indices do not coincide. Thus there are two \ types of
contributions: (i) All indices pairwise coincide and thus $\rho _{\alpha
^{\prime \prime }\beta ^{\prime \prime }}^{(0)}\rightarrow \rho _{\alpha
\beta }^{(0)}\sim \hat{V}(t)$ or (ii) $\alpha ^{\prime \prime }=\beta
^{\prime \prime }$ so that $\rho _{\alpha ^{\prime \prime }\beta ^{\prime
\prime }}^{(0)}\rightarrow \rho _{\alpha ^{\prime \prime }\alpha ^{\prime
\prime }}^{(0)\mathrm{eq}}$ and one of the pair of indices in projectors do
not coincide. This yields
\begin{eqnarray}
\frac{d}{dt}\rho _{\alpha \beta }^{(0)} &\cong &-\left( i\omega _{\alpha
^{\prime }\beta ^{\prime }}^{(0)}+\tilde{\Gamma}_{\alpha ^{\prime }\beta
^{\prime }}^{(0)}\right) \rho _{\alpha \beta }^{(0)}  \nonumber \\
&&-\sum_{\alpha ^{\prime }\beta ^{\prime }}\langle \chi _{\alpha
}^{(0)}\left| \chi _{\alpha ^{\prime }}\right\rangle \left( i\omega _{\alpha
^{\prime }\beta ^{\prime }}+\tilde{\Gamma}_{\alpha ^{\prime }\beta ^{\prime
}}\right) \rho _{\alpha ^{\prime }\alpha ^{\prime }}^{(0)\mathrm{eq}}\langle
\chi _{\alpha ^{\prime }}^{(0)}\left| \chi _{\beta ^{\prime }}\right\rangle
\langle \chi _{\beta ^{\prime }}\left| \chi _{\beta }^{(0)}\right\rangle
\nonumber \\
&&-\sum_{\alpha ^{\prime }\beta ^{\prime }}\langle \chi _{\alpha
}^{(0)}\left| \chi _{\alpha ^{\prime }}\right\rangle \left( i\omega _{\alpha
^{\prime }\beta ^{\prime }}+\tilde{\Gamma}_{\alpha ^{\prime }\beta ^{\prime
}}\right) \langle \chi _{\alpha ^{\prime }}\left| \chi _{\beta ^{\prime
}}^{(0)}\right\rangle \rho _{\beta ^{\prime }\beta ^{\prime }}^{(0)\mathrm{eq%
}}\langle \chi _{\beta ^{\prime }}\left| \chi _{\beta }^{(0)}\right\rangle
\nonumber \\
&\cong &-\left( i\omega _{\alpha ^{\prime }\beta ^{\prime }}^{(0)}+\tilde{%
\Gamma}_{\alpha ^{\prime }\beta ^{\prime }}^{(0)}\right) \rho _{\alpha \beta
}^{(0)}  \nonumber \\
&&-\langle \chi _{\alpha }^{(0)}\left| \chi _{\beta }\right\rangle \tilde{%
\Gamma}_{\beta \beta }\rho _{\beta \beta }^{(0)\mathrm{eq}}-\tilde{\Gamma}%
_{\alpha \alpha }\rho _{\alpha \alpha }^{(0)\mathrm{eq}}\langle \chi
_{\alpha }\left| \chi _{\beta }^{(0)}\right\rangle -\left( i\omega _{\alpha
\beta }+\tilde{\Gamma}_{\alpha \beta }\right) \rho _{\alpha \alpha }^{(0)%
\mathrm{eq}}\langle \chi _{\alpha }^{(0)}\left| \chi _{\beta }\right\rangle
\nonumber \\
&&-\langle \chi _{\alpha }^{(0)}\left| \chi _{\beta }\right\rangle \tilde{%
\Gamma}_{\beta \beta }\rho _{\beta \beta }^{(0)\mathrm{eq}}-\tilde{\Gamma}%
_{\alpha \alpha }\rho _{\alpha \alpha }^{(0)\mathrm{eq}}\langle \chi
_{\alpha }\left| \chi _{\beta }^{(0)}\right\rangle -\left( i\omega _{\alpha
\beta }+\tilde{\Gamma}_{\alpha \beta }\right) \langle \chi _{\alpha }\left|
\chi _{\beta }^{(0)}\right\rangle \rho _{\beta \beta }^{(0)\mathrm{eq}},
\end{eqnarray}
i.e.,
\begin{eqnarray}
\frac{d}{dt}\rho _{\alpha \beta }^{(0)} &\cong &-\left( i\omega _{\alpha
^{\prime }\beta ^{\prime }}^{(0)}+\tilde{\Gamma}_{\alpha ^{\prime }\beta
^{\prime }}^{(0)}\right) \rho _{\alpha \beta }^{(0)}  \nonumber \\
&&-2\langle \chi _{\alpha }^{(0)}\left| \chi _{\beta }\right\rangle \tilde{%
\Gamma}_{\beta \beta }\rho _{\beta \beta }^{(0)\mathrm{eq}}-2\langle \chi
_{\alpha }\left| \chi _{\beta }^{(0)}\right\rangle \tilde{\Gamma}_{\alpha
\alpha }\rho _{\alpha \alpha }^{(0)\mathrm{eq}}  \nonumber \\
&&-\left( i\omega _{\alpha \beta }+\tilde{\Gamma}_{\alpha \beta }\right)
\left( \langle \chi _{\alpha }^{(0)}\left| \chi _{\beta }\right\rangle \rho
_{\alpha \alpha }^{(0)\mathrm{eq}}+\langle \chi _{\alpha }\left| \chi
_{\beta }^{(0)}\right\rangle \rho _{\beta \beta }^{(0)\mathrm{eq}}\right) .
\end{eqnarray}
The small projections $\langle \chi _{\alpha }^{(0)}\left| \chi _{\beta
}\right\rangle $ can be calculated with the help of the perturbative
expansion
\begin{equation}
\left| \chi _{\alpha }\right\rangle \cong \left| \chi _{\alpha
}^{(0)}\right\rangle +\sum_{\alpha ^{\prime }}\left| \chi _{\alpha ^{\prime
}}^{(0)}\right\rangle \frac{\left\langle \chi _{\alpha ^{\prime
}}^{(0)}\left| \hat{V}(t)\right| \chi _{\alpha }^{(0)}\right\rangle }{%
\varepsilon _{\alpha }^{(0)}-\varepsilon _{\alpha ^{\prime }}^{(0)}},
\end{equation}
that yields
\begin{eqnarray}
\langle \chi _{\alpha }^{(0)}\left| \chi _{\beta }\right\rangle &\cong &%
\frac{\left\langle \chi _{\alpha }^{(0)}\left| \hat{V}(t)\right| \chi
_{\beta }^{(0)}\right\rangle }{\varepsilon _{\beta }^{(0)}-\varepsilon
_{\alpha }^{(0)}}=-\frac{V(t)_{\alpha \beta }^{(0)}}{\hbar \omega _{\alpha
\beta }^{(0)}}  \nonumber \\
\langle \chi _{\alpha }\left| \chi _{\beta }^{(0)}\right\rangle &\cong &%
\frac{\left\langle \chi _{\alpha }^{(0)}\left| \hat{V}(t)\right| \chi
_{\beta }^{(0)}\right\rangle }{\varepsilon _{\alpha }^{(0)}-\varepsilon
_{\beta }^{(0)}}=\frac{V(t)_{\alpha \beta }^{(0)}}{\hbar \omega _{\alpha
\beta }^{(0)}}.
\end{eqnarray}
Thus the DME for nondiagonal components has the form
\begin{equation}
\frac{d}{dt}\rho _{\alpha \beta }^{(0)}=-\left( i\omega _{\alpha \beta
}^{(0)}+\tilde{\Gamma}_{\alpha \beta }^{(0)}\right) \rho _{\alpha \beta
}^{(0)}+f_{\alpha \beta }^{(0)},  \label{DME-rhoeqSus0}
\end{equation}
where
\begin{equation}
f_{\alpha \beta }^{(0)}\equiv \left[ \left( i+\frac{\tilde{\Gamma}_{\alpha
\beta }^{(0)}}{\omega _{\alpha \beta }^{(0)}}\right) \left( \rho _{\alpha
\alpha }^{(0)\mathrm{eq}}-\rho _{\beta \beta }^{(0)\mathrm{eq}}\right) -2%
\frac{\tilde{\Gamma}_{\alpha \alpha }\rho _{\alpha \alpha }^{(0)\mathrm{eq}}-%
\tilde{\Gamma}_{\beta \beta }\rho _{\beta \beta }^{(0)\mathrm{eq}}}{\omega
_{\alpha \beta }^{(0)}}\right] \frac{V(t)_{\alpha \beta }^{(0)}}{\hbar }.
\end{equation}
As the secular approximation implies $\Gamma _{\alpha \beta }\ll \omega
_{\alpha \beta },$ relaxation terms in this equation can be neglected,
\begin{equation}
f_{\alpha \beta }^{(0)}\equiv i\left( \rho _{\alpha \alpha }^{(0)\mathrm{eq}%
}-\rho _{\beta \beta }^{(0)\mathrm{eq}}\right) V(t)_{\alpha \beta
}^{(0)}/\hbar .
\end{equation}
The conservative part of this driving term could actually be transformed
exactly using the relation between the two bases, Eqs. (\ref
{DME-RhoEqNatural}) and (\ref{DME-rhoEq-diagbasist}). Transformation of the
diagonal Eq. (\ref{DME-xbfhf-diag}) is as follows:
\begin{eqnarray}
\frac{d}{dt}\rho _{\alpha \alpha }^{(0)} &\cong &\sum_{\alpha ^{\prime
}}\langle \chi _{\alpha }^{(0)}\left| \chi _{\alpha ^{\prime }}\right\rangle
\dot{\rho}_{\alpha ^{\prime }\alpha ^{\prime }}\langle \chi _{\alpha
^{\prime }}\left| \chi _{\alpha }^{(0)}\right\rangle  \nonumber \\
&=&\sum_{\alpha ^{\prime }}\langle \chi _{\alpha }^{(0)}\left| \chi _{\alpha
^{\prime }}\right\rangle \left[ \sum_{\alpha ^{\prime \prime }}\left( \Gamma
_{\alpha ^{\prime }\alpha ^{\prime \prime }}\rho _{\alpha ^{\prime \prime
}\alpha ^{\prime \prime }}-\Gamma _{\alpha ^{\prime \prime }\alpha ^{\prime
}}\rho _{\alpha ^{\prime }\alpha ^{\prime }}\right) +f_{\alpha ^{\prime
}\alpha ^{\prime }}\right] \langle \chi _{\alpha ^{\prime }}\left| \chi
_{\alpha }^{(0)}\right\rangle  \nonumber \\
&=&\sum_{\alpha ^{\prime }}\left( \Gamma _{\alpha \alpha ^{\prime }}\rho
_{\alpha ^{\prime }\alpha ^{\prime }}-\Gamma _{\alpha ^{\prime }\alpha }\rho
_{\alpha \alpha }\right) +f_{\alpha \alpha }  \nonumber \\
&\cong &\sum_{\alpha ^{\prime }}\left( \Gamma _{\alpha \alpha ^{\prime
}}\sum_{\alpha ^{\prime \prime }\beta ^{\prime \prime }}\langle \chi
_{\alpha ^{\prime }}\left| \chi _{\alpha ^{\prime \prime
}}^{(0)}\right\rangle \rho _{\alpha ^{\prime \prime }\beta ^{\prime \prime
}}^{(0)}\langle \chi _{\beta ^{\prime \prime }}^{(0)}\left| \chi _{\alpha
^{\prime }}\right\rangle -\Gamma _{\alpha ^{\prime }\alpha }\sum_{\alpha
^{\prime \prime }\beta ^{\prime \prime }}\langle \chi _{\alpha }\left| \chi
_{\alpha ^{\prime \prime }}^{(0)}\right\rangle \rho _{\alpha ^{\prime \prime
}\beta ^{\prime \prime }}^{(0)}\langle \chi _{\beta ^{\prime \prime
}}^{(0)}\left| \chi _{\alpha }\right\rangle \right) +f_{\alpha \alpha }^{(0)}
\nonumber \\
&\cong &\sum_{\alpha ^{\prime }}\left( \Gamma _{\alpha \alpha ^{\prime
}}\rho _{\alpha ^{\prime }\alpha ^{\prime }}^{(0)}-\Gamma _{\alpha ^{\prime
}\alpha }\rho _{\alpha \alpha }^{(0)}\right) +f_{\alpha \alpha }^{(0)}.
\end{eqnarray}
Since there are no corrections to $\Gamma _{\alpha \alpha ^{\prime }}$
linear in $V(t),$ this finally yields
\begin{equation}
\frac{d}{dt}\rho _{\alpha \alpha }^{(0)}=\sum_{\alpha ^{\prime }}\left(
\Gamma _{\alpha \alpha ^{\prime }}^{(0)}\rho _{\alpha ^{\prime }\alpha
^{\prime }}^{(0)}-\Gamma _{\alpha ^{\prime }\alpha }^{(0)}\rho _{\alpha
\alpha }^{(0)}\right) +f_{\alpha \alpha }^{(0)}  \label{DME-xbfhf-diag-0}
\end{equation}
where
\begin{equation}
f_{\alpha \alpha }^{(0)}\equiv \frac{e^{-\varepsilon _{\alpha }^{(0)}/(k_{%
\mathrm{B}}T)}}{Z_{\mathrm{s}}}\sum_{\alpha ^{\prime }\neq \alpha }\Gamma
_{\alpha ^{\prime }\alpha }^{(0)}\frac{V(t)_{\alpha ^{\prime }\alpha
^{\prime }}^{(0)}-V(t)_{\alpha \alpha }^{(0)}}{k_{\mathrm{B}}T}.
\label{DME-falal0}
\end{equation}
One can see that this transformation was trivial. In the sequel we will drop
the superscript (0).

Let us now solve Eqs. (\ref{DME-rhoeqSus0})--(\ref{DME-falal0}). Similarly
to Sec. \ref{Sec-FE-secular}, one can write Eq. (\ref{DME-xbfhf-diag-0}) in
the vectorized form
\begin{equation}
\frac{d}{dt}\mathbf{\delta n}=\mathbf{\Phi }^{\sec }\mathbf{\cdot \delta n+f}%
^{\mathrm{diag}}\mathbf{,}
\end{equation}
where $\left( \mathbf{\delta n}\right) _{\alpha }\equiv \delta \rho _{\alpha
\alpha }$ and the elements of $\mathbf{\Phi }^{\sec }$ are given by Eq. (\ref
{DME-PsisecDeg}). Since $\mathbf{f}$ contains positively- and
negatively-rotating terms, the stationary solution of this equation has the
form
\begin{equation}
\pm i\omega \mathbf{\delta n}^{(\pm )}=\mathbf{\Phi }^{\sec }\mathbf{\cdot
\delta n}^{(\pm )}+\mathbf{f}^{\mathrm{diag,}(\pm )}
\label{DME-rhoEq-driven-diag}
\end{equation}
with
\begin{equation}
f_{\alpha \alpha }^{(\pm )}=\frac{e^{-\varepsilon _{\alpha }/(k_{\mathrm{B}%
}T)}}{Z_{\mathrm{s}}}\sum_{\alpha ^{\prime }\neq \alpha }\Gamma _{\alpha
^{\prime }\alpha }\frac{V_{\alpha ^{\prime }\alpha ^{\prime }}^{(\pm
)}-V_{\alpha \alpha }^{(\pm )}}{k_{\mathrm{B}}T}.
\label{DME-fplusminusDiagDef}
\end{equation}
Here $V_{0,\alpha \gamma }^{(+)}\equiv V_{0,\alpha \gamma }$ and $%
V_{0,\alpha \gamma }^{(-)}\equiv \left( \hat{V}_{0}^{\dagger }\right)
_{\alpha \gamma }=V_{0,\gamma \alpha }^{\ast }.$ The solution of this
equation can be expanded over the set of right eigenvectors defined by an
equation similar to Eq. (\ref{DME-DME-eigenproblem}),
\begin{equation}
\mathbf{\delta n}^{(\pm )}=\sum_{\mu }C_{\mu }^{(\pm )}\mathbf{R}_{\mu }.
\end{equation}
Inserting this into Eq. (\ref{DME-rhoEq-driven-diag}), multiplying from left
by the left eigenvector \ $\mathbf{L}_{\nu }$ and using orthogonality in Eq.
(\ref{DME-LRorthocompl}), one obtains
\begin{equation}
\pm i\omega C_{\nu }^{(\pm )}=-\Lambda _{\nu }C_{\nu }^{(\pm )}+\mathbf{L}%
_{\nu }\cdot \mathbf{f}^{\mathrm{diag,}(\pm )}
\end{equation}
and
\begin{equation}
C_{\nu }^{(\pm )}=\frac{\mathbf{L}_{\nu }\cdot \mathbf{f}^{\mathrm{diag,}%
(\pm )}}{\Lambda _{\nu }\pm i\omega }.
\end{equation}
Now the final result for the populations is
\begin{equation}
\mathbf{\delta n}^{(\pm )}=\sum_{\mu }\frac{\mathbf{L}_{\mu }\cdot \mathbf{f}%
^{\mathrm{diag,}(\pm )}}{\Lambda _{\mu }\pm i\omega }\mathbf{R}_{\mu }.
\label{DME-deltarhoslowRes}
\end{equation}
At equilibrium, $\omega =0,$ this expression should reduce to the static
result
\begin{equation}
\delta n_{\alpha }=\sum_{\pm }\sum_{\mu }\frac{\mathbf{L}_{\mu }\cdot
\mathbf{f}^{\mathrm{diag,}(\pm )}}{\Lambda _{\mu }}R_{\mu \alpha }=-\frac{%
e^{-\varepsilon _{\alpha }/(k_{\mathrm{B}}T)}}{Z_{\mathrm{s}}}\left( \frac{%
\delta \varepsilon _{\alpha }}{k_{\mathrm{B}}T}+\frac{\delta Z_{\mathrm{s}}}{%
Z_{\mathrm{s}}}\right)
\end{equation}
that can be proven to satisfy Eq. (\ref{DME-xbfhf-diag-0}) in the static
case. Using Eq. (\ref{DME-fplusminusDiagDef}) and $\sum_{\pm }V_{\alpha
\alpha }^{(\pm )}=\delta \varepsilon _{\alpha },$ one obtains the identity
\begin{equation}
\sum_{\mu }\frac{R_{\mu \alpha }}{\Lambda _{\mu }}\sum_{\alpha ^{\prime
}\alpha ^{\prime \prime }}\Gamma _{\alpha ^{\prime \prime }\alpha ^{\prime }}%
\frac{\delta \varepsilon _{\alpha ^{\prime \prime }}-\delta \varepsilon
_{\alpha ^{\prime }}}{k_{\mathrm{B}}T}L_{\mu \alpha ^{\prime }}=-\frac{%
\delta \varepsilon _{\alpha }}{k_{\mathrm{B}}T}-\frac{\delta Z_{\mathrm{s}}}{%
Z_{\mathrm{s}}}
\end{equation}
that should be satisfied by the matrix solution and can be used for
checking. Nondiagonal components of the DME satisfy the equations
\begin{equation}
\pm i\omega \delta \rho _{\alpha \beta }^{(\pm )}=-\left( i\omega _{\alpha
\beta }+\tilde{\Gamma}_{\alpha \beta }\right) \delta \rho _{\alpha \beta
}^{(\pm )}+f_{\alpha \beta }^{(\pm )},
\end{equation}
where
\begin{equation}
f_{\alpha \beta }^{(\pm )}\equiv -\frac{1}{\hbar }\left( i+\frac{\tilde{%
\Gamma}_{\alpha \beta }}{\omega _{\alpha \beta }}\right) V_{\alpha \beta
}^{(\pm )}\left( \rho _{\alpha \alpha }^{\mathrm{eq}}-\rho _{\beta \beta }^{%
\mathrm{eq}}\right) .
\end{equation}
These equations have the solution
\begin{equation}
\delta \rho _{\alpha \beta }^{(\pm )}=\frac{f_{\alpha \beta }^{(\pm )}}{%
\tilde{\Gamma}_{\alpha \beta }+i\omega _{\alpha \beta }\pm i\omega }.
\end{equation}
For a physical quantity $\hat{A}$ the linear response has the form
\begin{equation}
A(t)=e^{i\omega t}A^{(+)}(\omega )+e^{-i\omega t}A^{(-)}(\omega ),
\label{DME-At}
\end{equation}
where
\begin{equation}
A^{(\pm )}(\omega )=\sum_{\mu =2}^{N}\frac{\left( \mathbf{A}\cdot \mathbf{R}%
_{\mu }\right) \left( \mathbf{L}_{\mu }\cdot \mathbf{f}^{\mathrm{diag,}(\pm
)}\right) }{\Lambda _{\mu }\pm i\omega }+\sum_{\alpha \neq \beta }\frac{%
A_{\beta \alpha }f_{\alpha \beta }^{(\pm )}}{\tilde{\Gamma}_{\alpha \beta
}+i\omega _{\alpha \beta }\pm i\omega }  \label{DME-Apmomega}
\end{equation}
c.f. Eq. (\ref{DME-tauintSecGeneral}). Here $\left( \mathbf{A}\right)
_{\alpha }=A_{\alpha \alpha }$. The zero eigenvalue, $\mu =1$ and $\Lambda
_{1}=0,$ does not make a contribution to this formula since $\mathbf{L}_{1}$
given by Eq. (\ref{DME-L0R0sec}) is orthogonal to $\mathbf{f}^{\mathrm{diag,}%
(\pm )}$ defined by Eq. (\ref{DME-fplusminusDiagDef}). Physically relevant
are real and imaginary parts of the linear response
\begin{eqnarray}
A^{\prime }(\omega ) &=&\func{Re}A^{(+)}(\omega )+\func{Re}A^{(-)}(\omega )
\nonumber \\
A^{\prime \prime }(\omega ) &=&-\func{Im}A^{(+)}(\omega )+\func{Im}%
A^{(-)}(\omega ).
\end{eqnarray}
%

\newpage
\section{Application to molecular magnets}


\subsection{The model}

Let us consider, as an example, a magnetic molecule (MM) described as a
large spin $S$ with the Hamiltonian
\begin{equation}
\hat{H}_{\mathrm{s}}=\hat{H}_{S}=\hat{H}_{A}+\hat{H}_{\mathrm{Z}},
\label{DME-MMHam}
\end{equation}
where $\hat{H}_{A}$ is the crystal-field Hamitonian and $\hat{H}_{\mathrm{Z}%
} $ is the Zeeman Hamiltonian, $\hat{H}_{\mathrm{Z}}=-g\mu _{B}\mathbf{%
H\cdot S.}$ The crystal field $\hat{H}_{A}$ is usually dominated by the
uniaxial anisotropy,
\begin{equation}
\hat{H}_{A}=-DS_{z}^{2}-BS_{z}^{4}+\hat{H}_{A}^{\prime },  \label{DME-HA}
\end{equation}
whereas $\hat{H}_{A}^{\prime }$ contains smaller terms that do not commute
with $S_{z}.$ In the absence of the latter and of the transverse field$,$
the eigenstates of the spin are $\left| m\right\rangle ,$ $m=-S,\ldots ,S.$
The energy levels of the magnetic molecule are given by
\begin{equation}
\varepsilon _{m}=-Dm^{2}-Bm^{4}-gm_{B}H_{z}m.  \label{DME-epsilonm}
\end{equation}
The transition frequency for a pair of levels is
\begin{equation}
\hbar \omega _{mm^{\prime }}=\varepsilon _{m}-\varepsilon _{m^{\prime
}}=-\left( m-m^{\prime }\right) \left\{ \left( m+m^{\prime }\right) \left[
D+\left( m^{2}+m^{\prime 2}\right) B\right] +gm_{B}H_{z}\right\}
\label{DME-omegammpr}
\end{equation}
Condition $\hbar \omega _{mm^{\prime }}=0$ for $m\neq m^{\prime }$ (levels
on different sides of the barrier created by the uniaxial anisotropy)
defines the resonance values of the longitudinal field $H_{z}.$ For the
generic model of molecular magnets with $B=0$ the latter are given by
\begin{equation}
gm_{B}H_{z}=kD,\qquad k=0,\pm 1,\pm 2,\ldots  \label{DME-HzRes}
\end{equation}
For these fields \emph{all} levels in the right well
\begin{equation}
m^{\prime }=-m-k  \label{DME-mprDef}
\end{equation}
are at resonance with the corresponding levels in the left well $m<0.$ For
the realistic model with $B>0,$ the field creating resonances between
low-lying levels with large $m^{2}+m^{\prime 2}$ is greater than the
resonance fields for high levels. Transverse anisotropy and transverse field
$H_{x}$ that enter $\hat{H}_{A}^{\prime }$ result in the tunneling under the
barrier and tunneling splitting of the resonant levels $m,m^{\prime }.$

For Mn$_{12}$ used in illustrations below we adopt the values $D/k_{\mathrm{B%
}}=0.548$ K and $B/k_{\mathrm{B}}=1.1\times 10^{-3}$ K (Refs. \cite
{hilletal98prl,miretal99prl,barkenrumhencri03prl}) that makes up the barrier
of 66 K. $\hat{H}_{A}^{\prime }$ in Eq. (\ref{DME-HA}) can contain second-
and fourth order transverse anisotropy,
\begin{equation}
\hat{H}_{A}^{\prime }=E\left( S_{x}^{2}-S_{y}^{2}\right)
+C(S_{+}^{4}+S_{-}^{4}).  \label{DME-HAprDef}
\end{equation}
For Mn$_{12}$ $C/k_{\mathrm{B}}=3\times 10^{-5}$ K (Refs. \cite
{hilletal98prl,miretal99prl,barkenrumhencri03prl}), whereas $E=0$ in the
ideal case because of the tetragonal symmetry of the crystal. However, it
was shown \cite{barkenrumhencri03prl} that \emph{local} molecular
environments of Mn$_{12}$ molecules have a two-fold symmetry and rotated by
90$%
{{}^\circ}%
$ for different molecules. Although on average the four-fold symmetry of the
crystal is preserved, it gives rise to nonzero $E$ that will be set to $E/k_{%
\mathrm{B}}=2.5\times 10^{-3}$ K.

The spin Hamiltonian of molecular magnets can be easily numerically
diagonalized to yield eigenstates $\left| \alpha \right\rangle $ and
transition frequences $\omega _{\alpha \beta },$ $\alpha ,\beta =1,..,2S+1.$
The terms non-commuting with $S_{z}$ such as $\hat{H}_{A}^{\prime }$ and the
transverse magnetic field cause hybridization of the spin states in the two
wells that leads to spin tunneling.

\subsection{Spin-phonon interaction}

The magnetic molecule is embedded in the elastic matrix described by the
harmonic-phonon Hamiltonian
\begin{equation}
\hat{H}_{\mathrm{b}}=\hat{H}_{\mathrm{ph}}=\sum_{\mathbf{k}\lambda }\hbar
\omega _{\mathbf{k}\lambda }a_{\mathbf{k}\lambda }^{\dagger }a_{\mathbf{k}%
\lambda }.
\end{equation}
Describing the spin-phonon interaction, we will follow the approach
developed in Refs. \cite{chu04prl,chugarsch05prb} that allows to avoid using
unknown spin-phonon coupling constants and to greatly simplify the
formalism. As a magnetic molecule is more rigid than its ligand environment,
a good approximation is to consider this molecule rotated by transverse
phonons without distortion of its crystal field. This leads to the
spin-phonon interaction
\begin{equation}
\hat{V}=\hat{H}_{\mathrm{s-ph}}=\hat{R}\hat{H}_{A}\hat{R}^{-1}-\hat{H}%
_{A},\qquad \hat{R}=e^{-i\mathbf{S}\cdot \delta \mathbf{\phi }},
\label{DME-HsphR}
\end{equation}
where $\delta \mathbf{\phi }$ is a small rotation angle given by
\begin{equation}
\delta \mathbf{\phi =}\frac{1}{2}\nabla \times \mathbf{u}(\mathbf{r}),
\label{DME-deltaphi}
\end{equation}
$\mathbf{u}(\mathbf{r})$ being the lattice displacement due to phonons.
Expanding Eq. (\ref{DME-HsphR}) up to the second order in $\delta \mathbf{%
\phi }$ components yields
\begin{equation}
\hat{V}=\hat{V}^{(1)}+\hat{V}^{(2)}\mathbf{,}  \label{DME-V1V2}
\end{equation}
where
\begin{equation}
\hat{V}^{(1)}=i\left[ \hat{H}_{A},\mathbf{S}\right] \cdot \delta \mathbf{%
\phi }  \label{DME-V1}
\end{equation}
and
\begin{equation}
\hat{V}^{(2)}=\frac{i^{2}}{2!}\left[ \left[ \hat{H}_{A},S_{\xi }\right]
,S_{\xi ^{\prime }}\right] \delta \phi _{\xi }\delta \phi _{\xi ^{\prime
}},\qquad \xi ,\xi ^{\prime }=x,y,z  \label{DME-V2}
\end{equation}
with summation over repeated indices. We use canonical quantization of
phonons,
\begin{equation}
\mathbf{u}=\sqrt{\frac{\hbar }{2MN}}\sum_{\mathbf{k}\lambda }\frac{\mathbf{e}%
_{\mathbf{k}\lambda }e^{i\mathbf{k\cdot r}}}{\sqrt{\omega _{\mathbf{k}%
\lambda }}}\left( a_{\mathbf{k}\lambda }+a_{-\mathbf{k}\lambda }^{\dagger
}\right) ,  \label{DME-uQuantized}
\end{equation}
where $M$ is the mass of the unit cell, $N$ is the number of cells in the
crystal, $\mathbf{e}_{\mathbf{k}\lambda }$ are unit polarization vectors, $%
\lambda =t_{1},t_{2},l$ denotes polarization, and $\omega _{k\lambda
}=v_{\lambda }k$ is the phonon frequency. The operator $\delta \mathbf{\phi }
$ that follows from Eq. (\ref{DME-deltaphi}) is given by
\begin{equation}
\delta \mathbf{\phi }=\frac{1}{2}\sqrt{\frac{\hbar }{2MN}}\sum_{\mathbf{k}%
\lambda }\frac{\left[ i\mathbf{k}\times \mathbf{e}_{\mathbf{k}\lambda }%
\right] e^{i\mathbf{k\cdot r}}}{\sqrt{\omega _{\mathbf{k}\lambda }}}\left(
a_{\mathbf{k}\lambda }+a_{-\mathbf{k}\lambda }^{\dagger }\right) .
\label{DME-deltaphiPhiQuantized}
\end{equation}
Only transverse phonons, $\mathbf{e}_{\mathbf{k}\lambda }\bot \mathbf{k},$
survive in this formula. Whereas $\hat{V}^{(1)}$ is linear in phonon
operators and describes direct phonon processes, $\hat{V}^{(2)}$ is
quadratic and describes Raman processes. Relaxation rates due to Raman
processes are generally much smaller than that due to the direct processes
since they are the next order in the spin-phonon interaction. However, the
rates of direct processes can be small for special reasons, then Raman
processes become important. Processes of orders higher than Raman can be
always neglected.

It is important that the spin-phonon interaction above does not include any
poorly known spin-lattice coupling coefficients and it is entirely
represented by the crystal field $\hat{H}_{A}.$ Moreover, the relaxation
terms in the DME can be represented in the form that does not explicitly
contain $\hat{H}_{A},$ the information about it being absorbed in the spin
eigenstates $\left| \alpha \right\rangle $ and transition frequencies $%
\omega _{\alpha \beta }$ that can be found by numerical diagonalization of $%
\hat{H}_{S}.$ This can be achieved either by changing from the laboratory
frame to the local lattice frame in which $\hat{H}_{A}$ remains constant but
an effective rotation-generated magnetic field arises\cite
{chutej98book,chu04prl,chugarsch05prb} or by manipulating matrix elements of
the spin-phonon interaction with respect to spin states, $\left\langle
\alpha \left| \hat{V}\right| \beta \right\rangle $, Ref. \cite{chugarsch05prb}%
. Both methods are mathematically equivalent \cite{chugarsch05prb}. In
particular, for $\hat{V}^{(1)}$ one can use
\begin{equation}
\left[ \hat{H}_{A},\mathbf{S}\right] =\left[ \left( \hat{H}_{S}-\hat{H}_{%
\mathrm{Z}}\right) ,\mathbf{S}\right] =\left[ \hat{H}_{S},\mathbf{S}\right]
+i\mathbf{S\times }g\mu _{B}\mathbf{H}  \label{DME-HA-Identity}
\end{equation}
and the fact that $\left| \alpha \right\rangle $ are eigenstates of $\hat{H}%
_{S}$ to obtain the spin matrix element
\begin{equation}
\mathbf{\Xi }_{\alpha \beta }^{(1)}\equiv i\left\langle \alpha \left| \left[
\hat{H}_{A},\mathbf{S}\right] \right| \beta \right\rangle =i\hbar \omega
_{\alpha \beta }\left\langle \alpha \left| \mathbf{S}\right| \beta
\right\rangle -\left\langle \alpha \left| \mathbf{S}\right| \beta
\right\rangle \mathbf{\times }g\mu _{B}\mathbf{H}.  \label{DME-XiDef}
\end{equation}
For $\hat{V}^{(2)}$ one writes \cite{calchugar06prb}
\begin{eqnarray}
\left[ \left[ \hat{H}_{A},S_{\xi }\right] ,S_{\xi ^{\prime }}\right] &=&%
\left[ \left[ \left( \hat{H}_{S}-\hat{H}_{\mathrm{Z}}\right) ,S_{\xi }\right]
,S_{\xi ^{\prime }}\right]  \nonumber \\
&=&\left[ \left[ \hat{H}_{S},S_{\xi }\right] ,S_{\xi ^{\prime }}\right]
-g\mu _{B}\left( H_{\xi ^{\prime }}S_{\xi }-\delta _{\xi \xi ^{\prime
}}\sum_{\xi ^{\prime \prime }}H_{\xi ^{\prime \prime }}S_{\xi ^{\prime
\prime }}\right) .
\end{eqnarray}
Here the first term can be transformed as follows
\begin{eqnarray}
\left\langle \alpha \left| \left[ \left[ \hat{H}_{S},S_{\xi }\right] ,S_{\xi
^{\prime }}\right] \right| \beta \right\rangle &=&\left\langle \alpha \left|
\hat{H}_{S}S_{\xi }S_{\xi ^{\prime }}-S_{\xi }\hat{H}_{S}S_{\xi ^{\prime
}}-S_{\xi ^{\prime }}\hat{H}_{S}S_{\xi }+S_{\xi ^{\prime }}S_{\xi }\hat{H}%
_{S}\right| \beta \right\rangle  \nonumber \\
&=&\sum_{\gamma }\left[ \left( \varepsilon _{\alpha }-\varepsilon _{\gamma
}\right) \left\langle \alpha \left| S_{\xi }\right| \gamma \right\rangle
\left\langle \gamma \left| S_{\xi ^{\prime }}\right| \beta \right\rangle
+\left( \varepsilon _{\beta }-\varepsilon _{\gamma }\right) \left\langle
\alpha \left| S_{\xi ^{\prime }}\right| \gamma \right\rangle \left\langle
\gamma \left| S_{\xi }\right| \beta \right\rangle \right]  \nonumber \\
&\Rightarrow &\sum_{\gamma }\left( \varepsilon _{\alpha }+\varepsilon
_{\beta }-2\varepsilon _{\gamma }\right) \left\langle \alpha \left| S_{\xi
}\right| \gamma \right\rangle \left\langle \gamma \left| S_{\xi ^{\prime
}}\right| \beta \right\rangle ,
\end{eqnarray}
where on the last step the symmetry properties of Eq. (\ref{DME-V2}) were
used. Finally one obtains
\begin{eqnarray}
\Xi _{\alpha \beta ,\xi \xi ^{\prime }}^{(2)} &\equiv &i^{2}\left\langle
\alpha \left| \left[ \left[ \hat{H}_{A},S_{\xi }\right] ,S_{\xi ^{\prime }}%
\right] \right| \beta \right\rangle =-\sum_{\gamma }\left( \varepsilon
_{\alpha }+\varepsilon _{\beta }-2\varepsilon _{\gamma }\right) \left\langle
\alpha \left| S_{\xi }\right| \gamma \right\rangle \left\langle \gamma
\left| S_{\xi ^{\prime }}\right| \beta \right\rangle  \nonumber \\
&&\qquad \qquad +g\mu _{B}\left( H_{\xi ^{\prime }}\left\langle \alpha
\left| S_{\xi }\right| \beta \right\rangle -\delta _{\xi \xi ^{\prime
}}\sum_{\xi ^{\prime \prime }}H_{\xi ^{\prime \prime }}\left\langle \alpha
\left| S_{\xi ^{\prime \prime }}\right| \beta \right\rangle \right) .
\label{DME-Xi2Def}
\end{eqnarray}
Eqs. (\ref{DME-XiDef}) and (\ref{DME-Xi2Def}) provide a great simplification
of the formalism, since otherwise one would have to derive different forms
of the relaxation part of the DME for each particular $\hat{H}_{A}.$

\subsection{DME for molecular magnets}

\label{Sec-MM-DME}

\subsubsection{Secular vs non-secular}

Most of numerical work on molecular magnets used the secular form of the
DME, in fact reduced to the system of rate equations for the populations.
However, tunneling resonances can make the secular approximation invalid.
Indeed, if two levels $\alpha $ and $\alpha ^{\prime }$ of the small system
have very close energies, the density matrix element $\rho _{\alpha \alpha
^{\prime }}$ is oscillating with a very small frequency $\omega _{\alpha
\alpha ^{\prime }}$ in the absence of the coupling to the bath, see Eq. (\ref
{DME-rhoEqTensor}). If $\omega _{\alpha \alpha ^{\prime }}$ is smaller than
the relaxation rate between the neigboring energy levels, $\rho _{\alpha
\alpha ^{\prime }}$ does not decouple from the diagonal elements $\rho
_{\alpha \alpha }$ and $\rho _{\alpha ^{\prime }\alpha ^{\prime }},$ and the
secular approximation breaks down. It is easy to demonstrate that the
failure of the secular approximation at resonance may lead to unphysically
high escape rates out of the metastable state. Consider a MM exactly at $k$%
th resonance with $k>0$ and a very small $\hat{H}_{A}^{\prime }$ and the
transverse field, so that the metastable ground state $\left|
-S\right\rangle $ is at resonance with the excited state $\left|
S-k\right\rangle $ in the right well. The latter can decay into the
lower-lying state $\left| S-k+1\right\rangle $ with the rate $\Gamma
_{S-k+1,S-k}.$ Since the exact eigenstates at the tunneling resonance are $%
\left| \pm \right\rangle $ that are linear combinations of $\left|
-S\right\rangle $ and $\left| S-k\right\rangle $ (see Sec. \ref
{Sec-GS-tunneling}), both of these eigenstates are damped with the rate of
order $\Gamma _{S-k+1,S-k}$ (in fact, half of it). The secular DME uses rate
equations for $\rho _{++},$ $\rho _{--},$ etc., and the initial condition
spin in the state $\left| -S\right\rangle $ gives rise to the initial
conditions $\rho _{++}(0)=$ $\rho _{--}(0)=1/2.$ Since both $\rho _{++}$ and
$\rho _{--}$ relax with a rate of order $\Gamma _{S-k+1,S-k},$ the spin
quickly leaves the metastable state, even in the case of a vanishing tunnel
splitting, $\Delta \rightarrow 0.$ Indeed, such an unphysical behavior
follows from the analytical and numerical solution of the secular DME at
weak tunneling resonances. In contrast, coupling of $\rho _{++}$ and $\rho
_{--}$ to the slow non-diagonal DM elements $\rho _{+-}$ and $\rho _{-+}$ in
the non-secular DME leads to the physically expected vanishing of the escape
rate in the limit $\Delta \rightarrow 0$ and $T\rightarrow 0.$

Below the non-secular DME, Eq. (\ref{DME-rhoEqTensor}), will be used in the
development of the formalism. The secular and semi-secular reductions of it
can be obtained later. The relaxation tensor $R_{\alpha \beta ,\alpha
^{\prime }\beta ^{\prime }}$ is a sum of two contributions,
\begin{equation}
R_{\alpha \beta ,\alpha ^{\prime }\beta ^{\prime }}=R_{\alpha \beta ,\alpha
^{\prime }\beta ^{\prime }}^{(1)}+R_{\alpha \beta ,\alpha ^{\prime }\beta
^{\prime }}^{(2)},  \label{DME-RR1R2}
\end{equation}
that are due to the first- and second-order phonon processes. These
contributions will be calculated separately below.

\subsubsection{Initial condition for free relaxation}

Let us consider the question of the initial state of the spin in the case of
free evolution. In resonance experiments it is, typically, the first excited
state. Although, practically, in these experiments only a small portion of
the population is being transferred from the ground state to the excited
state, one can consider the system prepared fully in the excited state
because of the linearity of the DME. Preparing the spin in the metastable
energy minimum, one can study its thermal activation over the barrier and
tunneling under the barrier. In general, it is not easy to find the
quantum-mechanical state realizing or approximating this classical state,
and in the case of a tunneling resonance such a state does not exist. A good
practical way to create such an initial condition is to prepare the spin in
the coherent state $\left| \mathbf{n}(\theta ,\varphi )\right\rangle $
pointing in the direction of the metastable minimum found classically. The
spin coherent state is given by
\begin{equation}
\left| \mathbf{n}(\theta ,\varphi )\right\rangle =\sum_{m=-S}^{S}C_{m}\left|
m\right\rangle ,  \label{DME-CoherentState}
\end{equation}
where
\begin{equation}
C_{m}=\binom{2S}{S+m}^{1/2}\left( \cos \frac{\theta }{2}\right) ^{S+m}\left(
\sin \frac{\theta }{2}\right) ^{S-m}e^{-im\varphi }.  \label{DME-SpinCohCm}
\end{equation}

\subsubsection{Direct processes}

To compute matrix elements $V_{\alpha \varpi ,\gamma \varpi ^{\prime
}}^{(1)} $ etc., in Eq. (\ref{DME-Rdiag}) with respect to the phonon bath,
one can label the state $|\phi _{\varpi }\rangle $ by the numbers of phonons
$\nu _{\mathbf{k}\lambda }=0,1,2,\ldots $ in each phonon mode $\mathbf{k}%
\lambda $%
\begin{equation}
|\phi _{\varpi }\rangle =|\ldots ,\nu _{\mathbf{k}\lambda },\ldots \rangle
\Rightarrow |\nu _{\mathbf{k}\lambda }\rangle .
\end{equation}
In the direct processes, the state $|\phi _{\varpi ^{\prime }}\rangle $ is
not independent and it differs from $|\phi _{\varpi }\rangle $ by creation
or annihilation of one phonon, according to Eq. (\ref
{DME-deltaphiPhiQuantized}). We will make use of the phonon matrix elements
\begin{equation}
\mathbf{M}_{\pm }(\mathbf{k})=\left\langle \nu _{\mathbf{k}\lambda }\pm
1\left| \delta \mathbf{\phi }\right| \nu _{\mathbf{k}\lambda }\right\rangle
\end{equation}
and their conjugates. From Eq. (\ref{DME-deltaphiPhiQuantized}) one obtains
\begin{eqnarray}
\mathbf{M}_{-}(\mathbf{k}) &=&\left\langle \nu _{\mathbf{k}\lambda }-1\left|
\frac{1}{2}\sqrt{\frac{\hbar }{2MN}}\frac{\left[ i\mathbf{k}\times \mathbf{e}%
_{\mathbf{k}\lambda }\right] e^{i\mathbf{k\cdot r}}}{\sqrt{\omega _{\mathbf{k%
}}}}a_{\mathbf{k}\lambda }\right| \nu _{\mathbf{k}\lambda }\right\rangle =%
\frac{1}{2}\sqrt{\frac{\hbar }{2MN}}\frac{\left[ i\mathbf{k}\times \mathbf{e}%
_{\mathbf{k}\lambda }\right] e^{i\mathbf{k\cdot r}}}{\sqrt{\omega _{\mathbf{k%
}}}}\sqrt{\nu _{\mathbf{k}\lambda }}  \nonumber \\
\mathbf{M}_{-}^{\ast }(\mathbf{k}) &=&\left\langle \nu _{\mathbf{k}\lambda
}\left| \frac{1}{2}\sqrt{\frac{\hbar }{2MN}}\frac{\left[ -i\mathbf{k}\times
\mathbf{e}_{\mathbf{k}\lambda }\right] e^{-i\mathbf{k\cdot r}}}{\sqrt{\omega
_{\mathbf{k}}}}a_{\mathbf{k}\lambda }^{\dagger }\right| \nu _{\mathbf{k}%
\lambda }-1\right\rangle =\frac{1}{2}\sqrt{\frac{\hbar }{2MN}}\frac{\left[ -i%
\mathbf{k}\times \mathbf{e}_{\mathbf{k}\lambda }\right] e^{-i\mathbf{k\cdot r%
}}}{\sqrt{\omega _{\mathbf{k}}}}\sqrt{\nu _{\mathbf{k}\lambda }}  \nonumber
\\
\mathbf{M}_{+}(\mathbf{k}) &=&\left\langle \nu _{\mathbf{k}\lambda }+1\left|
\frac{1}{2}\sqrt{\frac{\hbar }{2MN}}\frac{\left[ -i\mathbf{k}\times \mathbf{e%
}_{\mathbf{k}\lambda }\right] e^{-i\mathbf{k\cdot r}}}{\sqrt{\omega _{%
\mathbf{k}}}}a_{\mathbf{k}\lambda }^{\dagger }\right| \nu _{\mathbf{k}%
\lambda }\right\rangle =\frac{1}{2}\sqrt{\frac{\hbar }{2MN}}\frac{\left[ -i%
\mathbf{k}\times \mathbf{e}_{\mathbf{k}\lambda }\right] e^{-i\mathbf{k\cdot r%
}}}{\sqrt{\omega _{\mathbf{k}}}}\sqrt{\nu _{\mathbf{k}\lambda }+1}  \nonumber
\\
\mathbf{M}_{+}^{\ast }(\mathbf{k}) &=&\left\langle \nu _{\mathbf{k}\lambda
}\left| \frac{1}{2}\sqrt{\frac{\hbar }{2MN}}\frac{\left[ i\mathbf{k}\times
\mathbf{e}_{\mathbf{k}\lambda }\right] e^{i\mathbf{k\cdot r}}}{\sqrt{\omega
_{\mathbf{k}}}}a_{\mathbf{k}\lambda }\right| \nu _{\mathbf{k}\lambda
}+1\right\rangle =\frac{1}{2}\sqrt{\frac{\hbar }{2MN}}\frac{\left[ i\mathbf{k%
}\times \mathbf{e}_{\mathbf{k}\lambda }\right] e^{i\mathbf{k\cdot r}}}{\sqrt{%
\omega _{\mathbf{k}}}}\sqrt{\nu _{\mathbf{k}\lambda }+1}.
\label{DME-MExplicit}
\end{eqnarray}
In Eq. (\ref{DME-Rdiag}) one has, e.g.,
\begin{eqnarray}
&&\frac{\pi }{\hbar Z_{\mathrm{b}}}\sum_{\varpi \varpi ^{\prime
}}e^{-E_{\varpi }/(k_{B}T)}\delta \left( \varepsilon _{\alpha ^{\prime
}}-\varepsilon _{\gamma }+E_{\varpi }-E_{\varpi ^{\prime }}\right) V_{\alpha
\varpi ,\gamma \varpi ^{\prime }}^{(1)}V_{\gamma \varpi ^{\prime },\alpha
^{\prime }\varpi }^{(1)}  \nonumber \\
&=&\sum_{\mathbf{k}\lambda }\frac{\pi }{\hbar Z_{\mathrm{b}}}\sum_{\nu _{%
\mathbf{k}\lambda },\ldots }e^{-E_{\nu _{\mathbf{k}\lambda },\ldots
}/(k_{B}T)}\delta \left( \varepsilon _{\alpha ^{\prime }}-\varepsilon
_{\gamma }+\hbar \omega _{\mathbf{k}}\right) \left( \mathbf{\Xi }_{\alpha
\gamma }^{(1)}\cdot \mathbf{M}_{-}^{\ast }(\mathbf{k})\right) \left( \mathbf{%
\Xi }_{\gamma \alpha ^{\prime }}^{(1)}\cdot \mathbf{M}_{-}(\mathbf{k})\right)
\nonumber \\
&&+\sum_{\mathbf{k}\lambda }\frac{\pi }{\hbar Z_{\mathrm{b}}}\sum_{\nu _{%
\mathbf{k}\lambda },\ldots }e^{-E_{\nu _{\mathbf{k}\lambda },\ldots
}/(k_{B}T)}\delta \left( \varepsilon _{\alpha ^{\prime }}-\varepsilon
_{\gamma }-\hbar \omega _{\mathbf{k}}\right) \left( \mathbf{\Xi }_{\alpha
\gamma }^{(1)}\cdot \mathbf{M}_{+}^{\ast }(\mathbf{k})\right) \left( \mathbf{%
\Xi }_{\gamma \alpha ^{\prime }}^{(1)}\cdot \mathbf{M}_{+}(\mathbf{k}%
)\right) .  \label{DME-DProcesses}
\end{eqnarray}
After averaging over phonon populations,
\begin{equation}
\frac{1}{Z_{\mathrm{b}}}\sum_{\nu _{\mathbf{k}\lambda },\ldots }e^{-E_{\nu _{%
\mathbf{k}\lambda },\ldots }/(k_{B}T)}\nu _{\mathbf{k}\lambda }\equiv
\left\langle \nu _{\mathbf{k}\lambda }\right\rangle =n_{\mathbf{k}\lambda
}=n_{\mathbf{k}}=\frac{1}{e^{\hbar \omega _{\mathbf{k}}/(k_{B}T)}-1},
\label{DME-nukAvrDef}
\end{equation}
this becomes
\begin{eqnarray}
&&\frac{\pi }{8MN}\sum_{\mathbf{k}\lambda }\left[ \left( \mathbf{\Xi }%
_{\alpha \gamma }^{(1)}\cdot \frac{\left[ \mathbf{k}\times \mathbf{e}_{%
\mathbf{k}\lambda }\right] }{\sqrt{\omega _{\mathbf{k}}}}\right) \left(
\mathbf{\Xi }_{\gamma \alpha ^{\prime }}^{(1)}\cdot \frac{\left[ \mathbf{k}%
\times \mathbf{e}_{\mathbf{k}\lambda }\right] }{\sqrt{\omega _{\mathbf{k}}}}%
\right) \delta \left( \varepsilon _{\alpha ^{\prime }}-\varepsilon _{\gamma
}+\hbar \omega _{\mathbf{k}}\right) n_{\mathbf{k}}\right.  \nonumber \\
&&\qquad \qquad +\left. \left( \mathbf{\Xi }_{\alpha \gamma }^{(1)}\cdot
\frac{\left[ \mathbf{k}\times \mathbf{e}_{\mathbf{k}\lambda }\right] }{\sqrt{%
\omega _{\mathbf{k}}}}\right) \left( \mathbf{\Xi }_{\gamma \alpha ^{\prime
}}^{(1)}\cdot \frac{\left[ \mathbf{k}\times \mathbf{e}_{\mathbf{k}\lambda }%
\right] }{\sqrt{\omega _{\mathbf{k}}}}\right) \delta \left( \varepsilon
_{\alpha ^{\prime }}-\varepsilon _{\gamma }-\hbar \omega _{\mathbf{k}%
}\right) \left( n_{\mathbf{k}}+1\right) \right] .
\end{eqnarray}
This can be further simplified using
\begin{equation}
\left[ \mathbf{k}\times \mathbf{e}_{\mathbf{k}t_{1}}\right] =\pm k\mathbf{e}%
_{\mathbf{k}t_{2}}  \label{DME-AnotherTrans}
\end{equation}
and, for the summation over the two transverse polarizations,
\begin{equation}
\sum_{t=t_{1},t_{2}}\left( \mathbf{e}_{\mathbf{k}t}\cdot \mathbf{a}\right)
\left( \mathbf{e}_{\mathbf{k}t}\cdot \mathbf{b}\right) =\left( \mathbf{%
a\cdot b}\right) -\frac{\left( \mathbf{k\cdot a}\right) \left( \mathbf{%
k\cdot b}\right) }{k^{2}}  \label{DME-transverseraus}
\end{equation}
with $\mathbf{a=b=e}_{x},$ and then $\mathbf{e}_{y.}$ Averaging over the
directions of the vector $\mathbf{k}$:
\begin{equation}
\left\langle \left( \mathbf{k}\cdot \mathbf{a}\right) \left( \mathbf{k}\cdot
\mathbf{b}\right) \right\rangle =\frac{k^{2}}{3}\left( \mathbf{a}\cdot
\mathbf{b}\right)  \label{DME-kAver}
\end{equation}
one obtains
\begin{eqnarray}
&&\left( \mathbf{\Xi }_{\alpha \gamma }^{(1)}\cdot \mathbf{\Xi }_{\gamma
\alpha ^{\prime }}^{(1)}\right) \frac{\pi }{12MN}\sum_{\mathbf{k}}\frac{k^{2}%
}{\omega _{\mathbf{k}}}\left[ \delta \left( \varepsilon _{\alpha ^{\prime
}}-\varepsilon _{\gamma }+\hbar \omega _{\mathbf{k}}\right) n_{\mathbf{k}%
}+\delta \left( \varepsilon _{\alpha ^{\prime }}-\varepsilon _{\gamma
}-\hbar \omega _{\mathbf{k}}\right) \left( n_{\mathbf{k}}+1\right) \right]
\nonumber \\
&=&\left( \mathbf{\Xi }_{\alpha \gamma }^{(1)}\cdot \mathbf{\Xi }_{\gamma
\alpha ^{\prime }}^{(1)}\right) \frac{\pi }{12\hbar MN}\sum_{\mathbf{k}}%
\frac{k^{2}}{\omega _{\mathbf{k}}}\left[ \delta \left( \omega _{\alpha
^{\prime }\gamma }+\omega _{\mathbf{k}}\right) n_{\mathbf{k}}+\delta \left(
\omega _{\alpha ^{\prime }\gamma }-\omega _{\mathbf{k}}\right) \left( n_{%
\mathbf{k}}+1\right) \right] .
\end{eqnarray}
Now recalling Eq. (\ref{DME-Rdiag}) one obtains
\begin{eqnarray}
R_{\alpha \beta ,\alpha ^{\prime }\beta ^{\prime }}^{(1)} &=&\frac{\pi D^{2}%
}{12\hbar MN}\sum_{\mathbf{k}}\frac{k^{2}}{\omega _{\mathbf{k}}}  \nonumber
\\
&&\left\{ -\sum_{\gamma }Q_{\alpha \alpha ^{\prime },\gamma \gamma }^{(1)}
\left[ \delta \left( \omega _{\alpha ^{\prime }\gamma }+\omega _{\mathbf{k}%
}\right) n_{\mathbf{k}}+\delta \left( \omega _{\alpha ^{\prime }\gamma
}-\omega _{\mathbf{k}}\right) \left( n_{\mathbf{k}}+1\right) \right] \delta
_{\beta ^{\prime }\beta }\right.  \nonumber \\
&&-\delta _{\alpha \alpha ^{\prime }}\sum_{\gamma }Q_{\beta ^{\prime }\beta
,\gamma \gamma }^{(1)}\left[ \delta \left( \omega _{\beta ^{\prime }\gamma
}+\omega _{\mathbf{k}}\right) n_{\mathbf{k}}+\delta \left( \omega _{\beta
^{\prime }\gamma }-\omega _{\mathbf{k}}\right) \left( n_{\mathbf{k}%
}+1\right) \right]  \nonumber \\
&&+\left. Q_{\alpha \beta ,\alpha ^{\prime }\beta ^{\prime }}^{(1)}\left[
\delta \left( \omega _{\alpha \alpha ^{\prime }}+\omega _{\mathbf{k}}\right)
\left( n_{\mathbf{k}}+1\right) +\delta \left( \omega _{\alpha \alpha
^{\prime }}-\omega _{\mathbf{k}}\right) n_{\mathbf{k}}+(\alpha \rightarrow
\beta )\right] \right\} ,
\end{eqnarray}
where
\begin{equation}
Q_{\alpha \beta ,\alpha ^{\prime }\beta ^{\prime }}^{(1)}\equiv \left(
\mathbf{\Xi }_{\alpha \alpha ^{\prime }}^{(1)}\cdot \mathbf{\Xi }_{\beta
^{\prime }\beta }^{(1)}\right) /D^{2}  \label{DME-QabaprbprDef}
\end{equation}
is a dimensionless combination that characterizes the spin. Next, it is
convenient to go over from summation to integration,
\begin{equation}
\frac{1}{N}\sum_{\mathbf{k}}\ldots \Rightarrow v_{0}\int \frac{d^{3}k}{%
\left( 2\pi \right) ^{3}}\ldots  \label{DME-Sum2Int}
\end{equation}
where $v_{0}$ is the unit-cell volume. Using $v_{0}/M=1/\rho $ and $\omega
_{k}=v_{t}k$ one can introduce the characteristic frequency $\Omega _{t}$
and the corresponding energy $E_{t}$ of the spin-phonon interaction
\begin{equation}
\Omega _{t}\equiv \left( \frac{\rho v_{t}^{5}}{\hbar }\right) ^{1/4},\qquad
E_{t}\equiv \left( \rho v_{t}^{5}\hbar ^{3}\right) ^{1/4}.
\label{DME-OmegatDef}
\end{equation}
As a result one obtains the characteristic relaxation rate
\begin{equation}
\frac{\pi D^{2}}{12\hbar MN}\sum_{\mathbf{k}}\frac{k^{2}}{\omega _{\mathbf{k}%
}}\delta \left( \omega _{\mathbf{k}}-\omega _{0}\right) =\frac{\omega
_{0}^{3}D^{2}}{24\pi \hbar ^{2}\Omega _{t}^{4}}\theta (\omega _{0})\equiv
\Gamma ^{(1)}(\omega _{0}),  \label{DME-GammaomegaDef}
\end{equation}
where
\begin{equation}
\theta (\omega )=\left\{
\begin{array}{cc}
0, & \omega \leq 0 \\
1, & \omega >0.
\end{array}
\right.
\end{equation}
that enters the relaxation terms. In terms of $\Gamma ^{(1)}(\omega _{0})$
and
\begin{equation}
n_{\omega }\equiv \frac{1}{e^{\hbar \omega /(k_{B}T)}-1}
\label{DME-nomegaDef}
\end{equation}
one obtains
\begin{eqnarray}
R_{\alpha \beta ,\alpha ^{\prime }\beta ^{\prime }}^{(1)} &=&-\sum_{\gamma
}Q_{\alpha \alpha ^{\prime },\gamma \gamma }^{(1)}\left[ \Gamma
^{(1)}(\omega _{\gamma \alpha ^{\prime }})n_{\omega _{\gamma \alpha ^{\prime
}}}+\Gamma ^{(1)}(\omega _{\alpha ^{\prime }\gamma })\left( n_{\omega
_{\alpha ^{\prime }\gamma }}+1\right) \right] \delta _{\beta ^{\prime }\beta
}  \nonumber \\
&&-\delta _{\alpha \alpha ^{\prime }}\sum_{\gamma }Q_{\beta ^{\prime }\beta
,\gamma \gamma }^{(1)}\left[ \Gamma ^{(1)}(\omega _{\gamma \beta ^{\prime
}})n_{\omega _{\gamma \beta ^{\prime }}}+\Gamma ^{(1)}(\omega _{\beta
^{\prime }\gamma })\left( n_{\omega _{\beta ^{\prime }\gamma }}+1\right)
\right]  \nonumber \\
&&+Q_{\alpha \beta ,\alpha ^{\prime }\beta ^{\prime }}^{(1)}\left[ \Gamma
^{(1)}\left( \omega _{\alpha ^{\prime }\alpha }\right) \left( n_{\omega
_{\alpha ^{\prime }\alpha }}+1\right) +\Gamma ^{(1)}\left( \omega _{\alpha
\alpha ^{\prime }}\right) n_{\omega _{\alpha \alpha ^{\prime }}}+(\alpha
\rightarrow \beta )\right]  \label{DME-RTensorGamma}.
\end{eqnarray}
Remember that here all $\Gamma ^{(1)}(\omega )$ with $\omega <0$ are zero.
Here $Q_{\alpha \beta ,\alpha ^{\prime }\beta ^{\prime }}^{(1)}$ is defined
by Eqs. (\ref{DME-QabaprbprDef}) and (\ref{DME-XiDef}).

Within the secular approximation, all relaxation terms in the DME are
defined by $\Gamma _{\alpha \alpha ^{\prime }}=\left. R_{\alpha \alpha
,\alpha ^{\prime }\alpha ^{\prime }}\right| _{\alpha ^{\prime }\neq \alpha
}, $ see Eqs. (\ref{DME-WaplaalphaprDef}), (\ref{DME-GammaalphabetaDef}),
and (\ref{DME-DME-final}), whereas for the direct processes considered here
one has $\bar{\Gamma}_{\alpha \beta }^{(1)}=0$. For $\Gamma _{\alpha \alpha
^{\prime }}^{(1)}$ from Eq. (\ref{DME-RTensorGamma}) one obtains
\begin{equation}
\Gamma _{\alpha \alpha ^{\prime }}^{(1)}=2\frac{\left| \mathbf{\Xi }_{\alpha
\alpha ^{\prime }}^{(1)}\right| ^{2}}{D^{2}}\left[ \Gamma ^{(1)}\left(
\omega _{\alpha ^{\prime }\alpha }\right) \left( n_{\omega _{\alpha ^{\prime
}\alpha }}+1\right) +\Gamma ^{(1)}\left( \omega _{\alpha \alpha ^{\prime
}}\right) n_{\omega _{\alpha \alpha ^{\prime }}}\right] ,
\label{DME-GammaaaprviaXi2}
\end{equation}
where we used
\begin{equation}
\left( \mathbf{\Xi }_{\alpha \alpha ^{\prime }}^{(1)}\cdot \mathbf{\Xi }%
_{\alpha ^{\prime }\alpha }^{(1)}\right) =\left( \mathbf{\Xi }_{\alpha
\alpha ^{\prime }}^{(1)}\cdot \mathbf{\Xi }_{\alpha \alpha ^{\prime
}}^{(1)\ast }\right) =\left| \mathbf{\Xi }_{\alpha \alpha ^{\prime
}}^{(1)}\right| ^{2}.  \label{DME-Xi2Sec}
\end{equation}

To compute the matrix elements of the spin operator components above, one
uses
\begin{eqnarray}
S_{\pm } &\equiv &S_{x}\pm iS_{y},\qquad S_{x}=\frac{1}{2}\left(
S_{-}+S_{+}\right) ,\qquad S_{y}=\frac{i}{2}\left( S_{-}-S_{+}\right)
\nonumber \\
S_{-} &=&\sum_{m=-S}^{S-1}\left| m\right\rangle l_{m,m+1}\left\langle
m+1\right| ,\qquad S_{+}=\sum_{m=-S}^{S-1}\left| m+1\right\rangle
l_{m+1,m}\left\langle m\right| ,\qquad S_{z}=\sum_{m=-S}^{S}\left|
m\right\rangle m\left\langle m\right| ,  \label{DME-lDef}
\end{eqnarray}
where $l_{m,m^{\prime }}=\sqrt{S(S+1)-mm^{\prime }}.$ Thus one obtains
\begin{eqnarray}
\left\langle \alpha \left| S_{z}\right| \alpha ^{\prime }\right\rangle
&=&\sum_{m=-S}^{S}\langle \alpha |m\rangle m\langle m|\alpha ^{\prime
}\rangle  \nonumber \\
\left\langle \alpha \left| S_{x}\right| \alpha ^{\prime }\right\rangle &=&%
\frac{1}{2}\sum_{m=-S}^{S-1}\left( \langle \alpha |m\rangle l_{m,m+1}\langle
m+1|\alpha ^{\prime }\rangle +\langle \alpha |m+1\rangle l_{m+1,m}\langle
m|\alpha ^{\prime }\rangle \right)  \nonumber \\
\left\langle \alpha \left| S_{y}\right| \alpha ^{\prime }\right\rangle &=&%
\frac{i}{2}\sum_{m=-S}^{S-1}\left( \langle \alpha |m\rangle l_{m,m+1}\langle
m+1|\alpha ^{\prime }\rangle -\langle \alpha |m+1\rangle l_{m+1,m}\langle
m|\alpha ^{\prime }\rangle \right) .  \label{DME-SxyzMatrEl}
\end{eqnarray}

\subsubsection{Raman processes}

Raman processes arise due to $V^{(2)}$ in Eq. (\ref{DME-V1V2}), as well as
due to $V^{(1)}$ in the second order of the perturbation theory (details can
be found in Ref. \cite{calchugar06prb}). The latter terms are nonessential
at higher temperatures where Raman processes can become non-negligible since
they contain a large thermal phonon frequency in the denominator. In Raman
processes a phonon $\mathbf{k}$ is absorbed and a phonon $\mathbf{q}$ is
emitted or vice versa. Processes with emission or absorption of two phonons
make a small contribution and they will be ignored here. Thus the relevant
phonon matrix elements are of the form
\begin{eqnarray}
&&\left\langle \nu _{\mathbf{k}\lambda }-1,\nu _{\mathbf{q}\lambda }+1\left|
\delta \phi _{\xi }\delta \phi _{\xi ^{\prime }}\right| \nu _{\mathbf{k}%
\lambda },\nu _{\mathbf{q}\lambda }\right\rangle  \nonumber \\
&=&\left\langle \nu _{\mathbf{q}\lambda }+1\left| \delta \phi _{\xi }\right|
\nu _{\mathbf{q}\lambda }\right\rangle \left\langle \nu _{\mathbf{k}\lambda
}-1\left| \delta \phi _{\xi ^{\prime }}\right| \nu _{\mathbf{k}\lambda
}\right\rangle +\left\langle \nu _{\mathbf{k}\lambda }-1\left| \delta \phi
_{\xi }\right| \nu _{\mathbf{k}\lambda }\right\rangle \left\langle \nu _{%
\mathbf{q}\lambda }+1\left| \delta \phi _{\xi ^{\prime }}\right| \nu _{%
\mathbf{q}\lambda }\right\rangle  \nonumber \\
&=&M_{-,\xi ^{\prime }}(\mathbf{k})M_{+,\xi }(\mathbf{q})+M_{-,\xi }(\mathbf{%
k})M_{+,\xi ^{\prime }}(\mathbf{q})\equiv 2\tilde{M}_{\xi \xi ^{\prime }}(%
\mathbf{k,q}),
\end{eqnarray}
where the matrix elements $M(\mathbf{k})$ are defined by Eq. (\ref
{DME-MExplicit}). Similarly to Eq. (\ref{DME-DProcesses}), for one of the
parts of $R_{\alpha \beta ,\alpha ^{\prime }\beta ^{\prime }}^{(2)}$ one
obtains
\begin{eqnarray}
&&\frac{\pi }{\hbar Z_{\mathrm{b}}}\sum_{\varpi \varpi ^{\prime
}}e^{-E_{\varpi }/(k_{B}T)}\delta \left( \varepsilon _{\alpha ^{\prime
}}-\varepsilon _{\gamma }+E_{\varpi }-E_{\varpi ^{\prime }}\right) V_{\alpha
\varpi ,\gamma \varpi ^{\prime }}^{(2)}V_{\gamma \varpi ^{\prime },\alpha
^{\prime }\varpi }^{(2)}  \nonumber \\
&=&\sum_{\mathbf{k}\lambda ,\mathbf{q}\lambda }\frac{\pi }{\hbar Z_{\mathrm{b%
}}}\sum_{\nu _{\mathbf{k}\lambda },\nu _{\mathbf{q}\lambda }\ldots
}e^{-E_{\nu _{\mathbf{k}\lambda },\nu _{\mathbf{q}\lambda },\ldots
}/(k_{B}T)}\delta \left( \varepsilon _{\alpha ^{\prime }}-\varepsilon
_{\gamma }+\hbar \omega _{\mathbf{q}}-\hbar \omega _{\mathbf{k}}\right)
\nonumber \\
&&\qquad \qquad \times \left( \Xi _{\alpha \gamma ,\xi \xi ^{\prime }}^{(2)}%
\tilde{M}_{\xi \xi ^{\prime }}^{\ast }(\mathbf{k,q})\right) \left( \Xi
_{\gamma \alpha ^{\prime },\zeta \zeta ^{\prime }}^{(2)}\tilde{M}_{\zeta
\zeta ^{\prime }}(\mathbf{k,q})\right)  \nonumber \\
&=&\frac{\pi }{\hbar }\sum_{\mathbf{k}\lambda ,\mathbf{q}\lambda }\delta
\left( \varepsilon _{\alpha ^{\prime }}-\varepsilon _{\gamma }+\hbar \omega
_{\mathbf{q}}-\hbar \omega _{\mathbf{k}}\right) \Xi _{\alpha \gamma ,\xi \xi
^{\prime }}^{(2)}\frac{1}{2}\left( M_{-,\xi ^{\prime }}^{\ast }(\mathbf{k}%
)M_{+,\xi }^{\ast }(\mathbf{q})+M_{-,\xi }^{\ast }(\mathbf{k})M_{+,\xi
^{\prime }}^{\ast }(\mathbf{q})\right)  \nonumber \\
&&\qquad \qquad \times \Xi _{\gamma \alpha ^{\prime },\zeta \zeta ^{\prime
}}^{(2)}\frac{1}{2}\left( M_{-,\zeta ^{\prime }}(\mathbf{k})M_{+,\zeta }(%
\mathbf{q})+M_{-,\zeta }(\mathbf{k})M_{+,\zeta ^{\prime }}(\mathbf{q})\right)
\nonumber \\
&=&\frac{\pi }{\hbar }\sum_{\mathbf{k}\lambda ,\mathbf{q}\lambda }\delta
\left( \varepsilon _{\alpha ^{\prime }}-\varepsilon _{\gamma }+\hbar \omega
_{\mathbf{q}}-\hbar \omega _{\mathbf{k}}\right) \bar{\Xi}_{\alpha \gamma
,\xi \xi ^{\prime }}^{(2)}M_{-,\xi }^{\ast }(\mathbf{k})M_{+,\xi ^{\prime
}}^{\ast }(\mathbf{q})\bar{\Xi}_{\alpha \gamma ,\xi \xi ^{\prime
}}^{(2)}M_{-,\zeta }(\mathbf{k})M_{+,\zeta ^{\prime }}(\mathbf{q}),
\end{eqnarray}
where
\begin{equation}
\bar{\Xi}_{\alpha \gamma ,\xi \xi ^{\prime }}^{(2)}=\frac{1}{2}\left( \Xi
_{\alpha \gamma ,\xi \xi ^{\prime }}^{(2)}+\Xi _{\alpha \gamma ,\xi ^{\prime
}\xi }^{(2)}\right)  \label{DME-Xi2Symm}
\end{equation}
and $M$ are averaged over the phonon populations, $\nu _{\mathbf{k}\lambda
}\rightarrow n_{\mathbf{k}\lambda },$ see Eq. (\ref{DME-nukAvrDef}).
Substituting the expressions for $M$ from Eq. (\ref{DME-MExplicit}), one
obtains
\begin{eqnarray}
&&\frac{\pi }{\hbar }\left( \frac{\hbar }{8MN}\right) ^{2}\sum_{\mathbf{k}%
\lambda ,\mathbf{q}\lambda }\delta \left( \varepsilon _{\alpha ^{\prime
}}-\varepsilon _{\gamma }+\hbar \omega _{\mathbf{q}}-\hbar \omega _{\mathbf{k%
}}\right) n_{\mathbf{k}}\left( n_{\mathbf{q}}+1\right) \bar{\Xi}_{\alpha
\gamma ,\xi \xi ^{\prime }}^{(2)}\bar{\Xi}_{\gamma \alpha ^{\prime },\zeta
\zeta ^{\prime }}^{(2)}  \nonumber \\
&&\qquad \qquad \times \frac{\left[ \mathbf{k}\times \mathbf{e}_{\mathbf{k}%
\lambda }\right] _{\xi }\left[ \mathbf{k}\times \mathbf{e}_{\mathbf{k}%
\lambda }\right] _{\zeta }}{\omega _{\mathbf{k}}}\frac{\left[ \mathbf{q}%
\times \mathbf{e}_{\mathbf{q}\lambda }\right] _{\xi ^{\prime }}\left[
\mathbf{q}\times \mathbf{e}_{\mathbf{q}\lambda }\right] _{\zeta ^{\prime }}}{%
\omega _{\mathbf{q}}}.
\end{eqnarray}
Now, using Eqs. (\ref{DME-AnotherTrans})--(\ref{DME-kAver}), one can
simplify this result to
\begin{eqnarray}
&&\frac{\pi }{\left( 8MN\right) ^{2}}\sum_{\mathbf{k}\lambda ,\mathbf{q}%
\lambda }\delta \left( \omega _{\alpha ^{\prime }\gamma }+\omega _{\mathbf{q}%
}-\omega _{\mathbf{k}}\right) n_{\mathbf{k}}\left( n_{\mathbf{q}}+1\right)
\bar{\Xi}_{\alpha \gamma ,\xi \xi ^{\prime }}^{(2)}\bar{\Xi}_{\gamma \alpha
^{\prime },\zeta \zeta ^{\prime }}^{(2)}\left( \frac{2}{3}\right) ^{2}\frac{%
k^{2}}{\omega _{\mathbf{k}}}\frac{q^{2}}{\omega _{\mathbf{q}}}\delta _{\xi
\zeta }\delta _{\xi ^{\prime }\zeta ^{\prime }}  \nonumber \\
&=&\frac{\pi }{\left( 12MN\right) ^{2}}\bar{\Xi}_{\alpha \gamma ,\xi \xi
^{\prime }}^{(2)}\bar{\Xi}_{\gamma \alpha ^{\prime },\xi \xi ^{\prime
}}^{(2)}\sum_{\mathbf{kq}}\frac{k^{2}}{\omega _{\mathbf{k}}}\frac{q^{2}}{%
\omega _{\mathbf{q}}}\delta \left( \omega _{\alpha ^{\prime }\gamma }+\omega
_{\mathbf{q}}-\omega _{\mathbf{k}}\right) n_{\mathbf{k}}\left( n_{\mathbf{q}%
}+1\right) .
\end{eqnarray}
Now recalling Eq. (\ref{DME-Rdiag}) one obtains
\begin{eqnarray}
R_{\alpha \beta ,\alpha ^{\prime }\beta ^{\prime }}^{(2)} &=&\frac{\pi D^{2}%
}{\left( 12MN\right) ^{2}}\sum_{\mathbf{kq}}\frac{k^{2}}{\omega _{\mathbf{k}}%
}\frac{q^{2}}{\omega _{\mathbf{q}}}n_{\mathbf{k}}\left( n_{\mathbf{q}%
}+1\right)  \nonumber \\
&&\times \left\{ -\sum_{\gamma }Q_{\alpha \alpha ^{\prime },\gamma \gamma
}^{(2)}\delta \left( \omega _{\alpha ^{\prime }\gamma }+\omega _{\mathbf{q}%
}-\omega _{\mathbf{k}}\right) \delta _{\beta ^{\prime }\beta }-\delta
_{\alpha \alpha ^{\prime }}\sum_{\gamma }Q_{\beta ^{\prime }\beta ,\gamma
\gamma }^{(2)}\delta \left( \omega _{\beta ^{\prime }\gamma }+\omega _{%
\mathbf{q}}-\omega _{\mathbf{k}}\right) \right.  \nonumber \\
&&+\left. Q_{\alpha \beta ,\alpha ^{\prime }\beta ^{\prime }}^{(2)}\left[
\delta \left( \omega _{\beta \beta ^{\prime }}+\omega _{\mathbf{q}}-\omega _{%
\mathbf{k}}\right) +\delta \left( \omega _{\alpha \alpha ^{\prime }}+\omega
_{\mathbf{q}}-\omega _{\mathbf{k}}\right) \right] \right\} ,
\end{eqnarray}
where
\begin{equation}
Q_{\alpha \beta ,\alpha ^{\prime }\beta ^{\prime }}^{(2)}\equiv \frac{1}{%
D^{2}}\sum_{\xi \xi ^{\prime }}\bar{\Xi}_{\alpha \alpha ^{\prime },\xi \xi
^{\prime }}^{(2)}\bar{\Xi}_{\beta ^{\prime }\beta ,\xi \xi ^{\prime }}^{(2)},
\label{DME-Q2Def}
\end{equation}
similarly to Eq. (\ref{DME-QabaprbprDef}). Since Raman processes can become
important only at high temperatures, one can drop spin transition
frequencies in the energy $\delta $-functions. Next, one can replace
summation by integration with the help of Eq. (\ref{DME-Sum2Int}) and
introduce the characteristic Raman rate
\begin{equation}
\Gamma ^{(2)}=\frac{D^{2}}{24^{2}\pi ^{3}\hbar ^{2}\Omega _{t}^{8}}\int
d\omega _{\mathbf{k}}\omega _{\mathbf{k}}^{6}n_{\mathbf{k}}\left( n_{\mathbf{%
k}}+1\right) .  \label{DME-Gamma2Def}
\end{equation}
Then $R_{\alpha \beta ,\alpha ^{\prime }\beta ^{\prime }}^{(2)}$ can be
written in the form
\begin{equation}
R_{\alpha \beta ,\alpha ^{\prime }\beta ^{\prime }}^{(2)}=\Gamma
^{(2)}\left( -\sum_{\gamma }Q_{\alpha \alpha ^{\prime },\gamma \gamma
}^{(2)}\delta _{\beta ^{\prime }\beta }-\delta _{\alpha \alpha ^{\prime
}}\sum_{\gamma }Q_{\beta ^{\prime }\beta ,\gamma \gamma }^{(2)}+2Q_{\alpha
\beta ,\alpha ^{\prime }\beta ^{\prime }}^{(2)}\right) .  \label{DME-R2Final}
\end{equation}
In the secular approximation one needs the rate
\begin{eqnarray}
\Gamma _{\alpha \alpha ^{\prime }}^{(2)} &=&\left. R_{\alpha \alpha ,\alpha
^{\prime }\alpha ^{\prime }}^{(2)}\right| _{\alpha ^{\prime }\neq \alpha
}=2\Gamma ^{(2)}Q_{\alpha \alpha ,\alpha ^{\prime }\alpha ^{\prime }}^{(2)}=%
\frac{2\Gamma ^{(2)}}{D^{2}}\sum_{\xi \xi ^{\prime }}\bar{\Xi}_{\alpha
\alpha ^{\prime },\xi \xi ^{\prime }}^{(2)}\bar{\Xi}_{\alpha ^{\prime
}\alpha ,\xi \xi ^{\prime }}^{(2)}  \nonumber \\
&=&\frac{2\Gamma ^{(2)}}{D^{2}}\sum_{\xi \xi ^{\prime }}\bar{\Xi}_{\alpha
\alpha ^{\prime },\xi \xi ^{\prime }}^{(2)}\bar{\Xi}_{\alpha \alpha ^{\prime
},\xi \xi ^{\prime }}^{(2)\ast }=\frac{2\Gamma ^{(2)}}{D^{2}}\sum_{\xi \xi
^{\prime }}\left| \bar{\Xi}_{\alpha \alpha ^{\prime },\xi \xi ^{\prime
}}^{(2)}\right| ^{2},  \label{DME-Gamma2aapr}
\end{eqnarray}
where from Eqs. (\ref{DME-Xi2Def}) and (\ref{DME-Xi2Symm}) follows
\begin{eqnarray}
\bar{\Xi}_{\alpha \beta ,\xi \xi ^{\prime }}^{(2)} &\equiv &-\frac{1}{2}%
\sum_{\gamma }\left( \varepsilon _{\alpha }+\varepsilon _{\beta
}-2\varepsilon _{\gamma }\right) \left( \left\langle \alpha \left| S_{\xi
}\right| \gamma \right\rangle \left\langle \gamma \left| S_{\xi ^{\prime
}}\right| \beta \right\rangle +\left\langle \alpha \left| S_{\xi ^{\prime
}}\right| \gamma \right\rangle \left\langle \gamma \left| S_{\xi }\right|
\beta \right\rangle \right)  \nonumber \\
&&+\frac{1}{2}g\mu _{B}\left( H_{\xi ^{\prime }}\left\langle \alpha \left|
S_{\xi }\right| \beta \right\rangle +H_{\xi }\left\langle \alpha \left|
S_{\xi ^{\prime }}\right| \beta \right\rangle \right) -\delta _{\xi \xi
^{\prime }}\sum_{\xi ^{\prime \prime }}g\mu _{B}H_{\xi ^{\prime \prime
}}\left\langle \alpha \left| S_{\xi ^{\prime \prime }}\right| \beta
\right\rangle .
\end{eqnarray}

Integration in Eq. (\ref{DME-Gamma2Def}) is limited by the Brillouin zone,
so that $\omega _{\mathbf{k}}$ does not exceed some maximal value. We will
use the Debye model in which the phonon spectrum continues in the same form
up to the Debye frequency $\Omega _{\mathrm{D}}$ that is the upper bound of
integration. Thus Eq. (\ref{DME-Gamma2Def}) can be represented in the form
\begin{equation}
\Gamma ^{(2)}=\frac{D^{2}}{24^{2}\pi ^{3}\hbar }\frac{\left( k_{\mathrm{B}%
}T\right) ^{7}}{E_{t}^{8}}G_{6}\left( \frac{\Theta _{\mathrm{D}}}{T}\right) ,
\end{equation}
where
\begin{equation}
G_{n}(y)\equiv \int_{0}^{y}dx\frac{x^{n}e^{x}}{(e^{x}-1)^{2}}.
\label{DME-GnDef}
\end{equation}
For $T\ll \Theta _{\mathrm{D}}$ the integration can be extended to infinity.
Using $G_{6}(\infty )=16\pi ^{6}/21,$ one obtains
\begin{equation}
\Gamma ^{(2)}=\frac{\pi ^{3}D^{2}\left( k_{B}T\right) ^{7}}{756\hbar
E_{t}^{8}}.  \label{DME-Gamma2T7}
\end{equation}
On the contrary, for $T\gg \Theta _{\mathrm{D}}$ one can use $G_{6}\left(
y\right) \cong y^{5}/5$ that yields
\begin{equation}
\Gamma ^{(2)}=\frac{D^{2}}{2880\pi ^{3}\hbar }\frac{\left( k_{\mathrm{B}%
}\Theta _{\mathrm{D}}\right) ^{5}\left( k_{\mathrm{B}}T\right) ^{2}}{%
E_{t}^{8}}.  \label{DME-Gamma2HT}
\end{equation}
The transition between these two regimes takes place at $T/\Theta _{\mathrm{%
D},1}=1/y\approx \left[ 21/\left( 5\times 16\pi ^{6}\right) \right]
^{1/5}\approx 0.2,$ i.e., much lower than the Debye temperature. For this
reason the contribution of Raman processes is small in comparison to that
of direct processes up to very high temperatures. Indeed, the contribution of Eq.
(\ref{DME-Gamma2T7}) could become essential at high temperatures but long
before it could happen the growth slows down to Eq. (\ref{DME-Gamma2HT}).
Then from Eqs. (\ref{DME-Gamma2HT}) and (\ref{DME-GammaomegaDef}) one obtains the
ratio
\begin{equation}
\eta =\frac{\Gamma ^{(2)}}{\Gamma ^{(1)}\coth \frac{\hbar \omega _{0}}{2k_{%
\mathrm{B}}T}}\cong \frac{1}{240\pi ^{2}}\frac{\left( k_{\mathrm{B}}\Theta _{%
\mathrm{D}}\right) ^{5}\left( k_{\mathrm{B}}T\right) }{\left( \hbar \omega
_{0}\right) ^{2}E_{t}^{4}}.  \label{DME-Gamma2overGamma1}
\end{equation}
For Mn$_{12}$ with $\Theta _{\mathrm{D}}\simeq 30$ K, $E_{t}/k_{\mathrm{B}%
}\simeq 150$ K, and $\hbar \omega _{0}/k_{\mathrm{B}}\simeq D\simeq 0.66$ K
(near the top of the barrier) the ratio $\eta =1$ requires $T=2\times 10^{4}$
K.

\subsection{The realistic phonon spectrum}

\label{Sec-realistic-phonons}

In the above derivations we have assumed that the crystal lattice possesses
two degenerate transverse phonon modes that contribute to the spin-lattice
relaxation. This is only the case for isotropic elastic bodies. In real
crystals all three acoustic phonon branches are different and they are
neither fully longitudinal nor fully transverse. Fitting the heat capacity
data for Mn$_{12}$ to the \emph{extended Debye model} of Ref. \cite
{gar08prbbr} has shown three acoustic modes with speeds $v_{1}=1541$ m/s, $%
v_{2}=2488$ m/s, and $v_{3}=3176$ m/s. Since $\Gamma ^{(1)}\varpropto \nu
_{t}^{-5},$ only the mode with the lowest speed should be taken into
account. This mode can be considered as approximately transverse. Thus we
introduce the factor 1/2 in Eq. (\ref{DME-GammaomegaDef}) and in all other
formulas for the spin-lattice rate due to direct processes. Similarly, the
factor 1/4 must be introduced in Eq. (\ref{DME-Gamma2Def}) and in all
subsequent formulas for Raman processes. The values of $\Theta _{\mathrm{D}}
$ and $E_{t}$ quoted below Eq. (\ref{DME-Gamma2overGamma1}) correspond to $%
v_{1}.$

\subsection{Ground-state tunneling and relaxation}

\label{Sec-GS-tunneling}

\subsubsection{The two-level model}

Consider the case in which $H_{x}$ and $H_{y}$ in Eq. (\ref{DME-MMHam}) are
small, so that, in the absence of tunneling, the spin eigenstates $|\alpha
\rangle $ are basically $|m\rangle $ that are only weakly hybridized with
the states in the same well. We will denote these states by
\begin{equation}
\left| \psi _{m}\right\rangle =\sum_{m^{\prime \prime }=-S}^{S}c_{mm^{\prime
\prime }}\left| m^{\prime \prime }\right\rangle ,
\label{DME-mmprimetildeDef}
\end{equation}
where $\left| c_{mm}\right| \cong 1$ and all other coefficients are small.
Near tunneling resonances, these states are strongly hybridized with
resonant states in the other well. Hybridization of the states $\left| \psi
_{m}\right\rangle $ and $\left| \psi _{m^{\prime }}\right\rangle $ can be
taken into account in the framework of the two-state model
\begin{eqnarray}
\left\langle \psi _{m_{i}}\left| \hat{H}_{S}\right| \psi
_{m_{i}}\right\rangle &=&\varepsilon _{m_{i}},\qquad m_{i}=m,m^{\prime }
\nonumber \\
\left\langle \psi _{m}\left| \hat{H}_{S}\right| \psi _{m^{\prime
}}\right\rangle &=&\frac{1}{2}\Delta e^{i\varphi },  \label{DME-TLS}
\end{eqnarray}
where $\Delta $ is the tunnel splitting of the levels $m$ and $m^{\prime }$
that can be calculated from the exact spin Hamiltonian $\hat{H}_{S}$ or
determined experimentally and $\varphi $ is a phase. Since one can multiply
the basis functions $\left| \psi _{m}\right\rangle $ by arbitrary phase
factors, we will set $\varphi =\pi $ for convenience. This will result in a
simpler form of the wave functions than in Ref. \cite{chugarsch05prb},
whereas all physical results remain the same. Then the model above can be
formulated as the pseudospin model
\begin{equation}
\hat{H}_{\mathrm{eff}}=-\frac{1}{2}\mathbf{\hat{\sigma}\cdot A+}\frac{1}{2}%
\left( \varepsilon _{m}+\varepsilon _{m^{\prime }}\right) ,
\label{DME-HeffDef}
\end{equation}
where components of $\mathbf{\hat{\sigma}}$\ are Pauli matrices,
\begin{equation}
\mathbf{A}\equiv \Delta \mathbf{e}_{x}+W\mathbf{e}_{z},  \label{DME-ADef}
\end{equation}
is the effective field, and $W$ is the energy bias or resonance detuning
\begin{equation}
W=\varepsilon _{m}-\varepsilon _{m^{\prime }},  \label{DME-WDef}
\end{equation}
defined by Eq. (\ref{DME-omegammpr}). We will need the direction angle $%
\theta $ of $\mathbf{A,}$
\begin{equation}
\cos \theta =\frac{W}{\sqrt{W^{2}+\Delta ^{2}}}.  \label{DME-thetaDef}
\end{equation}
The pseudospin acts on the states as
\begin{eqnarray}
\hat{\sigma}_{z}\left| \psi _{m}\right\rangle &=&-\left| \psi
_{m}\right\rangle ,\qquad \hat{\sigma}_{z}\left| \psi _{m^{\prime
}}\right\rangle =\left| \psi _{m^{\prime }}\right\rangle  \nonumber \\
\hat{\sigma}_{x}\left| \psi _{m}\right\rangle &=&\left| \psi _{m^{\prime
}}\right\rangle ,\qquad \hat{\sigma}_{x}\left| \psi _{m^{\prime
}}\right\rangle =\left| \psi _{m}\right\rangle .
\label{DME-pseudoSpinActing}
\end{eqnarray}
Of course, one also can calculate matrix elements of the physical spin $%
\mathbf{S}$ with respect to this basis.

Eigenstates of $\hat{H}_{\mathrm{eff}}$ are the states polarized parallel
and antiparallel to $\mathbf{A,}$ and the eigenvalues are given by
\begin{equation}
\varepsilon _{\alpha }=\frac{1}{2}\left( \varepsilon _{m}+\varepsilon
_{m^{\prime }}+\alpha \sqrt{W^{2}+\Delta ^{2}}\right) ,\qquad \alpha =\pm .
\label{DME-Epm}
\end{equation}
The transition frequency $\omega _{0}$ between the levels is defined by
\begin{equation}
\varepsilon _{+}-\varepsilon _{-}\equiv \hbar \omega _{0}\equiv \sqrt{%
W^{2}+\Delta ^{2}}.  \label{DME-hbaromegaDef}
\end{equation}
The eigenstates of $\hat{H}_{\mathrm{eff}}$ can be expanded over the natural
basis as
\begin{equation}
\left| \alpha \right\rangle \equiv \left| \chi _{\alpha }\right\rangle =%
\frac{1}{\sqrt{2}}\left( C_{\alpha }\left| \psi _{m}\right\rangle -\alpha
C_{-\alpha }\left| \psi _{m^{\prime }}\right\rangle \right) ,
\label{DME-Psipm}
\end{equation}
where
\begin{equation}
C_{\alpha }=\sqrt{1+\alpha \cos \theta }.  \label{DME-Cpm}
\end{equation}
One can directly check $\hat{H}_{\mathrm{eff}}\left| \chi _{\alpha
}\right\rangle =\varepsilon _{\alpha }\left| \chi _{\alpha }\right\rangle .$
The coefficients $C_{\alpha }$ satisfy $1/C_{\alpha }=C_{-\alpha }/\sin
\theta .$

Let us prepare the system in the state $\left| \psi _{-S}\right\rangle $,
the ground state in the left well that can be at resonance with the ground
or excited state in the right well, $\left| \psi _{m^{\prime }}\right\rangle
$ with $m^{\prime }\leq S.$ Near the resonance these states hybridyze into $%
\left| \psi _{\pm }\right\rangle $. At low temperatures, all levels above $%
\left| \psi _{-S}\right\rangle $ are unpopulated and do not contribute to
relaxation. The only relaxation processes are between $\left| \psi _{\pm
}\right\rangle $ and the levels in the right well below $\left| \psi
_{m^{\prime }}\right\rangle .$ Again, since the levels are only weakly
hybridyzed inside the wells, here the dominant process is decay $\left| \psi
_{m^{\prime }}\right\rangle \rightarrow \left| \psi _{m^{\prime
}+1}\right\rangle .$ The inverse process can be neglected at low
temperatures. In the case of the ground-state resonance, $\left| \psi
_{-S}\right\rangle $ with $\left| \psi _{S}\right\rangle ,$ this decay
process is, of course, absent. The full description of both ground-ground
and ground-excited resonances at low temperatures includes only two levels $%
\left| \chi _{\pm }\right\rangle ,$ so that the effective DM is a 2$\times 2$
matrix. The DME in the general case of time-dependent spin Hamiltonians is
Eq. (\ref{DME-rhoEqTensor}) with non-adiabatic terms from Eq. (\ref
{DME-rhoEq-diagbasist}) added, i.e.,
\begin{equation}
\frac{d}{dt}\rho _{\alpha \beta }=\sum_{\gamma }\left( \langle \dot{\chi}%
_{\alpha }\left| \chi _{\gamma }\right\rangle \rho _{\gamma \beta }+\rho
_{\alpha \gamma }\langle \chi _{\gamma }\left| \dot{\chi}_{\beta
}\right\rangle \right) -i\omega _{\alpha \beta }\rho _{\alpha \beta
}+\sum_{\alpha ^{\prime }\beta ^{\prime }}R_{\alpha \beta ,\alpha ^{\prime
}\beta ^{\prime }}\rho _{\alpha ^{\prime }\beta ^{\prime }},
\label{DME-DME-twolevel}
\end{equation}
where all indices take the values $\pm .$ If $\left| \psi _{m^{\prime
}}\right\rangle $ is an excited state, in general one cannot use the secular
approximation because $\omega _{0}$ can be comparable with the relaxation
rate.

Let us work out the non-adiabatic terms in Eq. (\ref{DME-DME-twolevel}).
Calculating the time derivative of $C_{\alpha }$ in Eq. (\ref{DME-Psipm}),
\begin{equation}
\dot{C}_{\alpha }=\frac{\alpha }{2C_{\alpha }}\frac{d}{dt}\cos \theta =-%
\frac{\alpha }{2C_{\alpha }}\sin \theta \,\dot{\theta}=-\alpha C_{-\alpha }%
\frac{\dot{\theta}}{2},
\end{equation}
one obtains
\begin{equation}
\left| \dot{\chi}_{\alpha }\right\rangle =\frac{1}{\sqrt{2}}\left( \dot{C}%
_{\alpha }\left| \psi _{-S}\right\rangle -\alpha \dot{C}_{-\alpha }\left|
\psi _{m^{\prime }}\right\rangle \right) =-\frac{\alpha }{\sqrt{2}}\left(
C_{-\alpha }\left| \psi _{-S}\right\rangle +\alpha C_{\alpha }\left| \psi
_{m^{\prime }}\right\rangle \right) \frac{\dot{\theta}}{2}=-\alpha \left|
\chi _{-\alpha }\right\rangle \frac{\dot{\theta}}{2}.
\end{equation}
Thus in Eq. (\ref{DME-DME-twolevel}) the scalar products are
\begin{equation}
\langle \dot{\chi}_{\alpha }\left| \chi _{\beta }\right\rangle =-\alpha
\frac{\dot{\theta}}{2}\langle \chi _{-\alpha }\left| \chi _{\beta
}\right\rangle =-\alpha \delta _{-\alpha ,\beta }\frac{\dot{\theta}}{2}.
\end{equation}
and
\begin{equation}
\sum_{\gamma }\left( \langle \dot{\chi}_{\alpha }\left| \chi _{\gamma
}\right\rangle \rho _{\gamma \beta }+\rho _{\alpha \gamma }\langle \chi
_{\gamma }\left| \dot{\chi}_{\beta }\right\rangle \right) =-\left( \alpha
\rho _{-\alpha ,\beta }+\beta \rho _{\alpha ,-\beta }\right) \frac{\dot{%
\theta}}{2}.  \label{DME-nonadi-viatheta}
\end{equation}

The density operator in the initial state typically is
\begin{equation}
\hat{\rho}(0)=\left| \psi _{-S}\right\rangle \left\langle \psi _{-S}\right| ,
\label{DME-rhoInit}
\end{equation}
so that the density matrix in the diagonal basis is given by
\begin{equation}
\rho _{\alpha \beta }(0)=\left\langle \alpha \left| \hat{\rho}(0)\right|
\beta \right\rangle =\left\langle \alpha \right| \psi _{-S}\rangle
\left\langle \psi _{-S}\right| \beta \rangle =\frac{1}{2}C_{\alpha }C_{\beta
},  \label{DME-rhoDiagInit}
\end{equation}
where Eq. (\ref{DME-Psipm}) with $m=-S$ was used. In particular,
\begin{eqnarray}
\rho _{++}(0) &=&\frac{1}{2}\left( 1+\cos \theta \right) ,\qquad \rho
_{--}(0)=\frac{1}{2}\left( 1-\cos \theta \right)  \nonumber \\
\rho _{+-}(0) &=&\rho _{-+}(0)=\frac{1}{2}\sin \theta .
\label{DME-rhoDiagInitParticlular}
\end{eqnarray}

\subsubsection{Ground-ground state resonance}

The results obtained above already allow to consider the dynamics at the
ground-state resonance, $m^{\prime }=S.$ In this case the relaxation terms
in Eq. (\ref{DME-DME-twolevel}) \ contain only $\Gamma (\omega _{0})\ll
\omega _{0},$ so that the secular approximation is applicable. Dropping
nonsecular terms in Eq. (\ref{DME-DME-twolevel}) one obtains
\begin{eqnarray}
\frac{d}{dt}\rho _{++} &=&-\left( \rho _{+-}+\rho _{-+}\right) \frac{\dot{%
\theta}}{2}+R_{++,++}\rho _{++}+R_{++,--}\rho _{--}  \nonumber \\
\frac{d}{dt}\rho _{+-} &=&-\left( \rho _{--}-\rho _{++}\right) \frac{\dot{%
\theta}}{2}-i\omega _{0}\rho _{+-}+R_{+-,+-}\rho _{+-},
\label{DME-DME-twolevel-secular}
\end{eqnarray}
whereas
\begin{equation}
\rho _{--}=1-\rho _{++},\qquad \rho _{-+}=\left( \rho _{+-}\right) ^{\ast }.
\label{DME-rhoRelations}
\end{equation}
Using Eq. (\ref{DME-RTensorGamma}), one obtains
\begin{eqnarray}
R_{++,++} &=&-2Q_{++,--}\Gamma (\omega _{0})\left( n_{\omega _{0}}+1\right)
\equiv -\Gamma _{-+}  \nonumber \\
R_{--,--} &=&-2Q_{--,++}\Gamma (\omega _{0})n_{\omega _{0}}=-\Gamma
_{+-}=-e^{-\hbar \omega _{0}/(k_{\mathrm{B}}T)}\Gamma _{-+}  \nonumber \\
R_{++,--} &=&2Q_{++,--}\Gamma (\omega _{0})n_{\omega _{0}}=\Gamma
_{+-}=e^{-\hbar \omega _{0}/(k_{\mathrm{B}}T)}\Gamma _{-+}  \nonumber \\
R_{--,++} &=&2Q_{--,++}\Gamma (\omega _{0})\left( n_{\omega _{0}}+1\right)
=\Gamma _{-+}  \label{DME-GammampDef}
\end{eqnarray}
and
\begin{equation}
R_{+-,+-}=-Q_{++,--}\Gamma (\omega _{0})\left( n_{\omega _{0}}+1\right)
-Q_{--,++}\Gamma (\omega _{0})n_{\omega _{0}}=-\frac{1}{2}\left( \Gamma
_{-+}+\Gamma _{+-}\right)
\end{equation}
etc. Thus Eq. (\ref{DME-DME-twolevel-secular}) takes the form
\begin{eqnarray}
\dot{\rho}_{++} &=&-\left( \rho _{+-}+\rho _{-+}\right) \frac{\dot{\theta}}{2%
}-\Gamma _{-+}\rho _{++}+\Gamma _{+-}\rho _{--}  \nonumber \\
\dot{\rho}_{+-} &=&\left( \rho _{++}-\rho _{--}\right) \frac{\dot{\theta}}{2}%
-\left[ i\omega _{0}+\frac{1}{2}\left( \Gamma _{-+}+\Gamma _{+-}\right) %
\right] \rho _{+-}.  \label{DME-GGRes-DME}
\end{eqnarray}
This system of equations can be rewritten as
\begin{eqnarray}
\dot{\rho}_{++} &=&-\dot{\theta}\func{Re}\rho _{+-}-\Gamma \left( \rho
_{++}-\rho _{++}^{\mathrm{eq}}\right)  \nonumber \\
\dot{\rho}_{+-} &=&\dot{\theta}\left( \rho _{++}-1/2\right) -\left( i\omega
_{0}+\Gamma /2\right) \rho _{+-},  \label{DME-GGRes-DME-pp}
\end{eqnarray}
where
\begin{equation}
\Gamma \equiv \Gamma _{-+}+\Gamma _{+-}  \label{DME-GammaTot}
\end{equation}
is the total relaxation rate between the $\pm $ states and
\begin{equation}
\rho _{++}^{\mathrm{eq}}=\frac{\Gamma _{+-}}{\Gamma _{-+}+\Gamma _{+-}}=%
\frac{e^{-\hbar \omega _{0}/(k_{\mathrm{B}}T)}}{1+e^{-\hbar \omega _{0}/(k_{%
\mathrm{B}}T)}}  \label{DME-rhoppeqDef}
\end{equation}
is the equilibrium population of the upper level.

\subsubsection{Dynamics of the ground-ground state resonance via effective
classical spin}

The DME for the ground-ground state resonance, Eq. (\ref{DME-GGRes-DME}),
can be conveniently formulated in terms of the averages the pseudospin $%
\mathbf{\hat{\sigma}}$ with the density operator. Using Eq. (\ref
{DME-Aavrrho}) one can write
\begin{equation}
\mathbf{\sigma }\equiv \left\langle \mathbf{\hat{\sigma}}\right\rangle
=\sum_{\alpha \beta }\rho _{\alpha \beta }\left\langle \chi _{\beta }\left|
\mathbf{\hat{\sigma}}\right| \chi _{\alpha }\right\rangle .
\end{equation}
Directing the axis $z^{\prime }$ along the total field $\mathbf{A}$, one has
\begin{eqnarray}
\mathbf{\hat{\sigma}} &\equiv &\left( \hat{\sigma}_{-}+\hat{\sigma}%
_{+}\right) \mathbf{e}_{x^{\prime }}+i\left( \hat{\sigma}_{-}-\hat{\sigma}%
_{+}\right) \mathbf{e}_{y^{\prime }}+\hat{\sigma}_{z^{\prime }}\mathbf{e}%
_{z^{\prime }}  \nonumber \\
&=&\left( \left| \chi _{+}\right\rangle \left\langle \chi _{-}\right|
+\left| \chi _{-}\right\rangle \left\langle \chi _{+}\right| \right) \mathbf{%
e}_{x^{\prime }}+i\left( \left| \chi _{+}\right\rangle \left\langle \chi
_{-}\right| -\left| \chi _{-}\right\rangle \left\langle \chi _{+}\right|
\right) \mathbf{e}_{y^{\prime }}+\left( \left| \chi _{-}\right\rangle
\left\langle \chi _{-}\right| -\left| \chi _{+}\right\rangle \left\langle
\chi _{+}\right| \right) \mathbf{e}_{z^{\prime }}.
\end{eqnarray}
Then one obtains
\begin{eqnarray}
\sigma _{x^{\prime }} &=&\rho _{+-}\left\langle \chi _{-}\left| \hat{\sigma}%
_{x^{\prime }}\right| \chi _{+}\right\rangle +\rho _{-+}\left\langle \chi
_{+}\left| \hat{\sigma}_{x^{\prime }}\right| \chi _{-}\right\rangle =\rho
_{-+}+\rho _{+-}=2\func{Re}\rho _{-+}  \nonumber \\
\sigma _{y^{\prime }} &=&\rho _{+-}\left\langle \chi _{-}\left| \hat{\sigma}%
_{y^{\prime }}\right| \chi _{+}\right\rangle +\rho _{-+}\left\langle \chi
_{+}\left| \hat{\sigma}_{y^{\prime }}\right| \chi _{-}\right\rangle =i\left(
\rho _{-+}-\rho _{+-}\right) =2\func{Im}\rho _{-+}  \nonumber \\
\sigma _{z^{\prime }} &=&\rho _{++}\left\langle \chi _{+}\left| \hat{\sigma}%
_{z^{\prime }}\right| \chi _{+}\right\rangle +\rho _{--}\left\langle \chi
_{-}\left| \hat{\sigma}_{z^{\prime }}\right| \chi _{-}\right\rangle =\rho
_{--}-\rho _{++}=1-2\rho _{++}.
\end{eqnarray}
Now Eq. (\ref{DME-GGRes-DME}) can be transformed as
\begin{eqnarray}
\dot{\sigma}_{x^{\prime }} &=&\dot{\rho}_{-+}+\dot{\rho}_{+-}=\left( \rho
_{++}-\rho _{--}\right) \dot{\theta}+i\omega _{0}\left( \rho _{-+}-\rho
_{+-}\right) +R_{x^{\prime }}=-\dot{\theta}\sigma _{z^{\prime }}+\omega
_{0}\sigma _{y^{\prime }}+R_{x^{\prime }}  \nonumber \\
\dot{\sigma}_{y^{\prime }} &=&i\left( \dot{\rho}_{-+}-\dot{\rho}_{+-}\right)
=i\left( i\omega _{0}\rho _{-+}+i\omega _{0}\rho _{+-}\right) +R_{y^{\prime
}}=-\omega _{0}\sigma _{x^{\prime }}+R_{y^{\prime }}  \nonumber \\
\dot{\sigma}_{z^{\prime }} &=&\left( \rho _{+-}+\rho _{-+}\right) \dot{\theta%
}+R_{z^{\prime }}=\dot{\theta}\sigma _{x^{\prime }}+R_{z^{\prime }}
\end{eqnarray}
or
\begin{equation}
\mathbf{\dot{\sigma}}=\left[ \mathbf{\sigma \times }\left( \mathbf{\omega }%
_{0}+\mathbf{\Omega }\right) \right] +\mathbf{R,\qquad \omega }_{0}=\omega
_{0}\mathbf{e}_{z^{\prime }},\qquad \mathbf{\Omega }=\dot{\theta}\mathbf{e}%
_{y^{\prime }}.  \label{DME-sigmaVecEq}
\end{equation}
This is a Larmor equation for the classical vector $\mathbf{\sigma }$ in the
frame rotating with frequency $\mathbf{\Omega }$ due to the time dependence
of the spin Hamiltonian. The relaxation vector is given by
\begin{eqnarray}
\mathbf{R} &=&\mathbf{-}\frac{1}{2}\left( \Gamma _{-+}+\Gamma _{+-}\right)
\sigma _{x^{\prime }}\mathbf{e}_{x^{\prime }}\mathbf{-}\frac{1}{2}\left(
\Gamma _{-+}+\Gamma _{+-}\right) \sigma _{y^{\prime }}\mathbf{e}_{y^{\prime
}}-2\left( -\Gamma _{-+}\rho _{++}+\Gamma _{+-}\rho _{--}\right) \mathbf{e}%
_{z^{\prime }}  \nonumber \\
&=&\mathbf{-}\frac{1}{2}\left( \Gamma _{-+}+\Gamma _{+-}\right) \sigma
_{x^{\prime }}\mathbf{e}_{x^{\prime }}\mathbf{-}\frac{1}{2}\left( \Gamma
_{-+}+\Gamma _{+-}\right) \sigma _{y^{\prime }}\mathbf{e}_{y^{\prime }}-%
\left[ -\Gamma _{-+}\left( 1-\sigma _{z^{\prime }}\right) +\Gamma
_{+-}\left( 1+\sigma _{z^{\prime }}\right) \right] \mathbf{e}_{z^{\prime }}
\nonumber \\
&=&\mathbf{-}\frac{1}{2}\left( \Gamma _{-+}+\Gamma _{+-}\right) \sigma
_{x^{\prime }}\mathbf{e}_{x^{\prime }}\mathbf{-}\frac{1}{2}\left( \Gamma
_{-+}+\Gamma _{+-}\right) \sigma _{y^{\prime }}\mathbf{e}_{y^{\prime }}-%
\left[ \left( \Gamma _{-+}+\Gamma _{+-}\right) \sigma _{z^{\prime }}-\left(
\Gamma _{-+}-\Gamma _{+-}\right) \right] \mathbf{e}_{z^{\prime }}
\end{eqnarray}
or
\begin{equation}
\mathbf{R}=\mathbf{-}\frac{\Gamma }{2}\left( \mathbf{\sigma }-\frac{\mathbf{%
\omega }_{0}\cdot \mathbf{\sigma }}{\omega _{0}^{2}}\mathbf{\omega }%
_{0}\right) -\Gamma \frac{\mathbf{\omega }_{0}}{\omega _{0}}\left( \frac{%
\mathbf{\omega }_{0}\cdot \mathbf{\sigma }}{\omega _{0}}-\sigma ^{\mathrm{eq}%
}\right) ,  \label{DME-RVec}
\end{equation}
where $\Gamma $ is given by Eq. (\ref{DME-GammaTot}) and
\begin{equation}
\sigma ^{\mathrm{eq}}=\frac{\Gamma _{-+}-\Gamma _{+-}}{\Gamma _{-+}+\Gamma
_{+-}}=\tanh \frac{\hbar \omega _{0}}{2k_{\mathrm{B}}T}
\label{DME-sigmaeqDef}
\end{equation}
is the equilibrium spin polarization.

Eq. (\ref{DME-sigmaVecEq}) can describe, in particular, the Landau-Zener
effect of transition between the energy levels of a two-level system as the
energy bias $W$ is swept though the resonance, $W=0.$ It is remarcable that
this essentially quantum phenomenon can be described by a classical
language. Studying the dynamical behavior of the classical spin $\mathbf{%
\sigma }$ helps a lot understanding the LZ effect. In particular, if the
sweep is slow, the non-adiabatic term $\mathbf{\Omega }$ in Eq. (\ref
{DME-sigmaVecEq}) is small and the spin remains nearly collinear to $\mathbf{%
\omega }_{0}$ at all times. In the adiabatic frame $\mathbf{\omega }_{0}=%
\mathrm{const}$ and the spin vector in the laboratory frame (i.e., in the
natural basis) can be obtained by a rotational transformation. One also can
rewrite Eq. (\ref{DME-sigmaVecEq}) in the laboratory frame by simply
dropping $\mathbf{\Omega }$ and making $\mathbf{\omega }_{0}$ time dependent
according to Eq. (\ref{DME-ADef}), $\hbar \mathbf{\omega }_{0}=\mathbf{A}.$
Because of the vector form of the equation of motion, its transformation
between the frames is easy, in contrast to the transformation of the DME
between the natural basis and the diagonal (adiabatic) basis described in
Sec. \ref{Sec-Trans-to-nat-bas}. As the vector $\mathbf{A}$ is rotating, $%
\mathbf{\sigma }$ is lagging behind it, depending on the sweep rate. For a
slow sweep rate it nearly follows the direction of $\mathbf{A}$ while for a
fast sweep it nearly remains in the initial state.

\subsubsection{Coherence in the ground-ground state resonance}

Let us consider the time evolution of a two-level system in the case of
time-independent spin Hamiltonian, $\dot{\theta}=0$. This can be done using
either Eq. (\ref{DME-GGRes-DME-pp}) or Eq. (\ref{DME-sigmaVecEq}). In
particular, the solution of Eq. (\ref{DME-GGRes-DME-pp}) with the initial
conditions of Eq. (\ref{DME-rhoDiagInitParticlular}) is
\begin{eqnarray}
\rho _{++}(t) &=&\rho _{++}^{\mathrm{eq}}+\left[ \rho _{++}(0)-\rho _{++}^{%
\mathrm{eq}}\right] e^{-\Gamma t}=\rho _{++}^{\mathrm{eq}}+\left[ \frac{1}{2}%
\left( 1+\frac{W}{\sqrt{W^{2}+\Delta ^{2}}}\right) -\rho _{++}^{\mathrm{eq}}%
\right] e^{-\Gamma t}  \nonumber \\
\rho _{+-}(t) &=&\rho _{+-}(0)e^{-\left( i\omega _{0}+\Gamma /2\right) t}=%
\frac{1}{2}\frac{\Delta }{\sqrt{W^{2}+\Delta ^{2}}}e^{-\left( i\omega
_{0}+\Gamma /2\right) t}.
\end{eqnarray}
Transformation back to the natural basis is done using Eq. (\ref
{DME-rgoviarhodiag}). The probability to remain in the initial state $\left|
\psi _{-S}\right\rangle $ is
\begin{eqnarray}
\rho _{-S,-S}(t) &=&\sum_{\alpha \beta }\langle \psi _{-S}|\alpha \rangle
\rho _{\alpha \beta }(t)\langle \beta |\psi _{-S}\rangle =\frac{1}{2}%
\sum_{\alpha \beta }C_{\alpha }C_{\beta }\rho _{\alpha \beta }(t)  \nonumber
\\
&=&\frac{1}{2}\left( 1+\frac{W}{\sqrt{W^{2}+\Delta ^{2}}}\right) \rho
_{++}(t)+\frac{1}{2}\left( 1-\frac{W}{\sqrt{W^{2}+\Delta ^{2}}}\right) \left[
1-\rho _{++}(t)\right] +\frac{\Delta }{\sqrt{W^{2}+\Delta ^{2}}}\func{Re}%
\rho _{+-}(t)  \nonumber \\
&=&\frac{1}{2}\left( 1-\frac{W}{\sqrt{W^{2}+\Delta ^{2}}}\right) +\frac{W}{%
\sqrt{W^{2}+\Delta ^{2}}}\rho _{++}(t)+\frac{\Delta }{\sqrt{W^{2}+\Delta ^{2}%
}}\func{Re}\rho _{+-}(t)  \nonumber \\
&=&\frac{1}{2}\left( 1-\frac{W}{\sqrt{W^{2}+\Delta ^{2}}}\right) +\frac{W}{%
\sqrt{W^{2}+\Delta ^{2}}}\left\{ \rho _{++}^{\mathrm{eq}}+\left[ \frac{1}{2}%
\left( 1+\frac{W}{\sqrt{W^{2}+\Delta ^{2}}}\right) -\rho _{++}^{\mathrm{eq}}%
\right] e^{-\Gamma t}\right\}  \nonumber \\
&&\qquad +\frac{1}{2}\frac{\Delta ^{2}}{W^{2}+\Delta ^{2}}e^{-\left( \Gamma
/2\right) t}\cos \left( \omega _{0}t\right)
\end{eqnarray}
or
\begin{eqnarray}
\rho _{-S,-S}(t) &=&\frac{1}{2}\left[ \frac{W^{2}}{W^{2}+\Delta ^{2}}%
e^{-\Gamma t}+\frac{\Delta ^{2}}{W^{2}+\Delta ^{2}}e^{-\left( \Gamma
/2\right) t}\cos \left( \omega _{0}t\right) +1-\frac{W}{\sqrt{W^{2}+\Delta
^{2}}}\left( 1-e^{-\Gamma t}\right) \right]  \nonumber \\
&&\qquad +\frac{W}{\sqrt{W^{2}+\Delta ^{2}}}\rho _{++}^{\mathrm{eq}}\left(
1-e^{-\Gamma t}\right) .  \label{DME-rhotgsgsres}
\end{eqnarray}
The time evolution described by this formula is a combination of relaxation
with rate $\Gamma $ and oscillations of frequency $\omega _{0}$ damped with
rate $\Gamma /2.$ It should be noted, however, that the ground-state tunnel
splitting $\Delta $ is typically very small, so that it is very difficult to
experimentally realize $\left| W\right| \lesssim \Delta $ to see coherent
oscillations of the spin between the two states.

\subsubsection{Relaxation rate between two tunnel-split states}

The relaxation rate $\Gamma _{-+}$ for the ground-state doublet can be found
analytically \cite{chugarsch05prb}. In particular, for the uniaxial model in
the presence of transverse field along the $x$ axis, with the help of the
high-order perturbation theory one obtains
\begin{eqnarray}
\left\langle \psi _{-}\left| S_{z}\right| \psi _{+}\right\rangle &=&-\frac{%
\Delta }{\sqrt{W^{2}+\Delta ^{2}}}\frac{m^{\prime }-m}{2}  \nonumber \\
\left\langle \psi _{-}\left| S_{x}\right| \psi _{+}\right\rangle &=&\frac{%
\Delta }{g\mu _{B}H_{x}}\frac{W}{\sqrt{W^{2}+\Delta ^{2}}}\frac{m^{\prime }-m%
}{2}  \nonumber \\
\left\langle \psi _{-}\left| S_{y}\right| \psi _{+}\right\rangle &=&-i\frac{%
\Delta }{g\mu _{B}H_{x}}\frac{m^{\prime }-m}{2}.
\label{DME-Spin-MatrixElements}
\end{eqnarray}
Then from Eq. (\ref{DME-XiDef}) in components,
\begin{eqnarray}
\Xi _{-+,x} &=&-i\hbar \omega _{0}\left\langle \psi _{-}\left| S_{x}\right|
\psi _{+}\right\rangle -\left\langle \psi _{-}\left| S_{y}\right| \psi
_{+}\right\rangle g\mu _{B}H_{z}+\left\langle \psi _{-}\left| S_{z}\right|
\psi _{+}\right\rangle g\mu _{B}H_{y}  \nonumber \\
\Xi _{-+,y} &=&-i\hbar \omega _{0}\left\langle \psi _{-}\left| S_{y}\right|
\psi _{+}\right\rangle -\left\langle \psi _{-}\left| S_{z}\right| \psi
_{+}\right\rangle g\mu _{B}H_{x}+\left\langle \psi _{-}\left| S_{x}\right|
\psi _{+}\right\rangle g\mu _{B}H_{z}  \nonumber \\
\Xi _{-+,z} &=&-i\hbar \omega _{0}\left\langle \psi _{-}\left| S_{z}\right|
\psi _{+}\right\rangle -\left\langle \psi _{-}\left| S_{x}\right| \psi
_{+}\right\rangle g\mu _{B}H_{y}+\left\langle \psi _{-}\left| S_{y}\right|
\psi _{+}\right\rangle g\mu _{B}H_{x},  \label{DME-XiComponents}
\end{eqnarray}
with $H_{y}=0$ one obtains
\begin{eqnarray}
\Xi _{-+,z} &=&0  \nonumber \\
\Xi _{-+,x} &=&-i\frac{\Delta }{g\mu _{B}H_{x}}\frac{m^{\prime }-m}{2}\left(
W-g\mu _{B}H_{z}\right)  \nonumber \\
\Xi _{-+,y} &=&\frac{\Delta }{\sqrt{W^{2}+\Delta ^{2}}}\frac{m^{\prime }-m}{2%
}g\mu _{B}H_{x}\left( 1-\frac{\Delta ^{2}}{\left( g\mu _{B}H_{x}\right) ^{2}}%
-\frac{W(W-g\mu _{B}H_{z})}{\left( g\mu _{B}H_{x}\right) ^{2}}\right) .
\label{DME-XiHx}
\end{eqnarray}
Taking into account $\Delta \ll g\mu _{B}H_{x}$ and using \ $W=(m^{\prime
}-m)g\mu _{B}H_{z}$ yields
\begin{eqnarray}
\left| \mathbf{\Xi }_{-+}\right| ^{2} &=&\left| \Xi _{-+,x}\right|
^{2}+\left| \Xi _{-+,y}\right| ^{2}  \nonumber \\
&=&\left( \frac{m^{\prime }-m}{2}\right) ^{2}\Delta ^{2}\left[ (m^{\prime
}-m-1)^{2}\frac{H_{z}^{2}}{H_{x}^{2}}+\frac{\left( g\mu _{B}H_{x}\right) ^{2}%
}{W^{2}+\Delta ^{2}}\left( 1-(m^{\prime }-m)(m^{\prime }-m-1)\frac{H_{z}^{2}%
}{H_{x}^{2}}\right) ^{2}\right] .
\end{eqnarray}
Now, from Eqs. (\ref{DME-GammampDef}), (\ref{DME-QabaprbprDef}), and (\ref
{DME-GammaomegaDef}) one obtains
\begin{eqnarray}
\Gamma _{-+} &=&2Q_{--,++}\Gamma \left( \omega _{0}\right) =\left| \mathbf{%
\Xi }_{-+}\right| ^{2}\frac{\omega _{0}^{3}}{12\pi \hbar ^{2}\Omega _{t}^{4}}
\nonumber \\
&=&\left( \frac{m^{\prime }-m}{2}\right) ^{2}\frac{\Delta ^{2}\omega
_{0}\left( g\mu _{B}H_{x}\right) ^{2}}{12\pi E_{t}^{4}}\left[ \frac{%
W^{2}+\Delta ^{2}}{\left( g\mu _{B}H_{x}\right) ^{2}}(m^{\prime }-m-1)^{2}%
\frac{H_{z}^{2}}{H_{x}^{2}}+\left( 1-(m^{\prime }-m)(m^{\prime }-m-1)\frac{%
H_{z}^{2}}{H_{x}^{2}}\right) ^{2}\right] .
\end{eqnarray}
Dropping again the small $\Delta ^{2}$ term in the square brackets one
obtains
\begin{equation}
\Gamma _{-+}=\left( \frac{m^{\prime }-m}{2}\right) ^{2}\frac{\Delta
^{2}\omega _{0}\left( g\mu _{B}H_{x}\right) ^{2}}{12\pi E_{t}^{4}}Q,
\label{DME-GammapmHx}
\end{equation}
where
\begin{eqnarray}
Q &=&\xi ^{4}+\left( 1-\xi ^{2}\right) ^{2}=1-2\xi ^{2}+2\xi ^{4},  \nonumber
\\
\xi &\equiv &\sqrt{(m^{\prime }-m)(m^{\prime }-m-1)}\frac{H_{z}}{H_{x}}.
\label{DME-QHx}
\end{eqnarray}
In particular, for the ground-state resonance, $m=-S,$ $m^{\prime }=S$, one
has $\xi =\sqrt{2S(2S-1)}H_{z}/H_{x}.$ One can see that $Q$ begins to
deviate from 1 only for $SH_{z}\gtrsim H_{x}$ that requires $W\gg \Delta .$
For such a strong bias, the dependence of $\Gamma _{-+}$ on the bias has the
form
\begin{equation}
\Gamma _{-+}\propto f(\xi )=\xi Q=\xi \left( 1-2\xi ^{2}+2\xi ^{4}\right) .
\end{equation}
The function $f(\xi )$ monotonically increases but its slope initially
decreases from 1, attaining the minimal value $f^{\prime }(\xi _{0})=1/10$
at $\xi _{0}=\sqrt{3/10}\simeq 0.5477.$ At this point $f(\xi _{0})=\sqrt{3/10%
}(29/50)\simeq 0.3177.$ After that the slope of $f(\xi )$ begins strongly to
increase, so that $\Gamma _{-+}\sim \xi ^{5}\sim H_{z}^{5}.$

\subsubsection{Ground-excited state resonance}

Let us consider the tunneling resonance between $\left| \psi
_{-S}\right\rangle $ and $\left| \psi _{m^{\prime }}\right\rangle $ with $%
m^{\prime }<S.$ The hybridyzed states $\left| \psi _{\pm }\right\rangle $
decay, predominantly, into $\left| \psi _{m^{\prime }+1}\right\rangle .$ The
corresponding relaxation rate is large, so that we will neglect the rate of
transitions between $\left| \psi _{\pm }\right\rangle $ that is small near
the resonance. At low temperatures one can neglect the energy-up processes $%
\left| \psi _{m^{\prime }+1}\right\rangle \rightarrow \left| \psi _{\pm
}\right\rangle .$ Thus, as in the case of the ground-ground state resonance,
it is sufficient to consider the 2$\times 2$ DME for the states $\left| \psi
_{\pm }\right\rangle .$ In this case, however, the normalization of the
effective DM of the two-level system is not conserved because of the decay
to the lower states.

The DME for the states $\left| \psi _{\pm }\right\rangle $ can be obtained
from the general formalism, as above, but this way is lengthy and the final
results can be written without any calculations. In fact, the system obeys
the damped Schr\"{o}dinger equation in the natural basis that is simpler
than the DME and has the form
\begin{eqnarray}
\frac{d}{dt}c_{-S} &=&-\frac{i}{2}\frac{W}{\hbar }c_{-S}-\frac{i}{2}\frac{%
\Delta }{\hbar }c_{m^{\prime }}  \nonumber \\
\frac{d}{dt}c_{m^{\prime }} &=&\left( \frac{i}{2}\frac{W}{\hbar }-\frac{1}{2}%
\Gamma _{m^{\prime }+1,m^{\prime }}\right) c_{m^{\prime }}-\frac{i}{2}\frac{%
\Delta }{\hbar }c_{-S},  \label{DME-DampedSchr}
\end{eqnarray}
where the level $m^{\prime }$ is damped and the level $-S$ is undamped. The
decay rate between the adjacent $m$-states in the generic MM model is given
by Eq. (A9) of Ref. \cite{chugarsch05prb} that can be rewritten as
\begin{equation}
\Gamma _{m+1,m}=\frac{\left( 2m+1\right) ^{2}l_{m+1,m}^{2}}{24\pi }\frac{%
\left( D/\hbar \right) ^{2}\left| \omega _{m+1,m}\right| ^{3}}{\Omega
_{t}^{4}}.  \label{DME-Gammammp1}
\end{equation}
Here $l_{m+1,m}$ is defined below Eq. (\ref{DME-lDef}) and $\Omega _{t}$ is
defined by Eq. (\ref{DME-OmegatDef}). In the case $m=S-1$ and $m+1=S,$ Eq. (%
\ref{DME-Gammammp1}) simplifies to the elegant form
\begin{equation}
\Gamma _{S,S-1}=\frac{S^{2}}{12\pi }\frac{\omega _{S-1,S}^{5}}{\Omega
_{t}^{4}},  \label{DME-GammaSSm1}
\end{equation}
where $\hbar \omega _{S-1,S}=\left( 2S-1\right) D.$

The DME can be obtained from Eq. (\ref{DME-DampedSchr}) by setting
\begin{equation}
\rho _{-S,-S}=c_{-S}c_{-S}^{\ast },\qquad \rho _{m^{\prime }m^{\prime
}}=c_{m^{\prime }}c_{m^{\prime }}^{\ast },\qquad \rho _{-S,m^{\prime
}}=c_{-S}c_{m^{\prime }}^{\ast }
\end{equation}
and calculating time derivatives. It has the form

\begin{eqnarray}
\frac{d}{dt}\rho _{-S,-S} &=&\frac{i}{2}\frac{\Delta }{\hbar }\left( \rho
_{-S,m^{\prime }}-\rho _{m^{\prime },-S}\right)  \nonumber \\
\frac{d}{dt}\rho _{m^{\prime },m^{\prime }} &=&-\Gamma _{m^{\prime
}+1,m^{\prime }}\rho _{m^{\prime },m^{\prime }}-\frac{i}{2}\frac{\Delta }{%
\hbar }\left( \rho _{-S,m^{\prime }}-\rho _{m^{\prime },-S}\right)  \nonumber
\\
\frac{d}{dt}\rho _{-S,m^{\prime }} &=&\left( -i\frac{W}{\hbar }-\frac{1}{2}%
\Gamma _{m^{\prime }+1,m^{\prime }}\right) \rho _{-S,m^{\prime }}+\frac{i}{2}%
\frac{\Delta }{\hbar }\left( \rho _{-S,-S}-\rho _{m^{\prime },m^{\prime
}}\right)  \label{DME-TwoLevsNaturalnearRes}
\end{eqnarray}
that coincides with the results of Ref. \cite{garchu97prb} (In the latter
the precession goes in the wrong direction, however). It should be stressed
once more that this tunneling DME is non-secular.

Of course Eq. (\ref{DME-DampedSchr}) is easier to solve than Eq. (\ref
{DME-TwoLevsNaturalnearRes}). We search for the solution of Eq. (\ref
{DME-DampedSchr}) in the form $e^{-\lambda t}.$ Eigenvalues $\lambda $ of
Eq. (\ref{DME-DampedSchr}) satisfy the equation
\begin{equation}
\left|
\begin{array}{cc}
-\lambda +\frac{i}{2}\frac{W}{\hbar } & \frac{i}{2}\frac{\Delta }{\hbar } \\
\frac{i}{2}\frac{\Delta }{\hbar } & -\lambda -\frac{i}{2}\frac{W}{\hbar }+%
\frac{1}{2}\Gamma _{m^{\prime }+1,m^{\prime }}
\end{array}
\right| =0
\end{equation}
or, with the use of Eq. (\ref{DME-hbaromegaDef}),
\begin{eqnarray}
0 &=&\left( -\lambda +\frac{i}{2}\frac{W}{\hbar }\right) \left( -\lambda -%
\frac{i}{2}\frac{W}{\hbar }+\frac{1}{2}\Gamma _{m^{\prime }+1,m^{\prime
}}\right) +\frac{1}{4}\left( \frac{\Delta }{\hbar }\right) ^{2}  \nonumber \\
&=&\lambda ^{2}-\frac{1}{2}\Gamma _{m^{\prime }+1,m^{\prime }}\lambda +\frac{%
1}{4}\omega _{0}^{2}+\frac{i}{4}\frac{W}{\hbar }\Gamma _{m^{\prime
}+1,m^{\prime }}
\end{eqnarray}
that yields
\begin{eqnarray}
\lambda _{\pm } &=&\frac{1}{2}\left( \frac{1}{2}\Gamma _{m^{\prime
}+1,m^{\prime }}\pm \sqrt{\frac{1}{4}\Gamma _{m^{\prime }+1,m^{\prime
}}^{2}-\omega _{0}^{2}-i\frac{W}{\hbar }\Gamma _{m^{\prime }+1,m^{\prime }}}%
\right)  \nonumber \\
&=&\frac{1}{2}\left( \frac{1}{2}\Gamma _{m^{\prime }+1,m^{\prime }}\pm i%
\sqrt{-\frac{1}{4}\Gamma _{m^{\prime }+1,m^{\prime }}^{2}+\omega _{0}^{2}+i%
\frac{W}{\hbar }\Gamma _{m^{\prime }+1,m^{\prime }}}\right) .
\label{DME-lampmEigen}
\end{eqnarray}
The solution of Eq. (\ref{DME-DampedSchr}) has the form
\begin{eqnarray}
c_{-S}(t) &=&a_{-S}^{(+)}e^{-\lambda _{+}t}+a_{-S}^{(-)}e^{-\lambda _{-}t}
\nonumber \\
c_{m^{\prime }}(t) &=&a_{m^{\prime }}^{(+)}e^{-\lambda _{+}t}+a_{m^{\prime
}}^{(-)}e^{-\lambda _{-}t}.  \label{DME-cSol}
\end{eqnarray}
The eigenvectors $\left( a_{-S}^{(\pm )},a_{m^{\prime }}^{(\pm )}\right) $
follow from Eq. (\ref{DME-DampedSchr}):
\begin{equation}
-\lambda _{\pm }a_{-S}^{(\pm )}=-\frac{i}{2}\frac{W}{\hbar }a_{-S}^{(\pm )}-%
\frac{i}{2}\frac{\Delta }{\hbar }a_{m^{\prime }}^{(\pm )},
\end{equation}
i.e.,
\begin{equation}
a_{m^{\prime }}^{(\pm )}=-\frac{\hbar }{\Delta }\left( \frac{W}{\hbar }%
+2i\lambda _{\pm }\right) a_{-S}^{(\pm )}.  \label{DME-ampr}
\end{equation}
Thus Eq. (\ref{DME-cSol}) can rewritten as
\begin{eqnarray}
c_{-S}(t) &=&a_{-S}^{(+)}e^{-\lambda _{+}t}+a_{-S}^{(-)}e^{-\lambda _{-}t}
\nonumber \\
c_{m^{\prime }}(t) &=&-a_{-S}^{(+)}\frac{W/\hbar +2i\lambda _{+}}{\Delta
/\hbar }e^{-\lambda _{+}t}-a_{-S}^{(-)}\frac{W/\hbar +2i\lambda _{-}}{\Delta
/\hbar }e^{-\lambda _{-}t}.
\end{eqnarray}
From the initial conditions $c_{-S}(0)=1$ and $c_{m^{\prime }}(0)=0$ one
obtains
\begin{eqnarray}
a_{-S}^{(+)}+a_{-S}^{(-)} &=&1  \nonumber \\
a_{-S}^{(+)}\left( W/\hbar +2i\lambda _{+}\right) +a_{-S}^{(-)}\left(
W/\hbar +2i\lambda _{-}\right) &=&0
\end{eqnarray}
that yields
\begin{eqnarray}
a_{-S}^{(+)} &=&\frac{1}{1-\frac{W/\hbar +2i\lambda _{+}}{W/\hbar +2i\lambda
_{-}}}=\frac{W/\hbar +2i\lambda _{-}}{2i\left( \lambda _{-}-\lambda
_{+}\right) }=\frac{iW/(2\hbar )-\lambda _{-}}{\lambda _{+}-\lambda _{-}}
\nonumber \\
a_{-S}^{(-)} &=&\frac{iW/(2\hbar )-\lambda _{+}}{\lambda _{-}-\lambda _{+}}
\end{eqnarray}
and, from Eq. (\ref{DME-ampr}),
\begin{eqnarray}
a_{m^{\prime }}^{(+)} &=&2i\frac{\hbar }{\Delta }\frac{\left( iW/(2\hbar
)-\lambda _{+}\right) \left( iW/(2\hbar )-\lambda _{-}\right) }{\lambda
_{+}-\lambda _{-}}  \nonumber \\
a_{m^{\prime }}^{(-)} &=&-a_{m^{\prime }}^{(+)}.
\end{eqnarray}
Finally one obtains
\begin{eqnarray}
c_{-S}(t) &=&\frac{1}{\lambda _{+}-\lambda _{-}}\left[ \left( \frac{i}{2}%
\frac{W}{\hbar }-\lambda _{-}\right) e^{-\lambda _{+}t}-\left( \frac{i}{2}%
\frac{W}{\hbar }-\lambda _{+}\right) e^{-\lambda _{-}t}\right]  \nonumber \\
c_{m^{\prime }}(t) &=&\frac{2i}{\lambda _{+}-\lambda _{-}}\frac{\hbar }{%
\Delta }\left( \frac{i}{2}\frac{W}{\hbar }-\lambda _{+}\right) \left( \frac{i%
}{2}\frac{W}{\hbar }-\lambda _{-}\right) \left( e^{-\lambda
_{+}t}-e^{-\lambda _{-}t}\right) .
\end{eqnarray}
Now the components of the density matrix are given by
\begin{eqnarray}
\rho _{-S,-S} &=&\frac{1}{\left| \lambda _{+}-\lambda _{-}\right| ^{2}}\left[
\left( \frac{i}{2}\frac{W}{\hbar }-\lambda _{-}\right) e^{-\lambda
_{+}t}-\left( \frac{i}{2}\frac{W}{\hbar }-\lambda _{+}\right) e^{-\lambda
_{-}t}\right]  \nonumber \\
&&\qquad \qquad \times \left[ \left( -\frac{i}{2}\frac{W}{\hbar }-\lambda
_{-}^{\ast }\right) e^{-\lambda _{+}^{\ast }t}-\left( -\frac{i}{2}\frac{W}{%
\hbar }-\lambda _{+}^{\ast }\right) e^{-\lambda _{-}^{\ast }t}\right]
\nonumber \\
&=&\frac{1}{\left| \lambda _{+}-\lambda _{-}\right| ^{2}}\left\{ \left|
\frac{i}{2}\frac{W}{\hbar }-\lambda _{-}\right| ^{2}e^{-2\func{Re}(\lambda
_{+})t}+\left| \frac{i}{2}\frac{W}{\hbar }-\lambda _{+}\right| ^{2}e^{-2%
\func{Re}(\lambda _{-})t}\right.  \nonumber \\
&&\qquad -\left. 2\func{Re}\left[ \left( \frac{i}{2}\frac{W}{\hbar }-\lambda
_{-}\right) \left( -\frac{i}{2}\frac{W}{\hbar }-\lambda _{+}^{\ast }\right)
e^{-\left( \lambda _{+}+\lambda _{-}^{\ast }\right) t}\right] \right\}
\label{DME-rhamSmStRes}
\end{eqnarray}
and
\begin{equation}
\rho _{m^{\prime }m^{\prime }}=\left( \frac{\hbar }{\Delta }\right) ^{2}%
\frac{4\left| \frac{i}{2}\frac{W}{\hbar }-\lambda _{-}\right| ^{2}\left|
\frac{i}{2}\frac{W}{\hbar }-\lambda _{+}\right| ^{2}}{\left| \lambda
_{+}-\lambda _{-}\right| ^{2}}\left[ e^{-2\func{Re}(\lambda _{+})t}+e^{-2%
\func{Re}(\lambda _{-})t}-2\func{Re}\left( e^{-\left( \lambda _{+}+\lambda
_{-}^{\ast }\right) t}\right) \right] .  \label{DME-rhomprmprRes}
\end{equation}
One can see that the relaxation is described by three relaxation rates, $2%
\func{Re}(\lambda _{\pm })$ and $\func{Re}(\lambda _{+}+\lambda _{-}).$ In
addition, there are oscillations with the frequency $\func{Im}(\lambda
_{+}-\lambda _{-})$ corresponding to quantum transitions $\left|
-S\right\rangle \rightleftharpoons \left| m^{\prime }\right\rangle .$

Spin polarization in our low-temperature tunneling process is given by
\begin{equation}
\left\langle S_{z}\right\rangle =-S\rho _{-S,-S}+\sum_{m=m^{\prime
}}^{S}m\rho _{mm}.
\end{equation}
As the states with $m=m^{\prime }+1,\ldots ,S-1$ decay faster than $\left|
m^{\prime }\right\rangle ,$ their contribution can be neglected. Then one
can write
\begin{eqnarray}
\left\langle S_{z}\right\rangle &=&-S\rho _{-S,-S}+m^{\prime }\rho
_{m^{\prime }m^{\prime }}+S\rho _{SS}  \nonumber \\
&=&-S\rho _{-S,-S}+m^{\prime }\rho _{m^{\prime }m^{\prime }}+S(1-\rho
_{-S,-S}-\rho _{m^{\prime }m^{\prime }})  \nonumber \\
&=&S-2S\rho _{-S,-S}-(S-m^{\prime })\rho _{m^{\prime }m^{\prime }}.
\end{eqnarray}
As both $\rho _{-S,-S}$ and $\rho _{m^{\prime }m^{\prime }}$ go
asymptotically to zero, the magnetization change in the process is $%
\left\langle S_{z}\right\rangle _{0}-\left\langle S_{z}\right\rangle
_{\infty }=-2S$, and
\begin{equation}
\left\langle S_{z}\right\rangle _{t}-\left\langle S_{z}\right\rangle
_{\infty }=-2S\rho _{-S,-S}-(S-m^{\prime })\rho _{m^{\prime }m^{\prime }}.
\end{equation}
Thus the integral relaxation time is given by
\begin{equation}
\tau _{\mathrm{int}}=\frac{\int_{0}^{\infty }dt\left( \left\langle
S_{z}\right\rangle _{t}-\left\langle S_{z}\right\rangle _{\infty }\right) }{%
\left\langle S_{z}\right\rangle _{0}-\left\langle S_{z}\right\rangle
_{\infty }}=\int_{0}^{\infty }dt\left( \rho _{-S,-S}(t)+\frac{S-m^{\prime }}{%
2S}\rho _{m^{\prime }m^{\prime }}(t)\right) .  \label{DME-tauintexcres}
\end{equation}
For a small bias, say, $m^{\prime }=S-1$ and large spin$,$ the contribution
of the second term is small. Using Eq. (\ref{DME-rhamSmStRes}), one obtains
\begin{equation}
\int_{0}^{\infty }dt\rho _{-S,-S}(t)=\frac{1}{\left| \lambda _{+}-\lambda
_{-}\right| ^{2}}\left\{ \frac{\left| \frac{i}{2}\frac{W}{\hbar }-\lambda
_{-}\right| ^{2}}{2\func{Re}(\lambda _{+})}+\frac{\left| \frac{i}{2}\frac{W}{%
\hbar }-\lambda _{+}\right| ^{2}}{2\func{Re}(\lambda _{-})}-2\func{Re}\frac{%
\left( \frac{i}{2}\frac{W}{\hbar }-\lambda _{-}\right) \left( -\frac{i}{2}%
\frac{W}{\hbar }-\lambda _{+}^{\ast }\right) }{\lambda _{+}+\lambda
_{-}^{\ast }}\right\}  \label{DME-intrhoSmSt}
\end{equation}
and
\begin{equation}
\int_{0}^{\infty }dt\rho _{m^{\prime }m^{\prime }}(t)=\left( \frac{\hbar }{%
\Delta }\right) ^{2}\frac{4\left| \frac{i}{2}\frac{W}{\hbar }-\lambda
_{-}\right| ^{2}\left| \frac{i}{2}\frac{W}{\hbar }-\lambda _{+}\right| ^{2}}{%
\left| \lambda _{+}-\lambda _{-}\right| ^{2}}\left\{ \frac{1}{2\func{Re}%
(\lambda _{+})}+\frac{1}{2\func{Re}(\lambda _{-})}-\func{Re}\frac{2}{\lambda
_{+}+\lambda _{-}^{\ast }}\right\} .  \label{DME-intrhomprmpr}
\end{equation}
Mathematica-aided simplification yields
\begin{equation}
\int_{0}^{\infty }dt\rho _{-S,-S}(t)=\frac{4W+\Delta ^{2}+\Gamma _{m^{\prime
}+1,m^{\prime }}^{2}}{\Gamma _{m^{\prime }+1,m^{\prime }}\Delta ^{2}}
\end{equation}
and
\begin{equation}
\int_{0}^{\infty }dt\rho _{m^{\prime }m^{\prime }}(t)=\frac{1}{\Gamma
_{m^{\prime }+1,m^{\prime }}}.
\end{equation}
Then from Eq. (\ref{DME-tauintexcres}) one obtains
\begin{equation}
\tau _{\mathrm{int}}=\frac{4W+\Delta ^{2}+\Gamma _{m^{\prime }+1,m^{\prime
}}^{2}}{\Gamma _{m^{\prime }+1,m^{\prime }}\Delta ^{2}}+\frac{S-m^{\prime }}{%
2S}\frac{1}{\Gamma _{m^{\prime }+1,m^{\prime }}}=\frac{4W+\left( 1+\frac{%
S-m^{\prime }}{2S}\right) \Delta ^{2}+\Gamma _{m^{\prime }+1,m^{\prime }}^{2}%
}{\Gamma _{m^{\prime }+1,m^{\prime }}\Delta ^{2}}.
\end{equation}
The corresponding rate can be written as
\begin{equation}
\Gamma _{\mathrm{int}}=\frac{\Delta ^{2}}{2\hbar ^{2}}\frac{\Gamma
_{m^{\prime }+1,m^{\prime }}/2}{\Omega ^{2}+\left( \Gamma _{m^{\prime
}+1,m^{\prime }}/2\right) ^{2}},\qquad \left( \hbar \Omega \right)
^{2}\equiv W^{2}+\frac{1}{4}\left( 1+\frac{S-m^{\prime }}{2S}\right) \Delta
^{2}.  \label{DME-Gammaint-ge}
\end{equation}
In comparison to the overdamped result of Ref. \cite{garchu97prb}, this
formula contains an additional term $\sim \Delta ^{2}$ in the denominator.
In the underdamped case $\Delta \gg \hbar \Gamma _{m^{\prime }+1,m^{\prime
}} $ at resonance $W=0,$ the rate is given by
\begin{equation}
\Gamma _{\mathrm{int}}=\Gamma _{m^{\prime }+1,m^{\prime }}/\left( 1+\frac{%
S-m^{\prime }}{2S}\right)
\end{equation}
that is of the order of thermal activation rate at high temperatures. This
means that the barrier is completely cut at resonance and the relaxation
rate does not significantly increase with temperature. Condition $\Delta \gg
\hbar \Gamma _{m^{\prime }+1,m^{\prime }}$ makes the secular approximation
valid, unlike the overdamped case $\Delta \lesssim \hbar \Gamma _{m^{\prime
}+1,m^{\prime }},$ where the secular approximation leads to unphysically
large relaxation rates at resonance.

\subsection{Numerical implementation and illustrations}

\begin{figure}[t]
\centerline{%
\includegraphics[angle=-90,width=12cm]{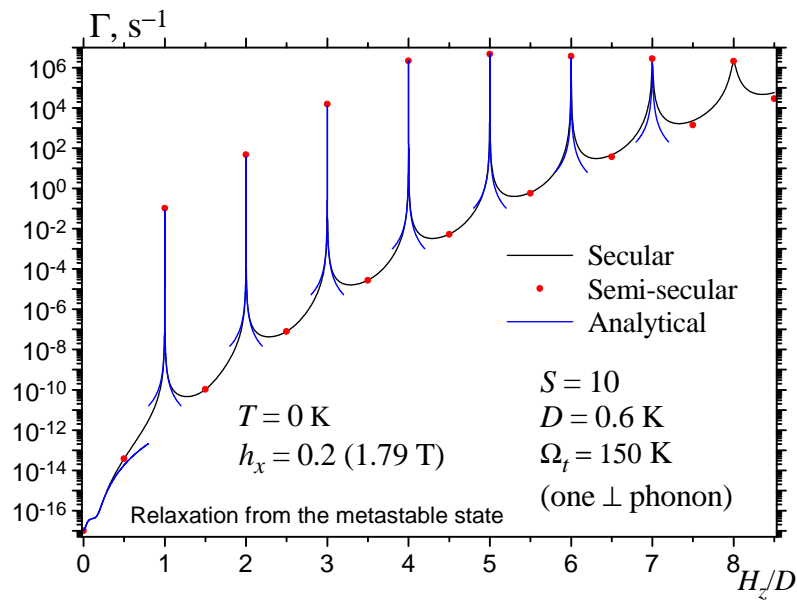}}
\caption{Zero-temperature escape rate from the metastable state vs the bias
field in the generic model of MM.}
\label{Fig-generic_hx=0.2_T=0_vs_Hz}
\end{figure}
\begin{figure}[t]
\centerline{%
\includegraphics[angle=-90,width=12cm]{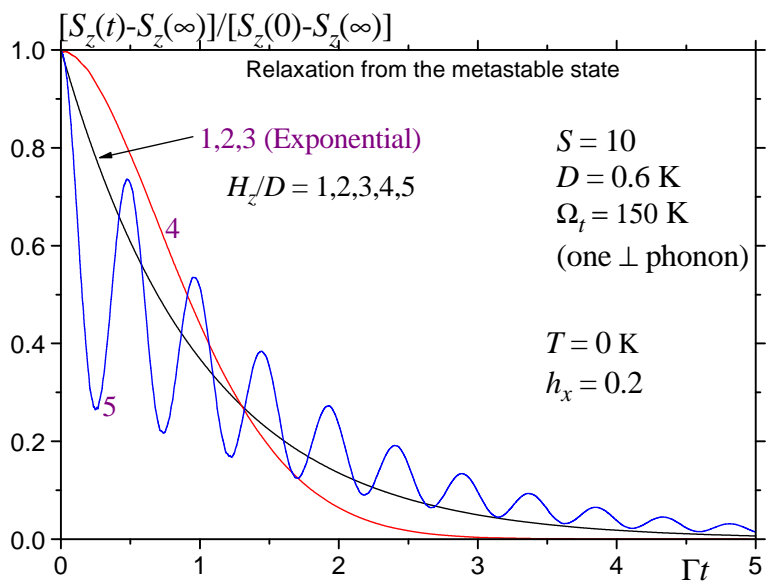}}
\caption{Time evolution of $\left\langle S_{z}\right\rangle $ at over- and
underdamped tunneling resonances in previous figure.}
\label{Fig-generic_hx=0.2_T=0_vs_t}
\end{figure}

In this section some representative results of the numerical solution of the
DME for molecular magnets are shown. The code implemented in Wolfram
Mathematica is available from the author. It is based on the diagonalization
of the dynamical matrix $\Phi $ of the DME for the free-evolution problem,
as described in Sec. \ref{Sec-Free-evolution}. Linear dynamical
susceptibility, Sec. \ref{Sec-linear-response}, can also be obtained on this
way. As the tunnel splitting $\Delta $ can be very small, especially in Mn$%
_{12}$ ($S=10)$ at small transverse fields, matrix algebra requires using a
high custom precision, typically 100 significant figures. For Ni$_{4}$ ($%
S=4) $ standard precision is sufficient. Custom precision makes computation
slower. Still, the DME in the secular approximation involving the $\left(
2S+1\right) \times \left( 2S+1\right) $ dynamical matrix solves fast enough
on a standard PC. Solution of the full non-secular DME is very slow for $%
S=10 $ because of the big size $\left( 2S+1\right) ^{2}\times \left(
2S+1\right) ^{2}$ of the dynamical matrix. It is very important to use the
semi-secular DME that is no less accurate than the full non-secular DME but
has the dynamical matrix of the size $\left( 6S+1\right) \times \left(
6S+1\right) .$ As a result, the solution, although slower than that of the
secular DME, is still realistically fast. The difference between the secular
and semi-secular versions of the method is confined to the close vicinity of
the overdamped tunneling resonances, $\Delta \lesssim \hbar \Gamma
_{m^{\prime }+1,m^{\prime }},$ while everywhere else the numerical results
are the same.

Numerical solution shows that the dynamical matrix $\Phi $ of the
non-secular DME has exactly $2S+1$ real eigenvalues out of the total $\left(
2S+1\right) ^{2}$ eigenvalues. One of real eigenvalues is zero and
corresponds to the thermal equilibrium. Complex eigenvalues occur in complex
conjugate pairs. This behavior is similar to that of the secular DME and
requires explanation. At low temperatures in the regime of thermal
activation and weak tunneling (overdamped resonances) there is one nonzero
real eigenvalue that is much smaller than all other real eigenvalues and
real parts of complex eigenvalues. In the case of underdamped resonances,
there are three eigenvalues, one real and two complex, that describe the
slow dynamics.

Fig. \ref{Fig-generic_hx=0.2_T=0_vs_Hz} shows the zero-temperature escape
rate vs the bias field $H_{z}$ in the generic MM model with $B=0$ in Eq. (%
\ref{DME-epsilonm}). The striking feature is the spin tunneling at resonance
fields that leads to the increase of the escape rate by many orders of
magnitude. Most of the points have been obtained from the secular DME, the
points at resonances and between them have been obtained from the
semi-secular DME, and the analytical result of Eq. (\ref{DME-Gammaint-ge})
is drawn in the vicinity of resonances. Near the zero-field resonance, Eq. (%
\ref{DME-GammapmHx}) is shown. The characteristic ``shoulder'' described by
this equation is well reproduced by the numerical result. As mentioned
above, the secular approximation can yield unphysically high escape rates at
resonances but the resonances are narrow and there are no secular points
that hit them. Resonances with $k=1,2,$ and 3 are overdamped and can be
approximately described by the DME solution based on effective resistances
method of Ref. \cite{garchu97prb}. As explained in Ref. \cite{garchu97prb},
the barrier at resonances is lower than the full classical barrier, and its
height corresponds to the so-called \emph{blocking level} for which $\Delta
_{mm^{\prime }}\sim \hbar \left( \Gamma _{m^{\prime }+1,m^{\prime }}+\Gamma
_{m-1,m}\right) .$ Resonances with $k\geq 4$ in Fig. \ref
{Fig-generic_hx=0.2_T=0_vs_Hz} are underdamped, so that the blocking level
is the ground state and the barrier is reduced to zero. Accordingly, the
escape rate at these resonances is of order $\Gamma \sim 3\times $10$^{6}$ s$%
^{-1}$ and it coincides with the spin-phonon rate between the adjacent
levels that is the highest possible rate achievable off resonance at
temperatures exceeding the energy barrier.

\begin{figure}[t]
\centerline{%
\includegraphics[angle=-90,width=12cm]{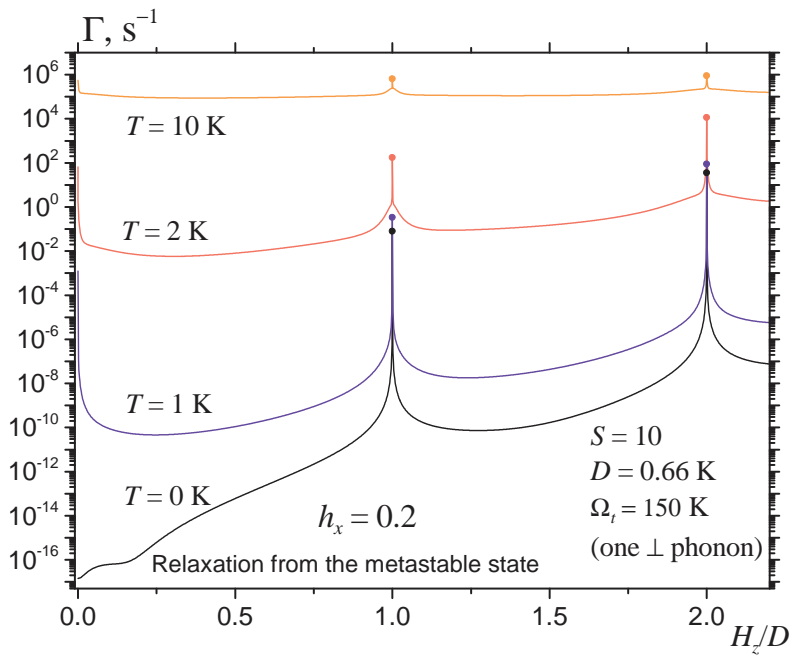}}
\caption{Escape rate vs bias field in the generic model of MM at different
temperatures.}
\label{Fig-Generic-Gamma_vs_Hz}
\end{figure}
\begin{figure}[t]
\centerline{%
\includegraphics[angle=-90,width=12cm]{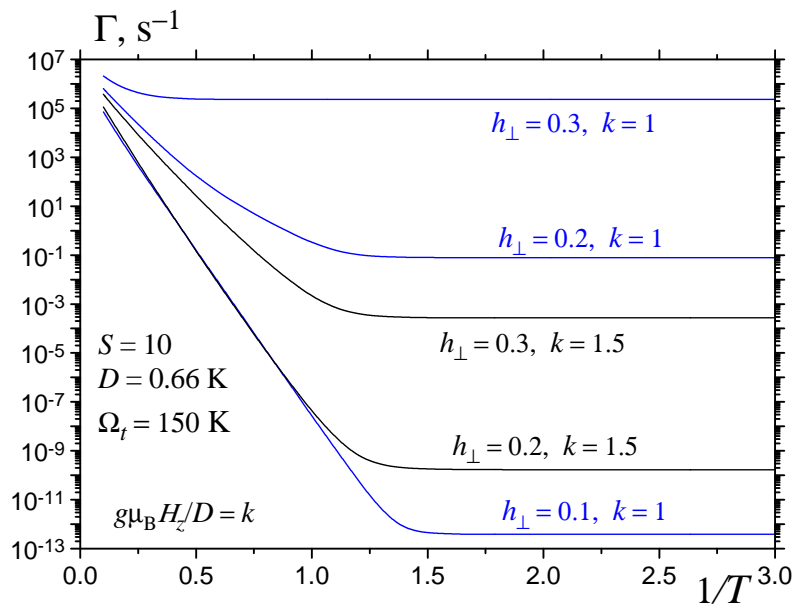}}
\caption{Temperature dependence of the escape rate in the generic model on
and off resonance.}
\label{Fig-generic_Gamma_vs_Tinv}
\end{figure}
\begin{figure}[t]
\centerline{%
\includegraphics[angle=-90,width=12cm]{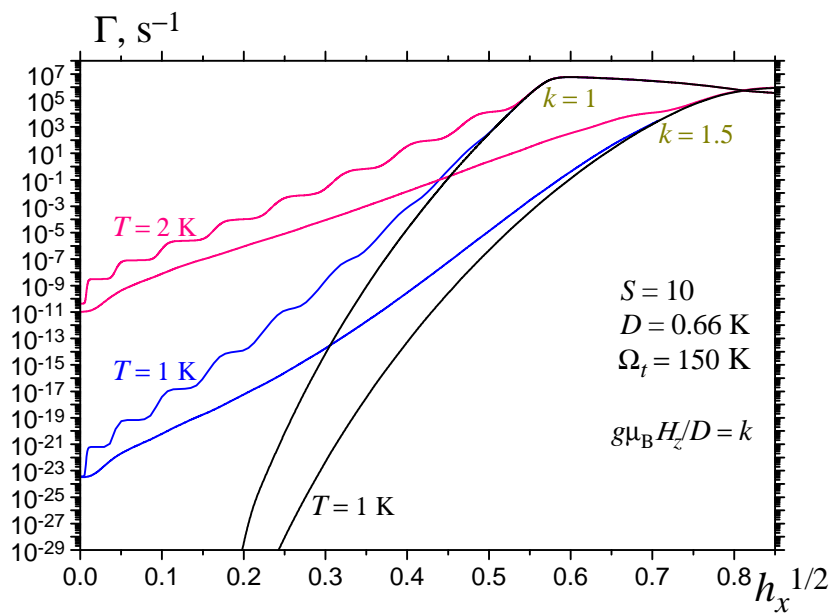}}
\caption{Escape rate vs transverse field in the generic model of MM at
different temperatures, on and off resonance.}
\label{Fig-generic_Gamma_vs_hx}
\end{figure}

Fig. \ref{Fig-generic_hx=0.2_T=0_vs_t} shows the time dependence of spin
polarization $\left\langle S_{z}\right\rangle $ at different resonances in
Fig. \ref{Fig-generic_hx=0.2_T=0_vs_Hz}. The relaxation is exponential for
the overdamped resonances, as well as off resonance (not shown). To the
contrast, at underdamped resonances with $k\geq 5$ there are damped
oscillations described by three different relaxation rates in Eq. (\ref
{DME-rhamSmStRes}). In the case of exponential relaxation, it is sufficient
to identify the escape rate with the smallest real eigenvalue of the
dynamical matrix. In the case of underdamped resonances there are three slow
eigenvalues, and obtaining the correct value of the escape rate requires
using the integral relaxation time.

Escape rate vs the bias field at different temperatures in the generic model
is shown in Fig. \ref{Fig-Generic-Gamma_vs_Hz}. All data were obtained from
the numerical solution of the semi-secular DME. The anisotropy value $D=0.66$
K has been chosen to fit the barrier height in Mn$_{12}$ (see below). As
expected, the escape rate increases with temperature, faster off resonance
than on resonance. One can see (especially clear for $T=2$ K and $k=1)$ that
at nonzero temperatures the tunneling peak may consist of several peaks of
different width on the top of each other \cite{garchu97prb}. Broad peaks
correspond to tunneling at high energy with a large splitting $\Delta $
while narrow peaks correspond to tunneling via a low-lying resonant pair of
levels with small $\Delta $. At zero temperatures the zero-bias tunneling
peak is very narrow because of the anomalously small damping of the
ground-state levels and it is not seen in the plot. However, for nonzero
temperatures this peak becomes broad as all other peaks because of tunneling
via excited levels that are regularly damped via decay to the lower-lying
levels.

Arrhenius plot in Fig. \ref{Fig-generic_Gamma_vs_Tinv} shows transition
between the thermal-activation and ground-state-tunneling regimes on and off
resonance for different transverse fields parametrized by $h_{x}=g\mu _{%
\mathrm{B}}H_{x}/\left( 2SD\right) .$ For small transverse fields, the
resonances are overdamped and ground-state tunneling is small. In this case
the activation part of the plot is nearly a straight line with the slope
corresponding to a particular effective barrier. The transition to the
horizontal line describing the ground-state tunneling has little rounding.
This is the so-called first-order transition between thermal activation and
ground-state tunneling, the two competing channels \cite
{garchu97prb,chugar97prl}. For $h_{x}=0.2$ at resonance, the activation part
of the plot is noticeably curved. This is a manifestation of the
second-order transition in which the dominating tunneling level gradually
moves down with lowering temperature, effectively decreasing the barrier
height and the slope of the curve. For $h_{x}=0.3$ at resonance, the barrier
is reduced to nearly zero and the ground-state tunneling is very strong.

Fig. \ref{Fig-generic_Gamma_vs_hx} shows the dependence of the escape rate
on the transverse field $h_{x}$ at different temperatures on and off
resonance. On resonance at nonzero temperatures, there are characteristic
steps arising as a result of moving the blocking level up or down the energy
\cite{garchu97prb}. This phenomenon can be seen in Fig. 5 of Ref. \cite
{garchu97prb}, obtained by the effective resistances method. One can see
that on resonance, $k=1,$ the barrier goes to zero with increasing $h_{x},$
so that above some critical value of $h_{x}$ curves corresponding to
different temperatures merge at the level of the highest possible rate. At
these transverse fields the barrier off resonance still exists since the
curves corresponding to different temperatures merge at higher values of $%
h_{x}.$

\begin{figure}[t]
\centerline{%
\includegraphics[angle=-90,width=12cm]{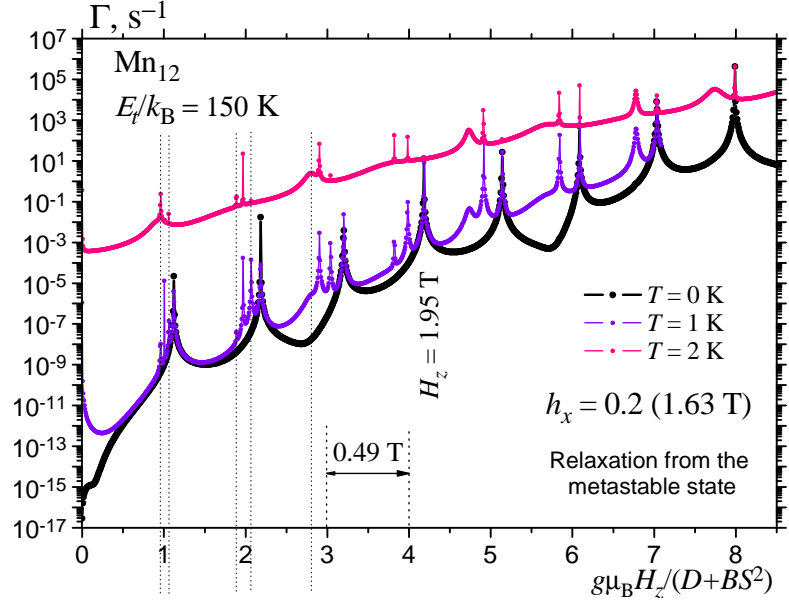}}
\caption{Escape rate vs bias field in Mn$_{12}$ at different temperatures.}
\label{Fig-Mn12_hx=0.2_vs_Hz}
\end{figure}
\begin{figure}[t]
\centerline{%
\includegraphics[angle=-90,width=12cm]{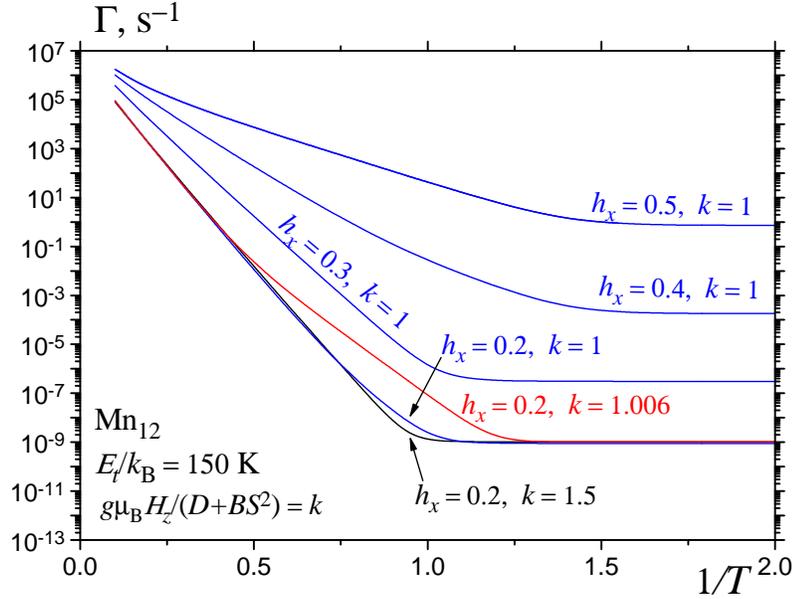}}
\caption{Temperature dependence of the escape rate in Mn$_{12}$.}
\label{Fig-Mn12_Gamma_vs_Tinv}
\end{figure}

Next figures show the numerical results for Mn$_{12}.$ Because of the
quartic uniaxial anisotropy $B,$ tunneling peaks in Fig. \ref
{Fig-Mn12_hx=0.2_vs_Hz} are split, as explained in the comment after Eq. (%
\ref{DME-mprDef}). The rightmost big peaks correspond to the ground-state
tunneling, and smaller peaks to the left of them, seen at nonzero
temperatures, are due to tunneling via excited states. Graphed results of
earlier calculations of this kind for Mn$_{12}$ can be found in Refs. \cite
{luibarfer98prb,leulos00prb}. In comparison to the results for the generic
model with the same barrier 66 K in Fig. \ref{Fig-generic_Gamma_vs_Tinv}, Mn$%
_{12}$ shows ground-state tunneling up to higher temperatures.

Temperature dependences of the escape rate in Fig. \ref
{Fig-Mn12_Gamma_vs_Tinv} are different for different bias fields. If for a
given $H_{z}$ there is a tunneling resonance at some energy between the top
of the classical barrier and the ground-state, thermally activated tunneling
via this resonant pair is a relaxation channel competing with the two
channels considered above. As a result, there are two different slopes in
the Arrhenius part of the plot, such as for $k=1.006.$ This value of $k$
corresponds to the high blue peak in Fig. \ref{Fig-Mn12_hx=0.2_vs_Hz} that
disappears at $T=0.$

\subsection{Discussion}

Existing work on molecular magnets using the density matrix equation can be
split up into two groups: (i) using the natural or $m$-basis and (ii) using
the diagonal basis. In all known cases the DME is reduced to the system of
rate equations for the diagonal DM elements, the level populations. Using
the natural basis is justified if the terms in the spin Hamiltonian that are
non-commuting with $S_{z}$ are a small perturbation. However, even a small
non-commiting perturbation can severely distort the levels near the top of
the barrier that are mostly inportant in thermal activation. On the other
hand, tunneling via robuster low-lying levels at low temperatures can be
well described perturbatively in the $m$-basis.

In Ref. \cite{gar97pre} the thermal activation rate of a generic MM was
calculated in the $m$-basis \emph{in the absence of tunneling} via the \emph{%
integral relaxation time}. Tunneling has been taken into account in Ref.
\cite{garchu97prb} by adiabatically eliminating fast nondiagonal DM elements
that amounts to using the high-order perturbation theory in calculating
tunnel splittings \cite{gar91jpa}. The resulting system of rate equations
with resonance tunneling was solved by the method of \emph{effective
resistances} \cite{garchu97prb} using the idea of the solution of the
Fokker-Planck equation at low temperatures in the classical case. Later the
system of rate equations in the $m$-basis was employed in Refs. \cite
{foretal98,leulos00prb,parketal01prb}.

In particular, Ref. \cite{leulos00prb} repeats the steps of Ref. \cite
{garchu97prb} using the realistic model of Mn$_{12}$ with $B\neq 0$ in Eq. (%
\ref{DME-HA}). A new element of Ref. \cite{leulos00prb} is the erroneous
consideration of spin-phonon interactions leading to the spin-phonon
coupling of the type $D\left( S_{+}^{2}+S_{-}^{2}\right) \left( \epsilon
_{xx}-\epsilon _{yy}\right) $, $\epsilon _{\alpha \alpha }$ being components
of the deformation tensor, because of tilting the easy axis by transverse
phonons at \emph{second} order in small tilting angle $\delta \varphi .$
This leads to nonexistent direct processes with changing $m$ by 2. In fact,
as we have seen above, second-order terms in $\delta \varphi ,$ Eq. (\ref
{DME-V2}), give rise to Raman processes rather than to direct processes. The
error made in Ref. \cite{leulos00prb}, neglection of a part of $\delta
\varphi ^{2}$ terms that cancel the result, has been explained in Ref. \cite
{chugar00epl}. Nevertheless, the appeal of $\Delta m=2$ direct processes has
been remaining strong, so that the relevance of Eq. (A12) of Ref. \cite
{leulos00prb} for explanation of experiments on molecular magnets is still
disputable. The recent examples are experimental works on Fe$_{8},$ Refs.
\cite{bahpetmoswer07prb,baletal08epl}. Whereas in Ref. \cite
{bahpetmoswer07prb} Eq. (A12) of Ref. \cite{leulos00prb} is used with
success, Ref. \cite{baletal08epl} states that direct processes with $\Delta m=2$
do not fit the data. On the other hand, for Fe$_{8}$ these
processes were shown to arise from rotations around the easy axis, the
corresponding coupling constant being the transverse anisotropy $E,$ see Eq.
(B5) of Ref. \cite{chugarsch05prb}.

Moving to the universal form of the DME proposed here can help to end the
confusion about what to include into the relaxation terms, since the latter
automatically follow from the spin Hamiltonian and there is no freedom to
make a mistake. Moreover, in the diagonal basis used in the universal DME
there are phonon-induced transitions between \emph{all} exact energy levels,
not only between the nearest or second-nearest neighbors. An example is the
relaxation rate between the levels of the ground-state doublet, Eq. (\ref
{DME-GammapmHx}) and Eq. (69) of Ref. \cite{chugarsch05prb}.

Among the works using the DME in the natural basis, in the secular
approximation, are Refs. \cite{luibarfer98prb,baletal08epl}. The authors say
that the advantage of this method is that spin tunneling is absorbed in the
exact basis states. This is overall true, although the full dynamical
description requires taking into account the decoupled nondiagonal DM
elements, in addition to the system \ of rate equations that was used. Also
in the case of weak tunneling (overdamped tunneling resonances) the secular
approximation fails and results in unphysically large escape rates at
resonances. This was explained at the beginning of Sec. \ref{Sec-MM-DME}, as
well as in the comments below Eq. (19) of Ref. \cite{luibarfer98prb} and
above Eq. (2) of Ref. \cite{baletal08epl}. Certainly something is missing if
spin tunneling is automatically incorporated into the exact basis states
but, in spite of it, one cannot approach tunneling resonances. The solution is to
use the non-secular or better semi-secular DME that takes into account the
dynamical coupling between the diagonal and slow nondiagonal DM elements and
is thus valid everywhere. It  should be stressed that the system of rate
equations with tunneling in the $m$-basis is essentially non-secular and for
this reason it does not fail at resonances.

\section*{Acknowledgments}

The author thanks Gr\'{e}goire de Loubens and Reem Jaafar for critically
reading the manuscript and E. M. Chudnovsky and J. R. Friedman for
stimulating discussions. This work has been supported by the NSF Grant No. DMR-0703639. 
D. A. Garanin is a Cottrell Scholar of Research Corporation.
%

\newpage
\bibliography{gar-own,chu-own,gar-tunneling,gar-relaxation,gar-oldworks,gar-books,gar-general}

\end{document}